\documentclass[final,paper]{jfm}

\usepackage{graphicx}
\usepackage{multirow}%
\usepackage{amsmath}
\usepackage{newtxtext}
\usepackage{overpic}
\usepackage{booktabs}%
\usepackage{newtxmath}
\usepackage{natbib}
\usepackage{bm}
\usepackage{hyperref}
\hypersetup{
    colorlinks = true,
    urlcolor   = blue,
    linkcolor = blue,
    citecolor  = blue,
}

\newcommand{\RomanNumeralCaps}[1]
\linenumbers

\graphicspath{{Figures/}}

\title{Numerical simulations of attachment-line boundary layer in hypersonic flow, Part I: roughness-induced subcritical transitions}

\author{Youcheng, Xi\aff{1}
  \corresp{\email{xiyc@mail.tsinghua.edu.cn}}, 
  Bowen, Yan\aff{2}, Guangwen Yang\aff{2},
  Xinguo, Sha\aff{3}, Dehua, Zhu\aff{3}, 
  \and Song, Fu\aff{1}}

\affiliation{\aff{1}School of Aerospace Engineering, Tsinghua University, 100084 Beijing, China
\aff{2}Institute of High Performance Computing, Department of Computer Science and Technology, Tsinghua University, 100084 Beijing, China
\aff{3}China Academy of Aerospace Aerodynamics, 100074 Beijing, China}

\begin{document}
\maketitle

\begin{abstract}
The attachment-line boundary layer is critical in hypersonic flows because of its significant impact on heat transfer and aerodynamic performance. 
In this study, high-fidelity numerical simulations are conducted to analyze the subcritical roughness-induced laminar-turbulent transition at the leading-edge attachment-line boundary layer of a blunt swept body under hypersonic conditions. 
This simulation represents a significant advancement by successfully reproducing the complete leading-edge contamination process induced by surface roughness elements in a realistic configuration, thereby providing previously unattainable insights.
Two roughness elements of different heights are examined. 
For the lower-height roughness element, additional unsteady perturbations are required to trigger a transition in the wake, suggesting that the flow field around the roughness element acts as a disturbance amplifier for upstream perturbations. 
Conversely, a higher roughness element can independently induce the transition. 
A low-frequency absolute instability is detected behind the roughness, leading to the formation of streaks. 
The secondary instabilities of these streaks are identified as the direct cause of the final transition.
\end{abstract}


{\bf MSC Codes }  {\it(Optional)} Please enter your MSC Codes here

\section{Introduction}\label{sec1}
The subcritical transition of leading-edge boundary layer near the attachment line of swept wings plays an important role in aerodynamic, which means the boundary layer may undergo transition to turbulence below the critical Reynolds number predicted by linear stability theory(LST). This phonomenon is especially critical because turbulent flow that starts at the leading edge of a swept wing can propagate downstream, affecting extensive regions of the wing's chord and compromising its overall aerodynamic performance.

As the actual flow is three-dimensional in nature, to simplify the problem, it is common to employ the swept Hiemenz boundary layer past a flat plate\citep{Rosenhead1963, Schlichting2017} as an approximation model for the actual three-dimensional boundary layer\citep{Poll1979,Hall1984,Theofilis1998,Theofilis2003} around the leading edge. 
Based on this model, the LST performed by \citet{Hall1984} gives a linear critical Reynolds number of $Re_{crit}\approx 583.1$, which is in good agreement with the previous experimental finding \citep{Poll1979} as well as the numerical simulation of  \citet{Spalart1988}. 
However, in many experimental tests\citep{Gaster1967,Poll1979,Arnal1997}, transitions are often observed at a significantly lower value $Re_{tr} \approx 250$, if the boundary layer is subject to sufficiently large external disturbulences. In order to understand the discrepancy between linear stability results and experimental findings, finite amplitude perturbations and nonlinear processes have to be taken into account. \citet{Obrist2012} and \citet{John2014,John2016} carried out direct numerical simulations on a swept Hiemenz boundary layer with a pair of stationary counter-rotating streamwise vortex-like structures with finite amplitude. A bypass transition scenario has been identified, which can explain the occurrence of subcritical transition in experiments. The initial pair of stationary counter-rotating vortex-like structures lead to the transient growth of streaks according to the lift-up effect, and then the damped primary vortices and streaks interacts with unsteady secondary perturbations, causing secondary instabilities and leading to the final transition to turbulence.

However, the aforementioned conclusions are based on simplified model in incompressible flow only. When compressible effects (such as Mach number, shock waves, wall temperature, etc.) are taken into account, the problem becomes significantly more complex. Based on previous studies\citep{Theofilis2006,LiFei2008,Mack2008,Xi2021a,Xi2021b,Fedorov2022}, for large sweep Mach numbers, the attachment-line mode is inviscid in
nature, while for lower sweep Mach numbers, the attachment-line instability exhibits the
behaviours of viscous Tollmien–Schlichting waves.
Detailed reviews for these research have been included in our previous studies\citep{Xi2021a,Xi2021b} and the connection between the linear stability features of the flow and the issues discussed in this study is not particularly direct. Therefore, we will not elaborate on them here.

In fact, experimental investigations of high-speed attachment-line flow date back to 1959. Initially, \citet{Beckwith1959} focused on the effects of sweep angles and heat flux along the attachment line in supersonic conditions. They detected the transition of attachment-line flow in their Mach 4.15 experiments, studying the effect of sweep angles over a relatively wide range. Later, \citet{Creel1986} and \citet{Chen1991} conducted experiments with a free-stream Mach number of 3.5 and various sweep angles, also detecting transition along the attachment line and finding transition Reynolds numbers around 650 (based on the boundary layer length scale at the leading edge). \citet{Skuratov1991} performed similar tests to validate Creel et al.'s results. \citet{Murakami1996} studied hypersonic attachment-line flow in a Ludwieg-tube wind tunnel. In some conditions, the bypass scenario is the most possible reasons for the transition. During the experiments, without the end plates and trip wires, the attachment-line boundary layer can keep laminar along the entire attachment line. 

\begin{figure}
  \centering
  \begin{overpic}[width=0.55\textwidth]{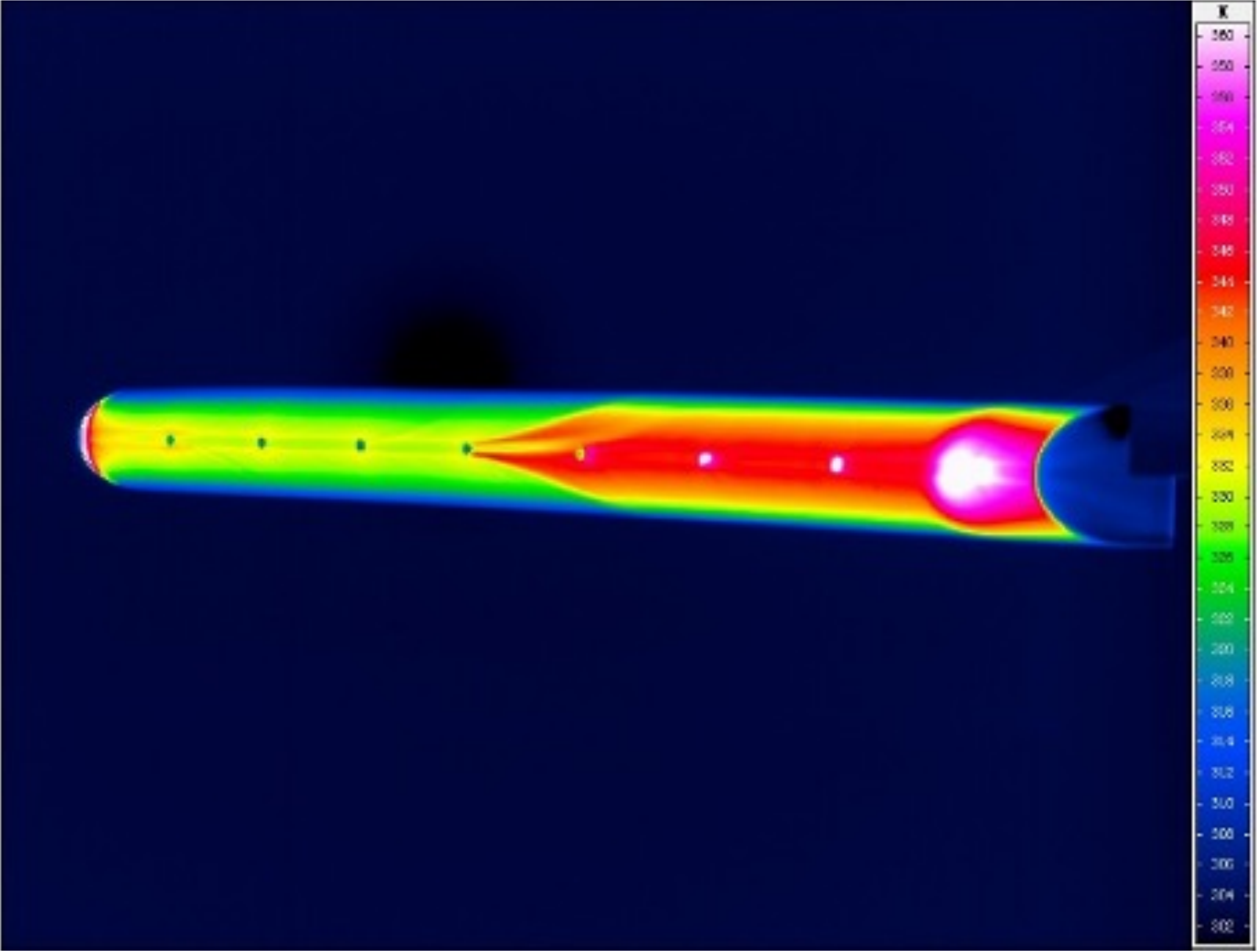}
  \end{overpic}
  \caption{Infrared measurements of the temperature distribution along the leading edge of the swept blunt body. Dots indicate the positions of the pressure sensors; pink represents high-temperature regions, while blue indicates low-temperature regions.}
  \label{Exp}
\end{figure}  

Recently, experimental tests are performed over a swept blunt leading edge, with a swept angle of $45^o$, in the FD-07 Mach 6.0 hypersonic wind tunnel of the China Academy of Aerospace Aerodynamics. 
During the experiment, despite high levels of external perturbations, in some cases, the attachment-line boundary layer remained laminar. 
When pressure sensors are mounted at the attachment-line position on the leading edge of a swept blunt body model, the surface of the model is no longer smooth. 
Due to unavoidable installation errors during the experiment, effective roughness elements, such as small protrusions or depressions, form along the attachment line. 
Experiments with this configuration have shown that disturbances induced by these roughness elements can effectively trigger the transition of the attachment-line boundary layer to turbulence, as shown in figure \ref{Exp}.

Previous studies indicate that the phenomenon of subcritical transition is highly significant in the context of attachment-line flows. However, most of these studies have been limited to incompressible flows or confined to linear analysis, leaving a significant gap in the understanding of subcritical compressible flows. Furthermore, even in incompressible flow scenarios, existing computational analyses have often employed simplified models or introduced artificial disturbances to facilitate numerical studies, raising questions about their validity under actual conditions. Therefore, it is imperative to conduct numerical investigations of the three-dimensional boundary layer at the leading edge of compressible blunt bodies under relative realistic conditions.
In this study, we perform numerical simulations of transitional high-speed attachment-line boundary layers that develop from finite amplitude initial disturbances caused by roughness element. These simulations correspond to experimental investigations of roughness-induced transition over a real blunt configuration, without assuming an infinite span. 
Unlike typical transitional studies, we have calculated the complete transition to turbulence over a real configuration. Our primary aim is to investigate the physical mechanisms of transition induced by roughness elements in three-dimensional attachment-line boundary layers at the leading edge. This paper is organized as follows. In section \ref{sec2}, the governing equations
are introduced as well as the details for numerical simulations. The results for transitional three-dimensional boundary layers are presented in section \ref{sec3} and the conclusions and some discussions are given in section \ref{sec4}.

\section{Methodology}\label{sec2}
\subsection{Governing equations}
The governing equations for all simulations in this work are the dimensionless compressible Navier–Stokes(NS) equations for a Newtonian fluid, which can be written as:
\begin{equation}
\frac{\partial Q}{\partial t}+\frac{\partial F_j}{\partial x_j} + \frac{\partial F_{j}^{v}}{\partial x_j}=0,
\end{equation}
\begin{equation}
Q = \left[
  \rho,{\rho {u_1}},{\rho {u_2}},{\rho {u_3}},{{E_t}} 
\right]^{T},
\end{equation}
\begin{equation}
{F_j} = \left[ {\begin{array}{c}
  {\rho {u_j}} \\ 
  {\rho {u_1}{u_j} + p{\delta _{1j}}} \\ 
  {\rho {u_2}{u_j} + p{\delta _{2j}}} \\ 
  {\rho {u_3}{u_j} + p{\delta _{3j}}} \\ 
  {\left( {{E_t} + p} \right){u_j}} 
\end{array}} \right],{F_{j}^{v}} = \left[ {\begin{array}{c}
  0 \\ 
  {{\tau _{1j}}} \\ 
  {{\tau _{2j}}} \\ 
  {{\tau _{3j}}} \\ 
  {{\tau _{jk}}{u_k} - {q_j}} 
\end{array}} \right].
\end{equation}
Throughout this work the coordinates $x_i, (i =1, 2, 3)$ are referred to as $x, y, z$, respectively, with corresponding velocity components $u_1 = u, u_2 = v, u_3 = w$. $F_j$ and $F_j^{v}$ stand for the inviscid and viscous flux.
The total energy \(E_t\) and the viscous stress \(\tau_{ij}\) are given as, respectively, 
\begin{equation}
\begin{aligned}
E_{t}&=\rho\left(\frac{T}{\gamma(\gamma-1) M^{2}_{\infty}}+\frac{u_{k} u_{k}}{2}\right), \\
\tau_{i j}&=\frac{\mu}{Re_{\infty}}\left(\frac{\partial u_{i}}{\partial x_{j}}+\frac{\partial u_{j}}{\partial x_{i}}-\frac{2}{3} \delta_{i j} \frac{\partial u_{k}}{\partial x_{k}}\right).
\end{aligned}
\end{equation}
The pressure \(p\) and heat flux \(q_i\) are obtained from:
\begin{equation}   \label{eq2_6}
p=\frac{\rho T}{\gamma M_{\infty}^{2}}, \quad q_{i}=-\frac{\mu}{(\gamma-1) M_{\infty}^{2} Re_{\infty} Pr} \frac{\partial T}{\partial x_{i}}.
\end{equation}
The viscosity is calculated using the Sutherland law
\begin{equation}   \label{eq2_7}
\mu = T^{3/2}\frac{T_{\infty} + C}{T\cdot T_{\infty} + C},
\end{equation}
with \(C = 110.4K\).
The free-stream Reynolds number $Re_{\infty}$, Mach number $M_{\infty}$ and Prandtl number $Pr$ are defined as
\begin{equation}
Re_{\infty} = \frac{\rho_{\infty}^* U_{\infty}^* l_0^*}{\mu_{\infty}^*}, \quad
M_{\infty} = \frac{U_{\infty}^*}{\sqrt{\gamma R_g^* T_{\infty}^*}}, \quad
Pr = 0.72,
\end{equation}
where $\rho_{\infty}^*$, $U_{\infty}^*$, $T_{\infty}^*$ and $\mu_{\infty}^*$  
stand for the freestream density, velocity, temperature and viscosity, respectively. $R_g^* = 287 \text{J}/(\text{K}\cdot \text{Kg})$ represents the gas constant and $\gamma$ stands for the ratio of specific heat. The length scale $l_0^*$ is chosen as $1$ millimeter in this research. The $*$ denotes dimensional flow parameters.

\subsection{Numerical method}

Two solvers have been employed in this study. 
The first code we use to perform computations is the high-order finite difference code developed recently at Tsinghua University. 
A shock-fitting (S-F) method \citep{Zhong1998} is used to compute steady hypersonic viscous flow together with the high-order accurate non-compact finite differences methods. 
The 5th-order upwind scheme (for inviscid flux \(F_{j}\)) and the 6th-order centre scheme (for viscous flux \(F_{vj}\)) are used to compute the flow field. 
A 4th-order Runge-Kutta method is applied for the time integration, and the simulations are performed until the maximum residual reaches a small value on the order of \(10^{-15}\). 
A full implicit scheme can also be used for fast convergence. 
Validations of the code and some applications for calorically perfect gas and thermal-chemical non-equilibrium flow can be found in our previous studies\citep{Xi2021a,Xi2021b,Chen2022}. 
The solver is used mainly to determing the location of the leading shock and give a high quality initial field. 

The second code, used in this study, is a well-validated fluid dynamic shock capture (S-C) solver OPENCFD, developed by \citet{Li2008}, which is mainly used to simulate the whole transition/turbulent processes. 
The code has been validated and verified in previous studies(\citet{Li2008,Liang2010,Li2010}). For three-dimensional calculations presented in this study, a hybrid high-order finite difference scheme, including the seventh-order upwind scheme, fifth-order and seventh-order WENO schemes\citep{Jiang1996}, together with a shock sensor\citep{Dang2022} is used for the inviscid flux in the characteristic form. Based on that formular, during the calculation, more than 98\% of the regions use the linear seventh-order upwind scheme, only a few regions corresponding to discontinuities use the nonlinear WENO schemes, which greatly increase the calculation efficiency. A standard sixth-order central difference scheme is used for viscous flux. Time advancement is achieved by the explicit third-order total variation diminishing Runge-Kutta scheme.

\subsection{Models}

\begin{figure}
  \centering
  \begin{overpic}[width=\textwidth]{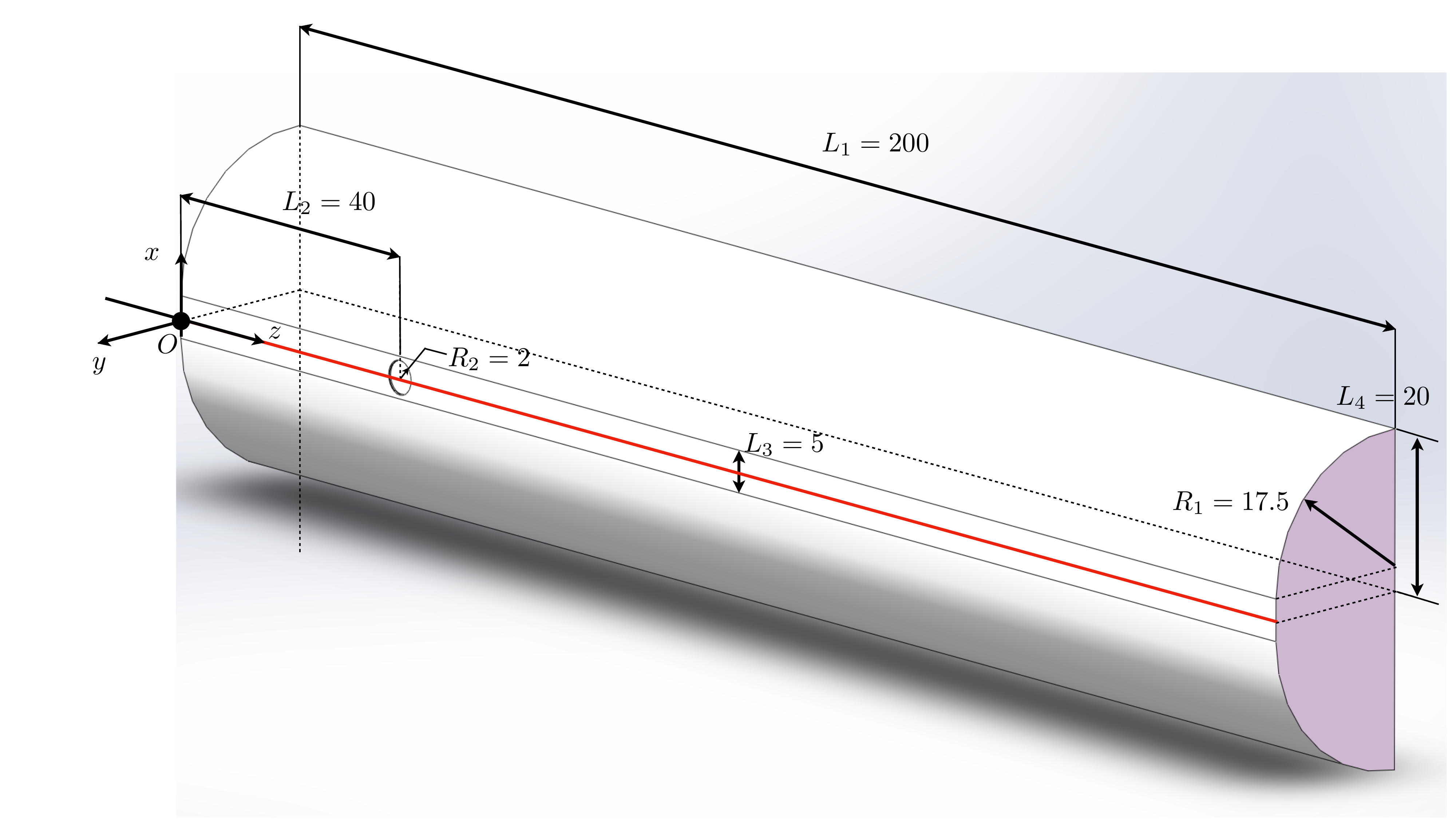}
  \end{overpic}
  \caption{Schematics of the swept blunt leading edge used for numerical simulations.}
  \label{Schematics}
\end{figure}

The computational model comes from recent experimental tests in the FD-07 Mach 6.0 hypersonic wind tunnel of the China Academy of Aerospace Aerodynamics. The experimental model is the front part of a delta wing with a swept angle $\Lambda$ of 45 degrees. The thickness of the wing is $2L_4 = 40$. The spanwise length along the attachment line is $425\text{mm}$. An asymptotic state can be reached at around half the position of the model. The front part of the wing is polished and can be seen as a plate. In this study, we have established a coordinate system, as in figure \ref{Schematics} wherein the $z$-axis aligns with the leading edge of the swept blunt model, coinciding with the attachment line and extending in the corresponding spanwise direction. The normal direction on the corresponding attachment line and swept blunt body is defined as the $y$-axis. Finally, the $x$-axis is defined to complete the typical Cartesian coordinate system in conjunction with these two axes. As usual, a body fitted coordinate $(\xi,\eta,z)$ is also established with the same spanwise direction as the Cartesian coordinate system, while the $\xi$-axis is defined along the chordwise direction and the $\eta$-axis is defined along the surface normal directions.

Based on that geometry, a computational model is designed as in figure \ref{Schematics}. The computational model can be likened to a sandwich-like configuration, where the top and bottom layers consist of semicircles with a radius of $R_1 = 17.5$, and the intermediate layer is a flat plate with a width of $5\text{mm}$. Together, these three layers form the complete swept blunt body configuration. The roughness elements is located at $z = L_2 = 40$, at the center of the leading plate. The radius of the roughness is $R_2 = 2$. The length of the whole model is designed as $L_1 = 200$. Based on experiments, the surface temperature is set to $T_w^* = 370\text{K}$ , other relative flow parameters are listed in table \ref{BasicParameters}. 
   
\begin{table}
\centering
\begin{tabular}{c|cccccc}
\toprule
Flow conditions & $M_{\infty}$ & $Re_{\infty}$ & $T_{\infty}^*$ & $T_w^*$ & $\Lambda$ & $\gamma$ \\

& 6.0  & $1.8\times10^4$ & $56.58$K & $370$K & $45^o$ & 1.4 \\
\midrule
Parameters for roughness elements & $S_r$ & $k_h$ & $k_h/\delta_{bl}^*$ & $d$ & $Re_{kk}$ & $N_k$\\
case H0100 & 1.0 & 0.1mm & $\approx 0.5$ & 4mm & $\approx 678$ & $\approx 87$ \\
case H0200 & 1.0 & 0.2mm & $\approx 1$ & 4mm & $\approx 2776$ & $\approx 125$ \\
\bottomrule
\end{tabular}
\caption{Basic parameters for flow and roughness at basic grid. $N_k$ is the
number of wall normal points for $0 \leqslant y \leqslant k_h$. $\delta_{bl}^* = 0.2\text{mm}$ is the thickness of the laminar boundary layer at the attachment-line boundary layer.}
\label{BasicParameters}
\end{table}%

\begin{figure}
  \centering
  \begin{overpic}[width=0.9\textwidth]{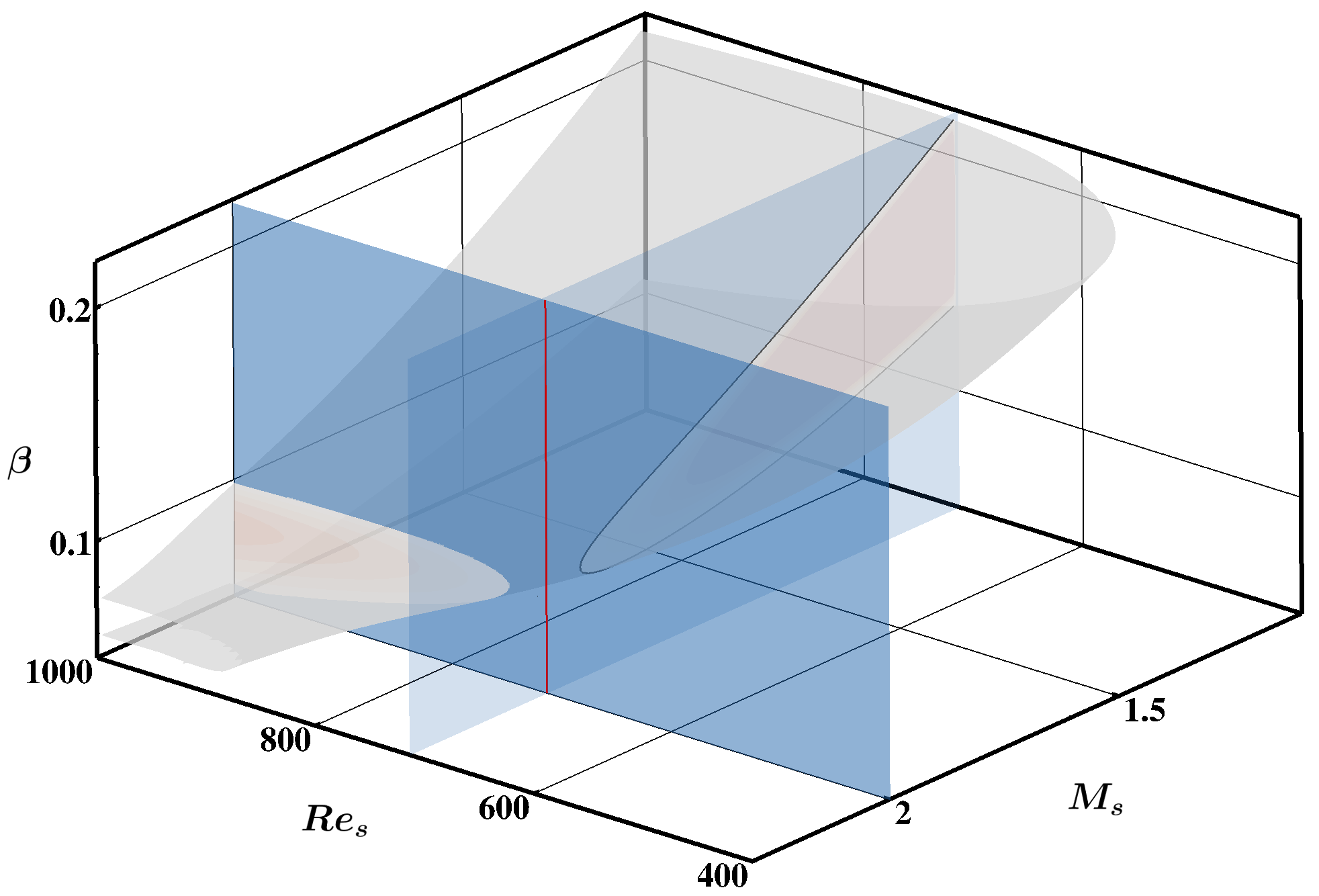}
  \put(50,56){Unstable}
  \put(84,36){Stable}
  \end{overpic}
  \caption{The neutral surface of the most dangerous discrete temporal mode over the $Re_s-M_s-\beta$ plane. The growth rate space is divided into stable and unstable regions by the neutral surface.}
  \label{NeutralSurface}
\end{figure}

As previous analysis around the attachment-line boundary layers, we define the sweep Mach number $M_s$ and the sweep Reynolds number $Re_s$ as
\begin{equation}
Re_s = \frac{w^*_{\infty} \delta_{al}^* }{\nu^*_r} \approx 714, M_s = \frac{w^*_{\infty}}{c^*_s} \approx 2
\end{equation}
based on the length scale $\delta_{al}^* = \sqrt{\nu^*_r {\partial u^*_e}/{\partial x^*}}$ at exact attachment line $x^*=0$ and the variables outside of the attachment-line boundary layer. $c^*_s$ is the sound speed after the leading shock, $\nu^*_r$ stands for the kinematic viscosity at recovery temperature $T^*_r\approx 433K$ and the temperature at the edge of the leading attachment-line boundary layer is $T_{at, e}^*\approx 260K$. The value ${\partial u^*_e}/{\partial x^*}$ at exact attachment line $x^*=0$ is not known a priori for the present case, the potential flow around a circular cylinder with equivalent radius $L_4$ is thus used to evaluate the derivative. By using the linear stability theory over two dimensional domains, the neutral surface of the most dangerous discrete mode are presented in figure \ref{NeutralSurface}, over a $Re_s - M_s - \beta$ coordinate.
Detailed settings and calculations can be found in many previous studies\citep{Theofilis2006, Gennaro2013, Xi2021a}. Here, $\beta$ is the normal spanwise wave numbers. The red line in figure \ref{NeutralSurface} indicates the case we focus on in this study. 
It is found that the critical Reynolds number increases with increasing sweep Mach number, which indicates that the leading viscous mode (Görtler-Hämmerlin mode) is supressed by the compressible effects.   
Also, it is clearly shown that the present case (the red line) is locate at the stable or subcritical region, which means that the transition to turbulence at the present case is not triggered by a linear instability. 

\subsection{Roughness elements}

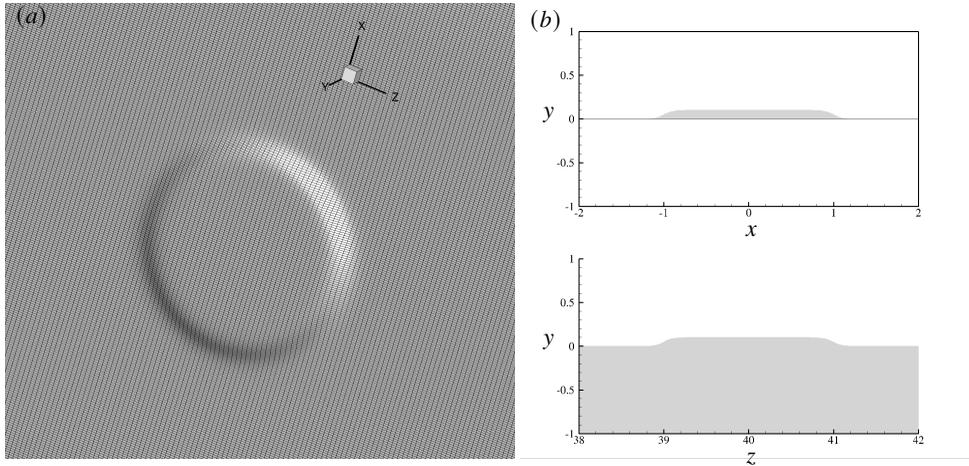
\begin{figure}
  \centering
  \begin{overpic}[scale=0.126]{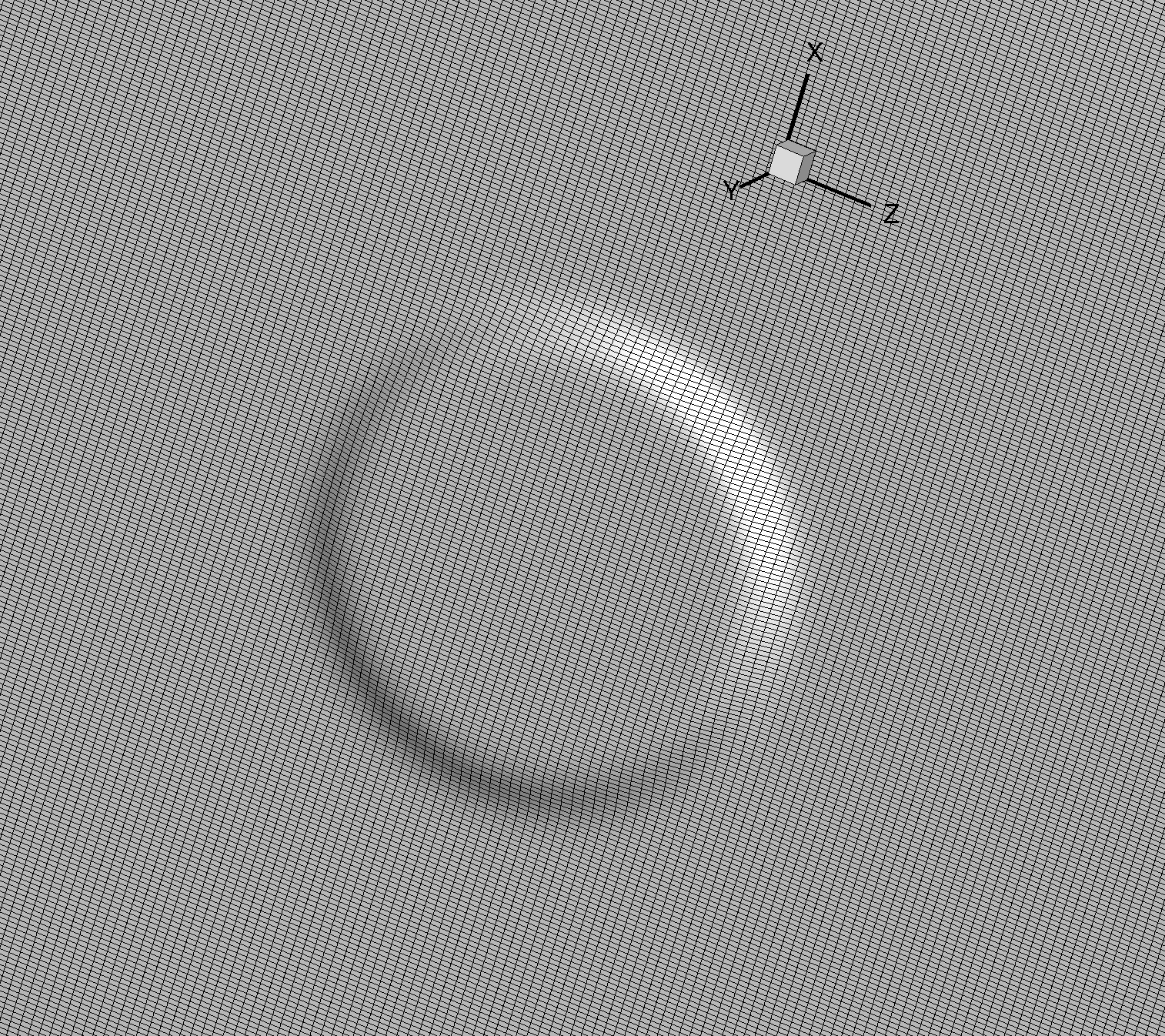}
  \put(2,85){$(a)$}
  \end{overpic}
  \begin{overpic}[scale=0.131]{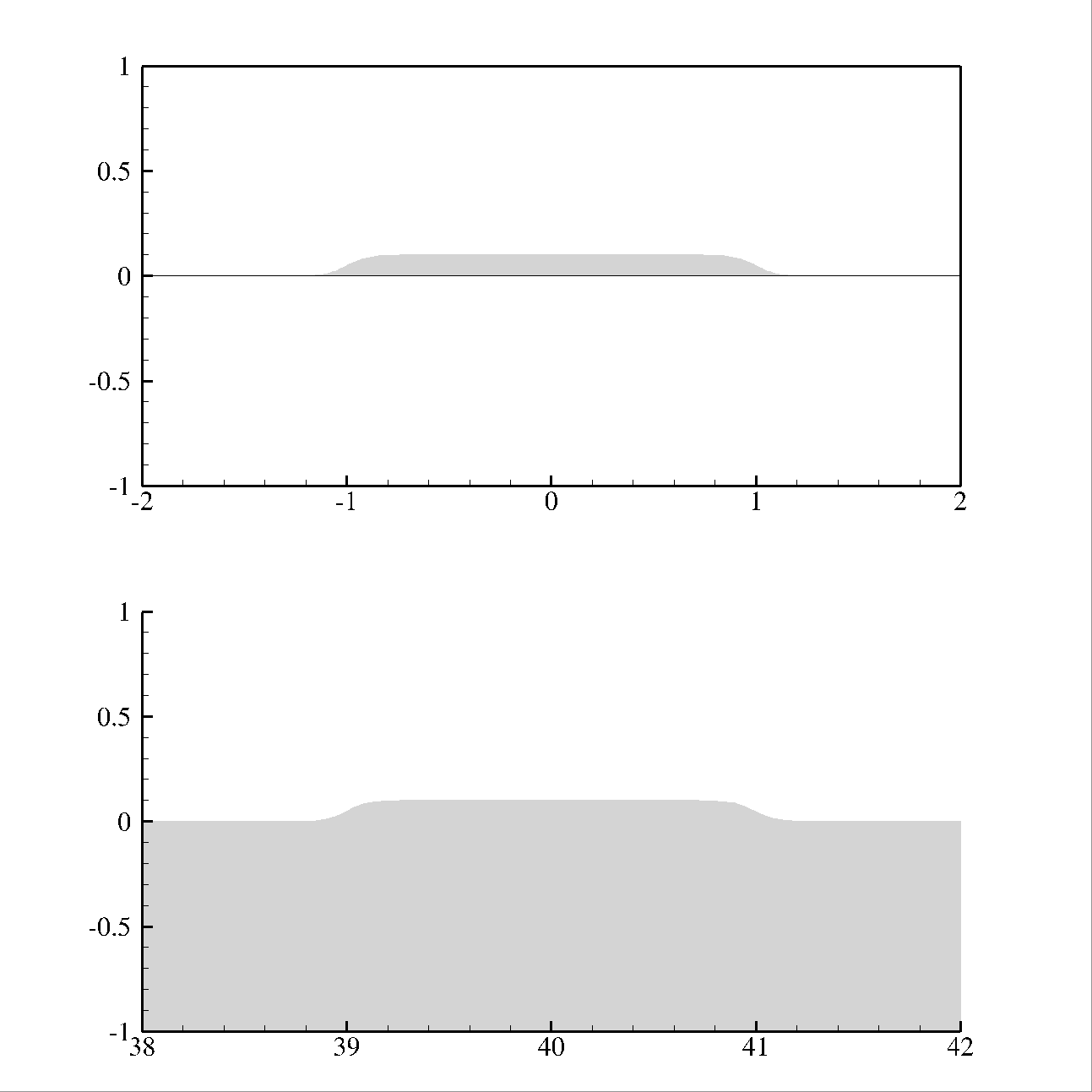}
  \put(2,95){$(b)$}
  \put(50,49){$x$}
  \put(5,75){$y$}
  \put(50,-1){$z$}
  \put(5,25){$y$}
  \end{overpic}
  \caption{$(a)$ The grid distributions arount the roughness with the roughness height $k_h = 0.1$mm in full resolution. $(b)$ The shape of roughness in two cross sections}
  \label{RoughnessShape}
\end{figure}

The roughness element, shown in figure \ref{RoughnessShape}, designed to simulate a pressure sensor with a circular disk configuration during the experiment, is characterized by the  function expressed in polar coordinates $(r, \phi)$, with the shape of the shoulders being defined by a hyperbolic tangent function in similar ways as in previous studies\citep{Kurz2014, Kurz2016}. The function is defined as
\begin{equation}
h(r, \phi) =\frac{k_h}{2} + \frac{k_h}{2}\tanh\left[\frac{S_r}{k_h}\left(\frac{d}{2} - r\right)\right], 
\end{equation}
with $k_h$ and $d$ being the height and diameter of the roughness. The slope factor $S_r$ is set to 1.0 for all cases in the present study. In general, the center of the roughness is locate at the points $(x_c, z_c) = (0, 40)$, the diameter is $d = 2R_2 = 4$. 

Another important parameter for roughness induced transition is the Roughness reynolds number $Re_{kk}$, characterised based on the height $(k_h)$ and the velocity $(w)$ in the
undisturbed laminar flow with respect to the position of the roughness. This roughness reynolds number is defined as a function of 
\begin{equation}
Re_{kk} = \frac{\rho(k_h) w(k_h) k_h }{\mu(k_h)},
\end{equation}
and listed in table \ref{BasicParameters} based on the laminar boundary layer.

\subsection{Simulation strategy}

\begin{figure}
  \centering
  \begin{overpic}[width=\textwidth]{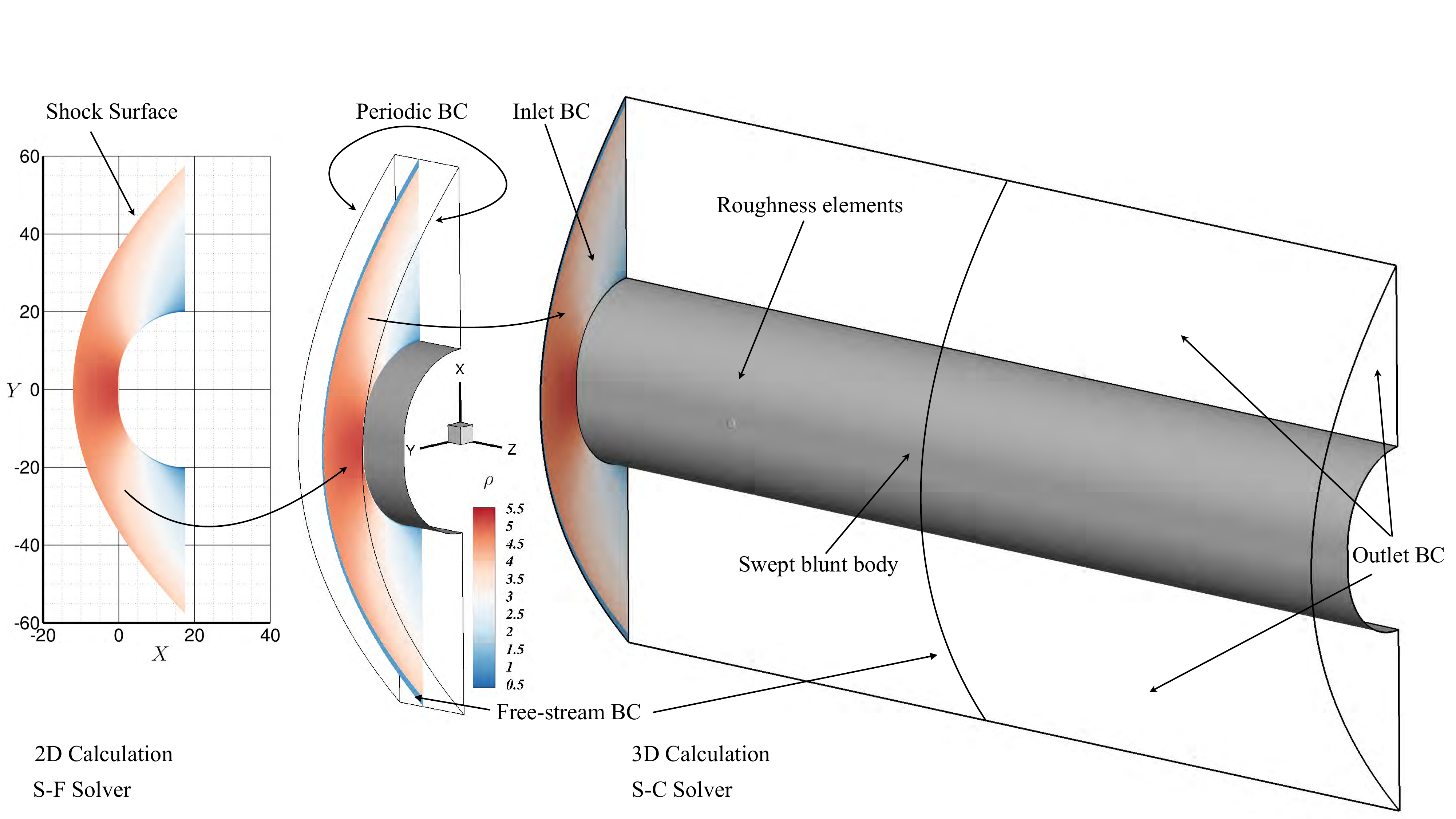}
  \put(5,52){$(a)$ step 1}
  \put(25,52){$(b)$ step 2}
  \put(45,52){$(c)$ step 3}
  \end{overpic}
  \caption{The outline of calculation processes. S-F: Shock Fitting, S-C: Shock Capture}
  \label{Schematics2}
\end{figure}
The calculation process for this kind of problem is divided into three steps, with two kinds of compressible solvers\citep{Li2008, Xi2021b} as shown in figure \ref{Schematics2}. 
Assuming the incoming flow has reached an asymptotic state, a two-dimensional calculation with infinity span assumption $(\partial/\partial z = 0)$ is performed first using a shock-fitting solver. With the exact location of the shock revealed, alignment and clustering of the mesh along the bow shockwave can be easily achieved for the following shock-capture calculation.
To diminish the numerical perturbations between the two solvers, a three-dimensional domain is further designed for pre-calculation with a periodic boundary condition along the spanwise direction, with the newly built grid and the initial field from the fitting solver. 
When the calculation is converged, the solution from the middle slice of the periodic three-dimensional domain is used for the fully three-dimensional calculation. 
As the boundary layer develops along the attachment line and the chordwise direction, non-reflection outlet boundary conditions are used further downstream along the attachment line and the chordwise direction. 
Away from the surface, as the shock is embeded in the computational domain, freestream boundary conditions are used at the outside.

To closely mimic the conditions observed during experimental tests, the generation of unsteady perturbations is implemented in two distinct phases. In the first phase, random velocity perturbations, with maximum amplitude constituting approximately $2\%$ of the free-stream velocity, are introduced upstream of the leading shock waves. This procedure aims to replicate the perturbations measured in wind tunnel experiments. In the second phase, to simulate the disturbances inherent to upstream boundary layers along the attachment line, random wall normal blowing and suction are executed via a hole on the wall. These disturbances also possess an maximum amplitude of roughly $2\%$ of the free-stream velocity. The specified hole is positioned at coordinates $(z_c, x_c) = (30, 0)$ and defined with a radius of $2$. 

In the computational analyses conducted within the scope of this study, two distinct cases were examined. In the first scenario, characterized by a roughness element height of 0.1, unsteady perturbations were deliberately introduced to facilitate the onset of transition. Otherwise, the transition would not occur within the wake flow induced by the roughness elements.
Conversely, the scenario involving a roughness element height of 0.2, presented a fundamentally different dynamic. The inherent absolute instability associated with this configuration led to a spontaneous disruption of flow symmetry. This natural progression towards asymmetry effectively initiates the transition process, obviating the need for the introduction of external perturbations. 

\subsection{Computational grids}

\subsubsection{Grid distributions for shock-fitting solver}
The basic grid number used for shock-fitting simulations is $N_{\xi} \times N_{\eta} = 801 \times 401$. The distribution of the grid points in the wall normal direction is controlled through a function that provides clustering towards the wall, with two parameters $h_{im}$ and $\sigma_s$. The distribution function, which maps $\eta$ to $h$, can be expressed as
\begin{equation}
\begin{aligned}
h = H_{shk}\frac{a_y (1 + Y)}{(b_y - Y)}, b_y =1 + 2 a_y,\quad a_y =\frac{h_{im}}{1 - 2 h_{im}},
Y =2 \frac{\left[1 - \tanh{(\sigma_s)}\right]\frac{1+\eta}{2}}{ 1 - \tanh{\left(\sigma_s\frac{1+\eta}{2}\right)}} - 1,
\end{aligned}
\end{equation}
where $H_{shk}$ is the local shock height and is solved as a dependent variable with the flow field in shock-fitting methods. $\eta$ is a uniform grid distribution along the region $\left[-1.0,1.0\right]$, $h$ is the actual wall normal grid distributions. The values of $h_{im}$ and $\sigma_s$ are chosen to be $0.3$ and $0.95$ for the fitting simulations presented in this paper. Along the chordwise direction, at the wall surface, the surface grid $s(x,y)$ is clustered at the round head with the function
\begin{equation}
\frac{s}{S} = \frac{a_{\xi}(1 - \xi)}{b_{\xi} - \xi}, b_{\xi}= 1 + 2 a_{\xi}, a_{\xi}=\frac{s_{im}}{1 - 2s_{im}},
\end{equation}
where $s_{im}$ is chosen to be $0.2$ for the simulations based on shock-fitting methods, $s$ is the local surface curve length along the model surface, and $S$ is the total model surface curve length. Note that the surface grid does not change during the calculation. 

\subsubsection{Grid distributions for shock-capture solver}
\begin{figure}
  \centering
  \begin{overpic}[width=0.95\textwidth]{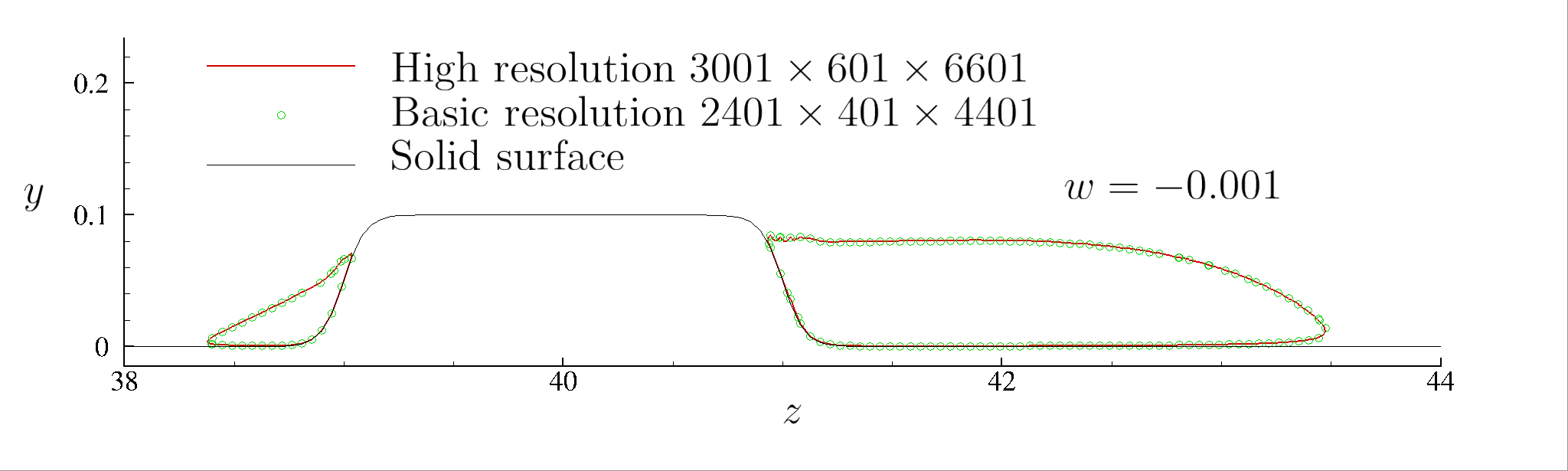}
  \end{overpic} 
  \caption{Regions of flow separations in a cross-cut plane through the centre of a roughness for case H0100, marked by contours of $w=-0.001$, for basic and high resolutions. The axis is stretched for clarity.}
  \label{GridIndependentStudy}
\end{figure}

The basic grid numbers used for shock-capture simulations are $N_{\xi} \times N_{\eta} \times N_{z} = 801 \times 401 \times 8$ and $N_{\xi} \times N_{\eta} \times N_{z} =2401 \times 401 \times 4401$. More grids points with $N_{\xi} \times N_{\eta} \times N_{z} =3001 \times 601 \times 6601$ are used to validate the independence of the solutions, as shown in figure \ref{GridIndependentStudy}. Also, the independence of the solutions have be further verificated by the Reynold stress as well as the balance of the kinetic energy in the full developed turbulent region.
Unlike the fitting solver, for the capture solver, the wall normal grid needs clustering towards the wall as well as the shock region. Therefore, the wall normal grid is divided into three parts. 
The first part is at the region $\left[0, 0.105H_{shk}\right]$, where $H_{shk}$ is the local shock height.
The second part is a transition region which connects the near-wall region and the shock region and is located at the region $\left[0.105H_{shk}, 0.945H_{shk}\right]$. 
The final region stands for the part used to capture the shock and is located at the region $\left[0.945H_{shk},1.05H_{shk}\right]$.
As the local shock height $H_{shk}$ is solved by the fitting solver in advance, the grid lines are adapted well to both the body and the shock shape. Unlike the usual calculations of the transitional/turbulent boundary layer, the shock regions in the present simulation have been taken into account with enough accuracy.

\begin{figure}
  \centering
  \begin{overpic}[width=0.6\textwidth]{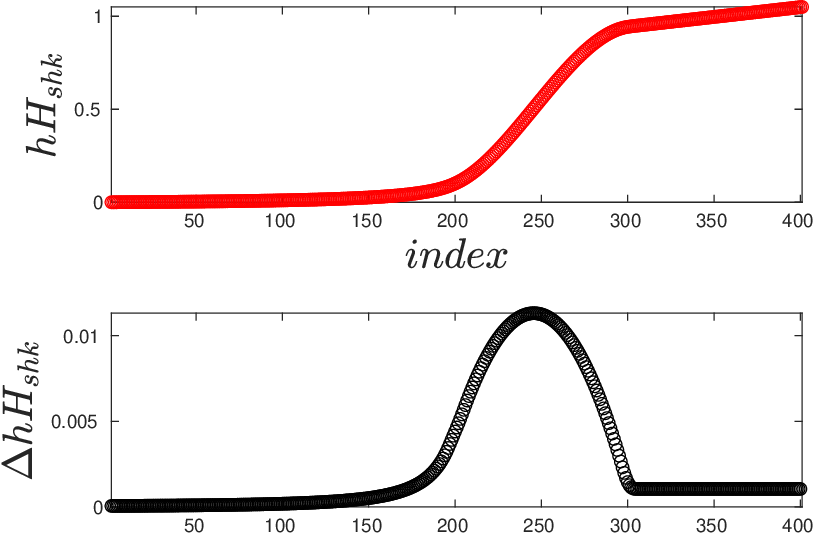}
  \put(1,62){$(a)$}
  \put(1,32){$(b)$}
  \end{overpic}
  \caption{Basic grid distributions with 401 grid points along the wall normal directions. $(a)$ The scaled grid distributions. $(b)$ The scaled grid spacing along the wall normal direction.}
  \label{WallNormalGrids}
\end{figure}

For basic grid, the first region has 201 grid points, with the distributions $h_1$ as
\begin{equation}
h_1 = \frac{21}{1600}\frac{1 + \eta_1}{3.5 -\eta_1}H_{shk},
\end{equation}
where $\eta_1$ is the uniform distributions in region $\left[-1,1\right]$. The third region has 101 grid points with a uniform distribution. The second region has 101 grid points and is used to link grids with different grid spacings using a Hermite
function obtained by imposing C3 continuity of the resulting stretching function. The grid distributions and relative grid spacing is shown in figure \ref{WallNormalGrids}. 
For the step 2 calculation of the periodic box, 8 grids are used to cover a spanwise region of length $20$. For the final full three-dimensional calculation, 4401 grids are distributed uniformly along the spanwise direction. The roughness element and unsteady hole are captured with 101 grids points.

By using the above approach, for the basic grid numbers, around the attachment-line boundary layer, $125$ grid points are located inside the boundary layer(based on laminar boundary layer thickness) in the wall normal $\eta$-axis direction. 
About $230$ grid points are located at the leading-edge plate along the $x$-axis.
As the boundary layers develop along the chordwise direction as well, $136$ grid points are located inside the boundary layer at the chordwise outlet boundary. Detailed grid numbers that are used to resolve roughness are shown in table \ref{BasicParameters}. Comparing with the existing researches\citep{Mayer2011,Tullio2013,Tullio2015,Groskopf2016,Giovanni2018,Hader2019,Shrestha2019} on numerical simulations of high-speed boundary layer transitions, these grid numbers and distributions can reveal the detailed behaviour of linear and nonlinear waves. To further identify the possible features of the turbulence at the further downstreams, grids points along the wall normal directions are increased to 601 in the final simulations presented in this paper. The grid sizes in wall units in the turbulent region are shown and listed in Table \ref{BasicGridWallUnits}. 
Based on that grid size, as there are no additional stress and heat flux terms in the present simulations, resolutions for an implicit large-eddy simulations (ILES) can be recognized for the possible turbulent region.

\begin{table}
\centering
\begin{tabular}{c}
\begin{overpic}[width=0.55\textwidth]{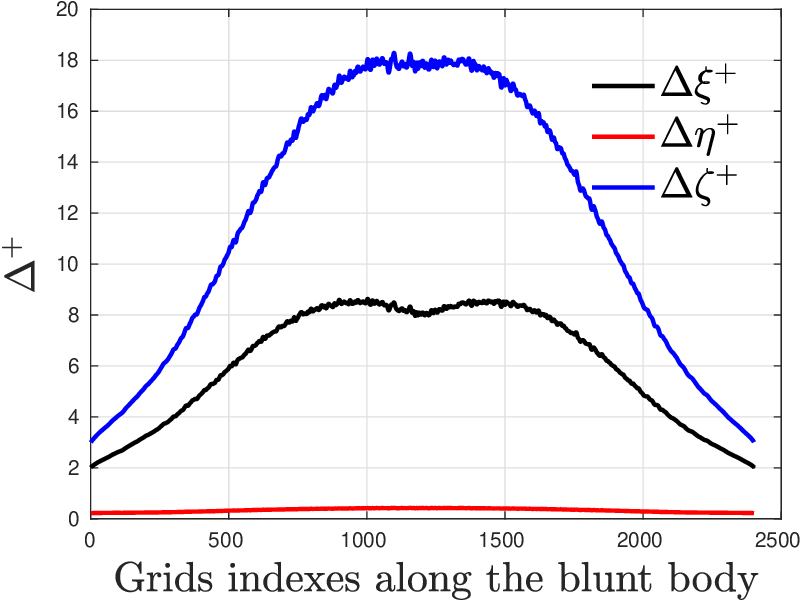}
\end{overpic} \\
\begin{tabular}{ccccccc}
\toprule
 $\xi$ & $\eta$ & $\zeta$ & Total points & $\Delta \xi_{max}^+$ & $\Delta \eta_{max}^+$ & $\Delta \zeta_{max}^+$ \\
 \midrule
 2401 & 601 & 4401 & $6.4\times10^9$ & 8.5 & 0.4 & 18.2 \\ 
\bottomrule
\end{tabular}
\end{tabular}
\caption{Grid points and maximum grid sizes in wall units at the turbulent boundary region.}
\label{BasicGridWallUnits}
\end{table}%

\section{Results}\label{sec3}
\subsection{Verification of the asymptotic assumption}
\begin{figure}
  \centering
  \begin{overpic}[width=\textwidth]{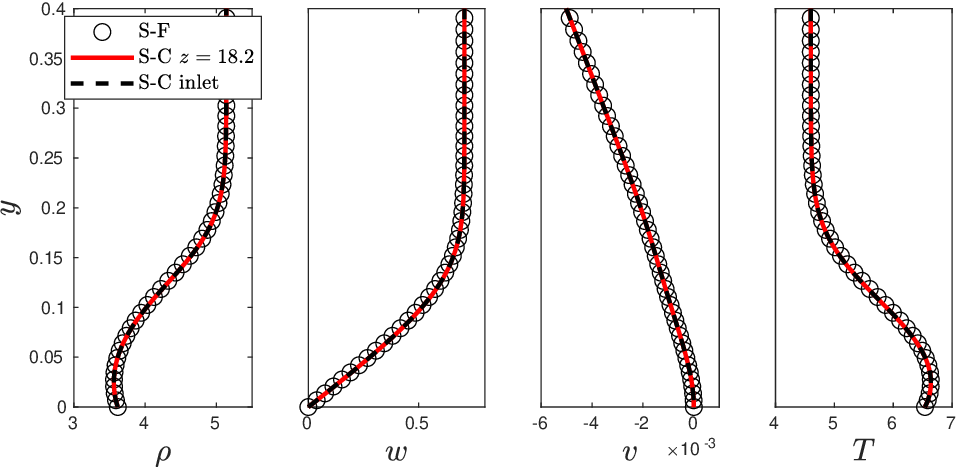}
    \put(-2,45){$(a)$}
  \put(28,45){$(b)$}
  \put(52,45){$(c)$}
  \put(76,45){$(d)$}
  \end{overpic} 
  \caption{Profiles of major variables at $x=0$ plane, for case H0100. }
  \label{H0100_ahead_x0}
\end{figure}

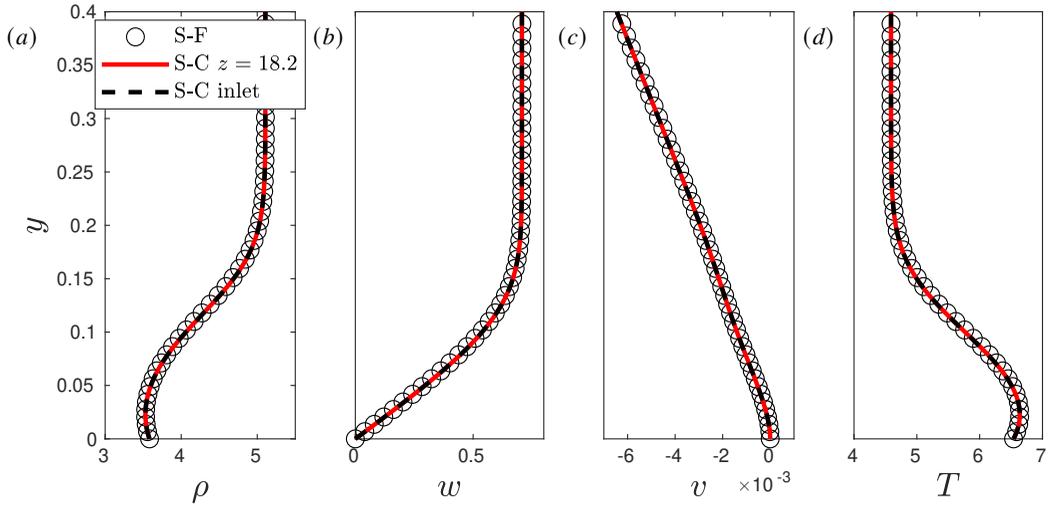
\begin{figure}
  \centering
   \begin{overpic}[width=\textwidth]{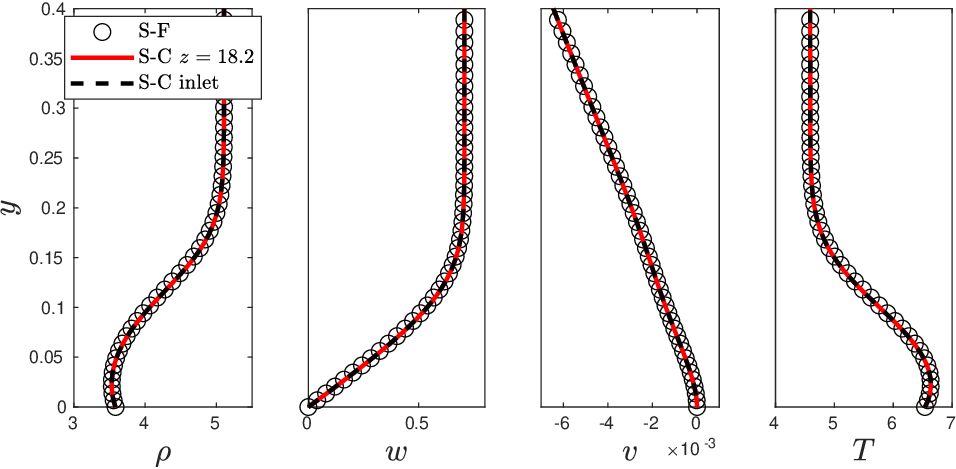}
    \put(-2,45){$(a)$}
  \put(28,45){$(b)$}
  \put(52,45){$(c)$}
  \put(76,45){$(d)$}
  \end{overpic} 
  \caption{Profiles of major variables at $x=2.5$ plane, for case H0100.}
  \label{H0100_ahead_x25}
\end{figure}

In the current study, as the three-dimensional boundary layer develops along the spanwise direction, the commonly employed asymptotic assumption ($\partial/\partial z = 0$) for attachment-line and crossflow analyses can be verified by examining the profiles upstream of the roughness.
Taking the case H0100 as an example, the profiles of major variables such as density $\rho$, spanwise velocity $w$, and temperature $T$ in the attachment-line region $x\in[-2.5,2.5]$ are depicted in figure \ref{H0100_ahead_x0} and \ref{H0100_ahead_x25}.
A perfect alignment of these profiles is observed, which confirms two critical points. 
Firstly, the exact correspondence of results obtained from the S-C and S-F solvers corroborates the precision of the solvers utilized in this research. 
Secondly, and more significantly, because the agreements of the profiles at inlet and at $z=18.2\text{mm}$, the asymptotic assumptions are still applicable to a spatially-developing attachment-line boundary layer.

\begin{figure}
  \centering
   \begin{overpic}[width=\textwidth]{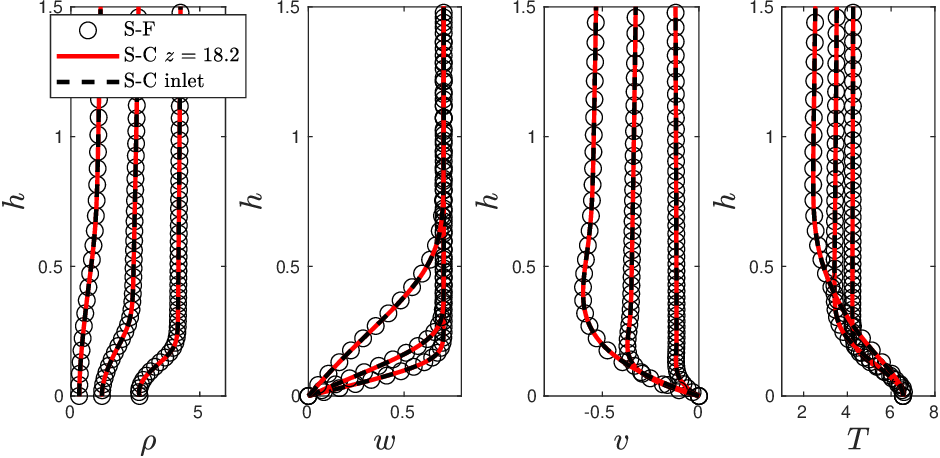}
    \put(-2,43){$(a)$}
  \put(28,43){$(b)$}
  \put(52,43){$(c)$}
  \put(76,43){$(d)$}
  \end{overpic} 
  \caption{Profiles of major variables at three locations ($30^o, 60^o, 90^o$ ) over cylinder surface, for case H0100.}
  \label{H0100_ahead_cross}
\end{figure}
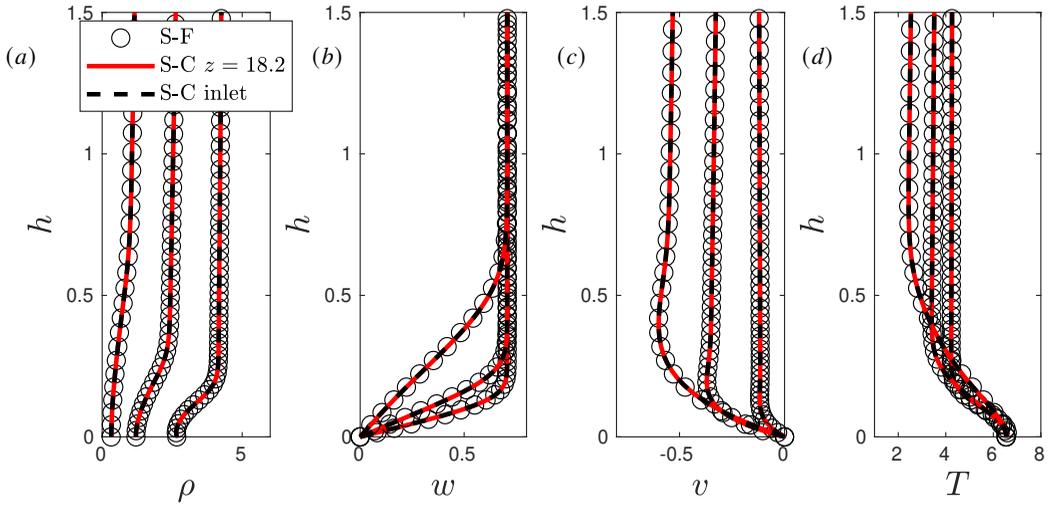

Upon broadening our perspective further, it becomes apparent that even in regions where crossflow effects are pronounced, this assumption remains valid, as shown in figure \ref{H0100_ahead_cross}. It is imperative to emphasize, however, that the upstream flow has reached an asymptotic state, and that this assumption is applicable to fully developed laminar boundary layers, provided that the spanwise extent is sufficiently large. 

\begin{figure}
  \centering
  \begin{overpic}[width=\textwidth]{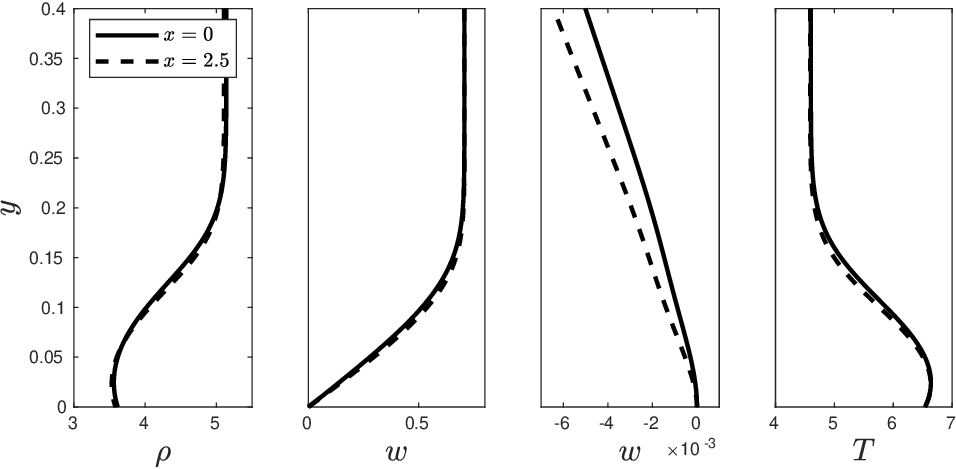}
    \put(-2,45){$(a)$}
  \put(28,45){$(b)$}
  \put(52,45){$(c)$}
  \put(76,45){$(d)$}
  \end{overpic} 
  \caption{Comparison of variables at $x=0$ and $x=2.5$ planes, for case H0100.}
  \label{Compare_H0100_ahead_x}
\end{figure}

We further compare the profiles (at $x=2.5$), a little bit away from the attachment line at $x=0$, as presented in figure \ref{Compare_H0100_ahead_x}.
As the flow develops along the chordwise direction, the basic feature of the variables are kept the same, but the boundary layer has become slightly thinner.

\subsection{General features of the flow fields}

\begin{figure}
  \centering
  \begin{overpic}[width=\textwidth]{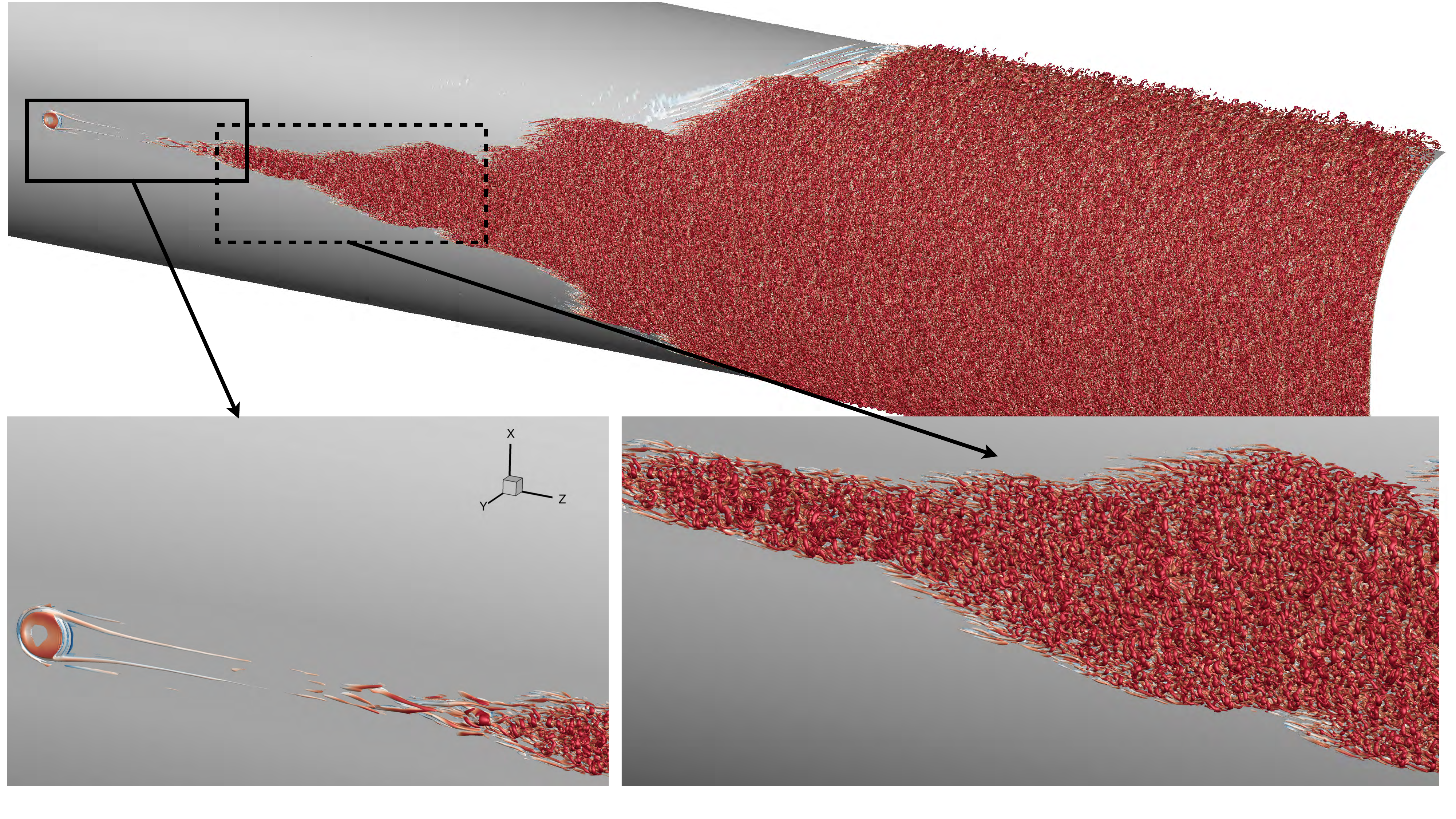}
  \end{overpic} 
  \caption{Instantaneous iso-surface of $\lambda_2 = -0.035$, colour indicates $w$, for the first part of case H0100.}
  \label{H0100_Structure_part1}
\end{figure}

\begin{figure}
  \centering
  \begin{overpic}[width=\textwidth]{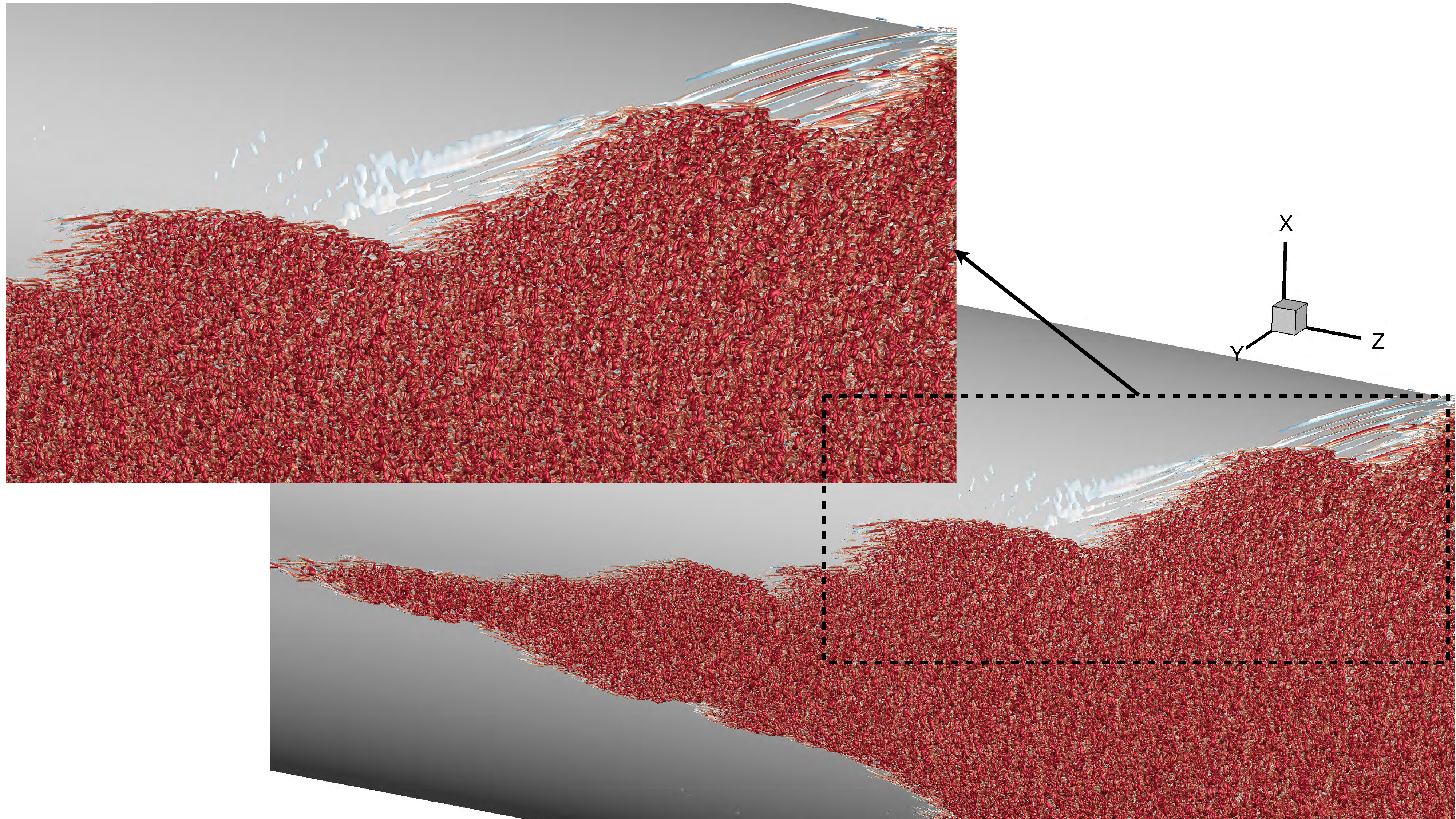}
  \end{overpic} 
  \caption{Instantaneous iso-surface of $\lambda_2 = -0.035$, colour indicates $w$, for the second part of case H0100.}
  \label{H0100_Structure_part2}
\end{figure} 

\begin{figure}
  \centering
  \begin{overpic}[width=\textwidth]{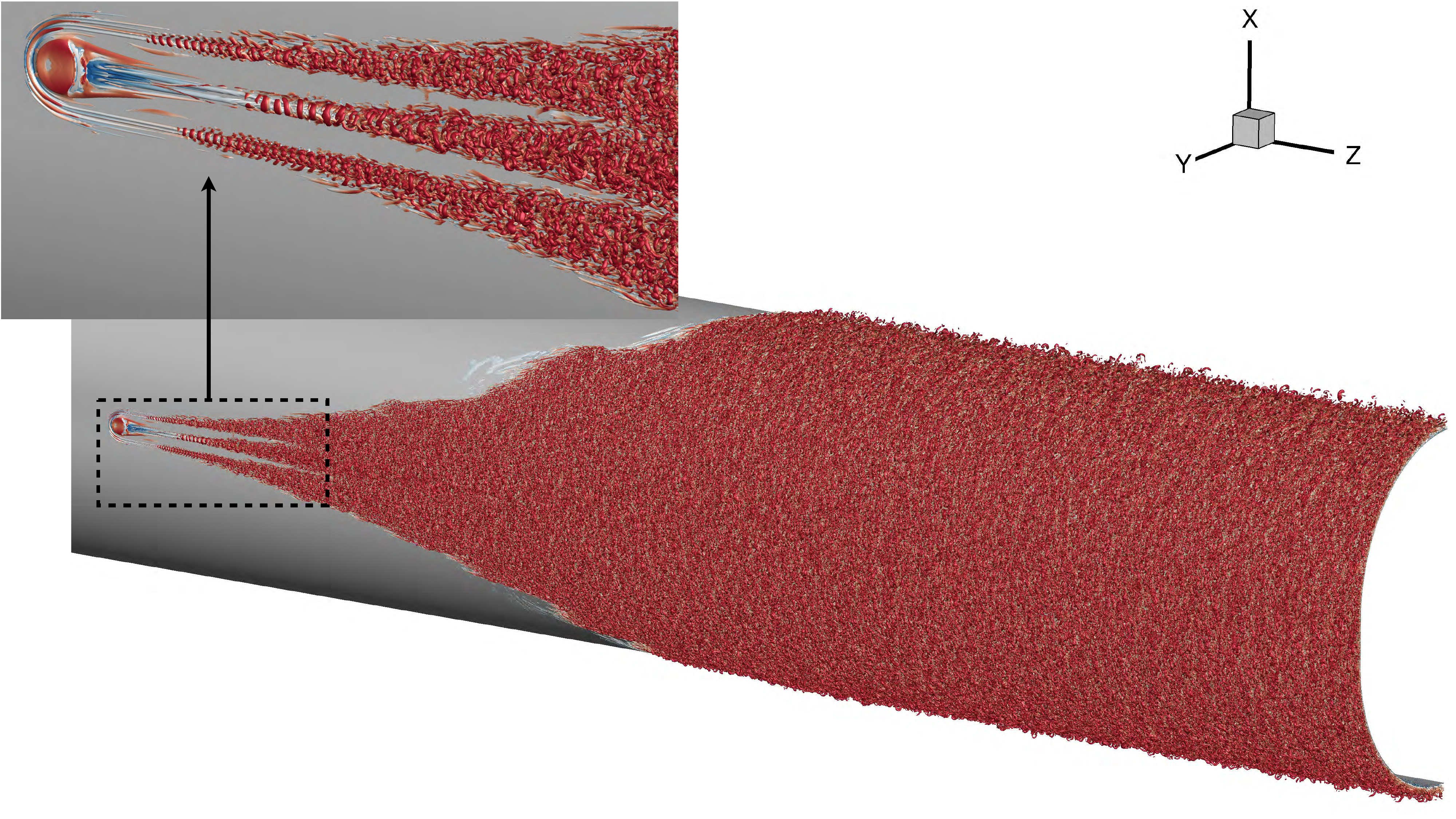}
  \end{overpic} 
  \caption{Instantaneous iso-surface of $\lambda_2 = -0.035$, colour indicates $w$, for the case H0200.}
  \label{H0200_Structure_part1}
\end{figure}

The general features of the whole flow fields are shown in figure \ref{H0100_Structure_part1} and figure \ref{H0100_Structure_part2}, with the iso-surfaces of $\lambda_2 = -0.035$ transient fields. 
The iso-surfaces are colored with spanwise velocity $w$. 
The whole flow fields can be divided into three parts. 
The first part is the roughness region, in which the initial laminar flow is perturbed by the surface deformation. 
Typical vortex structures are formed behind the roughness. 
The breakdowns of the vortex structures lead to typical turbulent structures, along the attachment line, and small vortexes structures are shown.
The second part is the transitional region, in which the initial turbulences at the attachment line develop along the spanwise direction as well as the chordwise direction. 
At the very beginning, the turbulences are located around the attachment line and the turbulent region expands along the chordwise direction slightly. 
Then, as the flows develop further downstream, the turbulent structures are flushed from attachment-line region to chordwise outlet. The final part is the fully turbulent region, where the fully developed turbulences cover the whole region of the leading-blunt body, ranging from attachment line to chordwise outflow.

\begin{figure}
  \centering
  \begin{tabular}{cc}
  \begin{overpic}[width=0.48\textwidth]{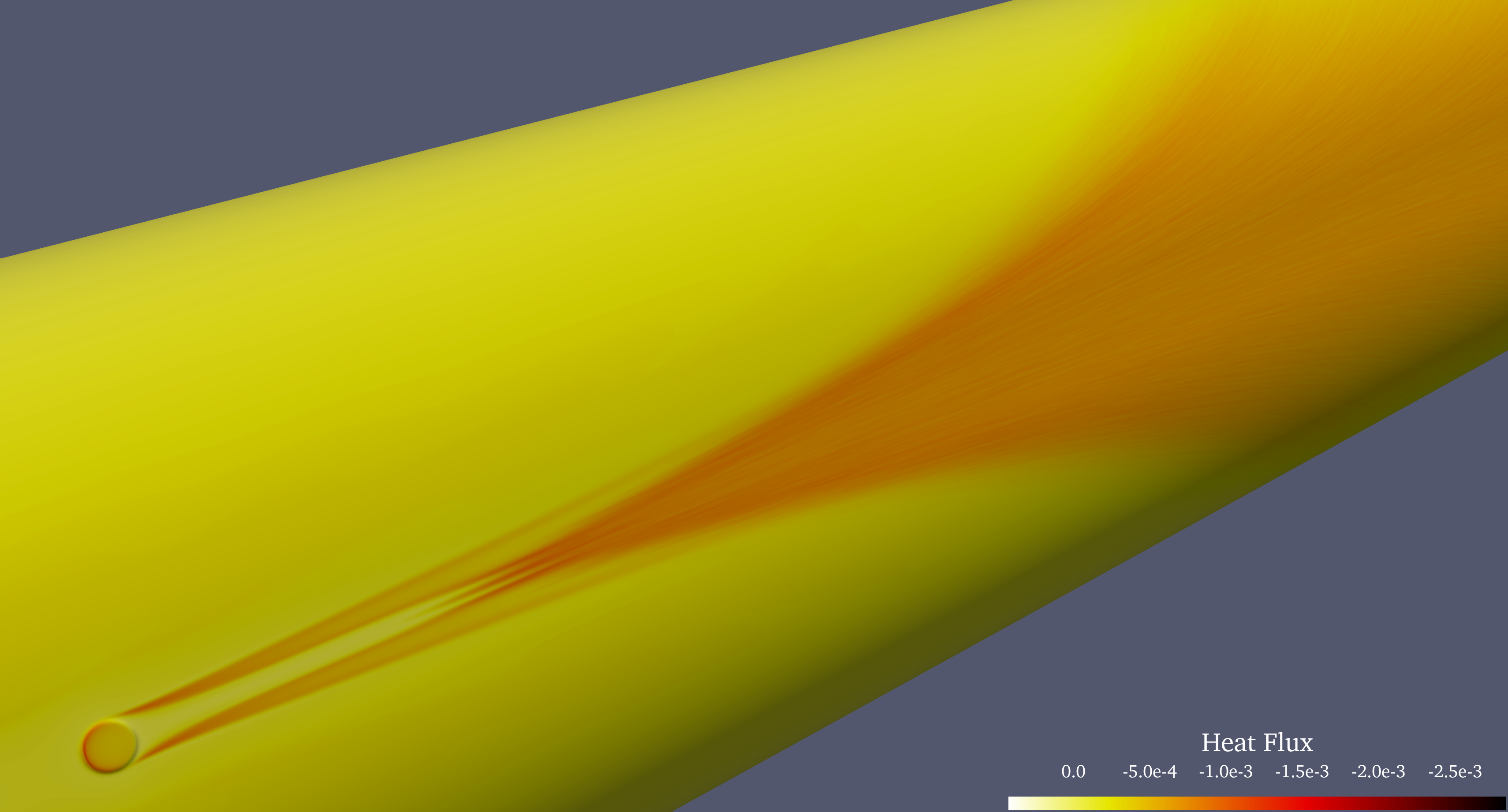}
  \put(1,30){$(a)$}
  \end{overpic} & 
  \begin{overpic}[width=0.48\textwidth]{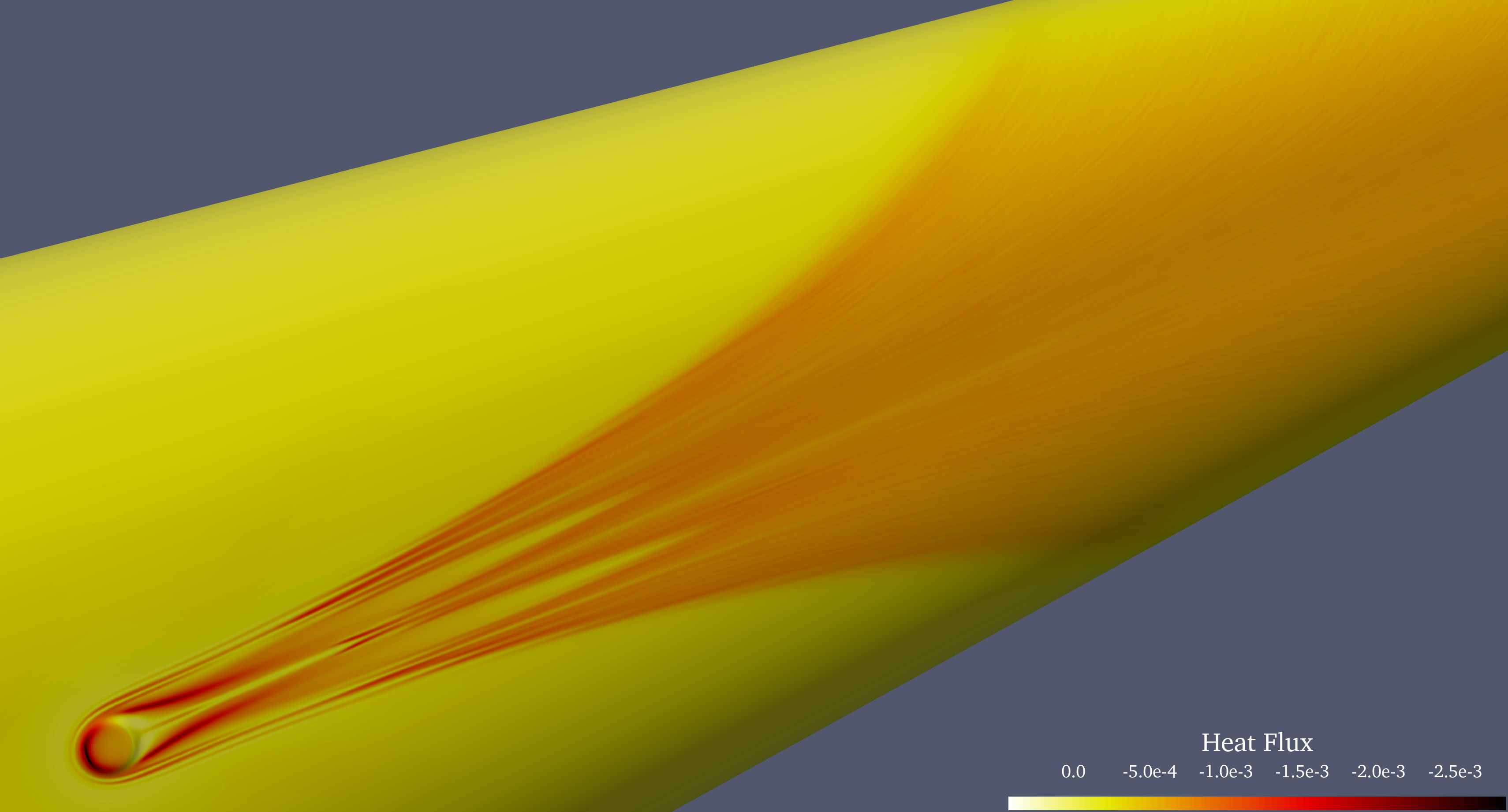}
  \put(1,30){$(b)$}
  \end{overpic}
  \end{tabular} 
  \caption{The surface heat fluxes ${\theta}_{tw}$ distributions for two cases. $(a)$ for H0100, $(b)$ for H0200.}
  \label{Srf_HF}
\end{figure}

\begin{figure}
  \centering
  \begin{tabular}{cc}
  \begin{overpic}[width=0.48\textwidth]{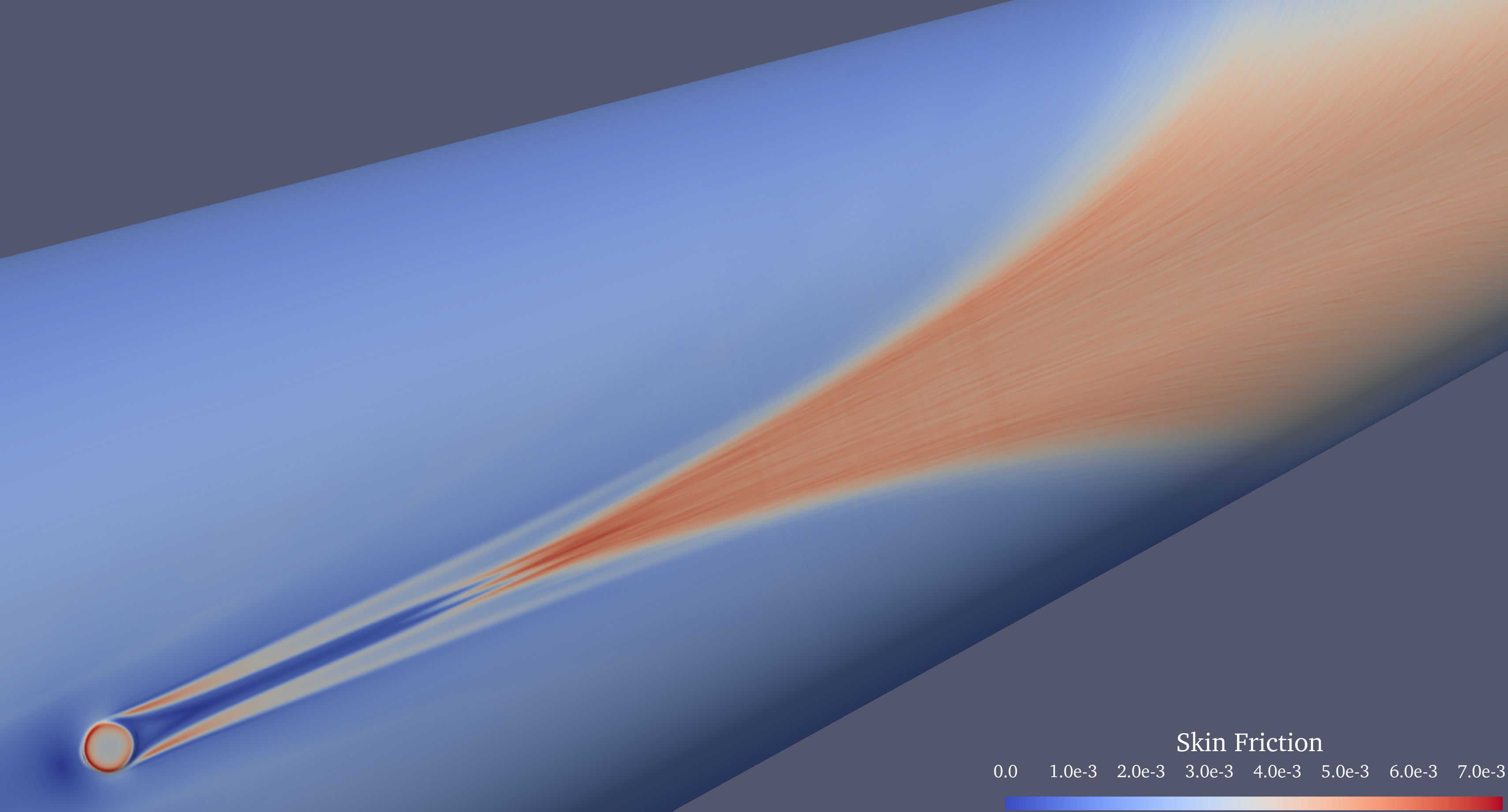}
  \put(1,30){$(a)$}
  \end{overpic} &
  \begin{overpic}[width=0.48\textwidth]{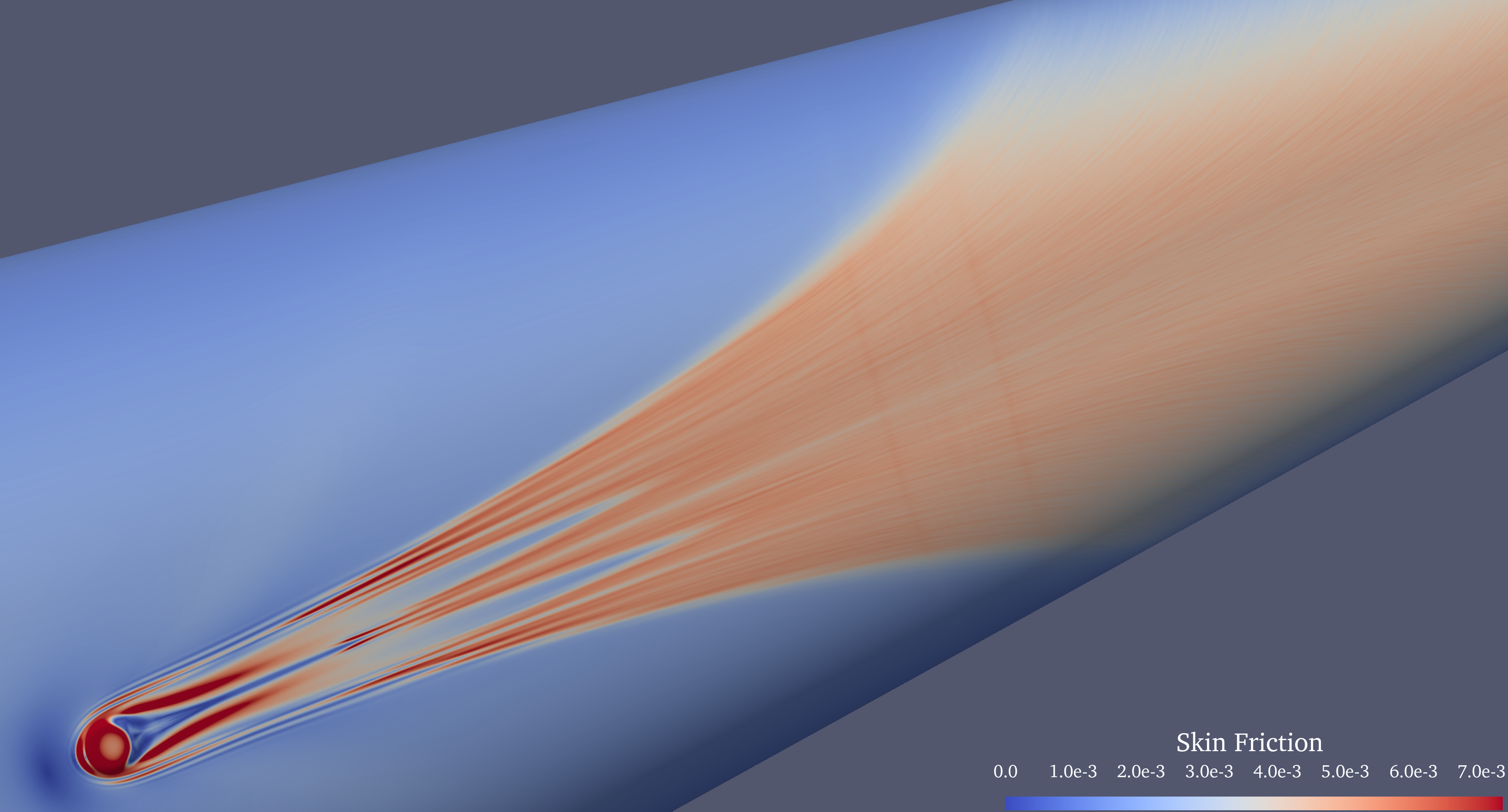}
  \put(1,30){$(b)$}
  \end{overpic}
  \end{tabular} 
  \caption{The surface skin friction $\overline{\tau}_w$ distributions for two cases. $(a)$ for H0100, $(b)$ for H0200.}
  \label{Srf_SF}
\end{figure}

When the height of the roughness element is increased from 0.1 to 0.2,  there are significant differences in the vortex structures formed behind the roughness, as shown in figure \ref{H0100_Structure_part1} and \ref{H0200_Structure_part1}. These differences can be seen more directly in the contours of surface average heat fluxes $\theta_{tw}$ and skin frictions $\overline{\tau}_w$, as depicted in figures \ref{Srf_HF} and \ref{Srf_SF}. These metrics essentially serve as "footprints" of the boundary layer dynamics, providing insights into the complex interactions and flow structures present within the boundary layers.
Here, as the usual boundary layers in previously, we define the velocity $\overline{u}^{+}$, based on inner scale as
\begin{equation}
\left.
\begin{aligned}
h^+ &= \frac{\overline{\rho}_w \overline{u}_{\tau} h}{\overline{\mu}_w}, \quad \overline{u}^+ = \frac{|\overline{u}_p|}{\overline{u}_\tau}, \quad
|\overline{u}_p| = \sqrt{\overline{u}_{\xi}^2 + \overline{w}^2}, \\
\overline{u}_{\tau} &= \sqrt{\frac{\overline{\tau}_w}{\overline{\rho}_w}}, \quad
\overline{\tau}_w = \frac{\overline{\mu}}{Re}\left. \frac{\partial \overline{u}_p}{\partial h} \right|_{h=0},
\end{aligned}
\right\},
\end{equation}  
where, $\overline{u}_{p}$ is the velocity parallel to the surface. The skin-friction coefficient $C_f$ and surface heat-flux ${\theta}_{tw}$ for this kind of flow are defined as
\begin{equation}
C_f = \frac{2 \overline{\mu}_w^*}{\rho_{\infty}^* U_{\infty}^{*2}} = \frac{2 \overline{\mu}_w}{Re}\frac{\partial \overline{u}_{p}}{\partial h}
= 2 \overline{\tau}_w, {\theta}_{tw} = - \left|\kappa \nabla \overline{T} \cdot \boldsymbol{n}\right|.
\end{equation}
The derivatives of surface normals, denoted as ${\partial}/{\partial h}$, for arbitrary variables $f_{\psi}$, are determined through a two-step process. Initially, the gradients of the variables $f_{\psi}$ are computed utilizing the identical scheme adopted for the calculation of viscous fluxes during the simulations. Subsequently, the derivatives of the surface normals ${\partial f_{\psi}}/{\partial h}$ are obtained by projecting the calculated gradients $\nabla f_{\psi}$ onto the surface normal vectors $\bm{n}$.

\begin{figure}
  \centering
  \begin{overpic}[width=\textwidth]{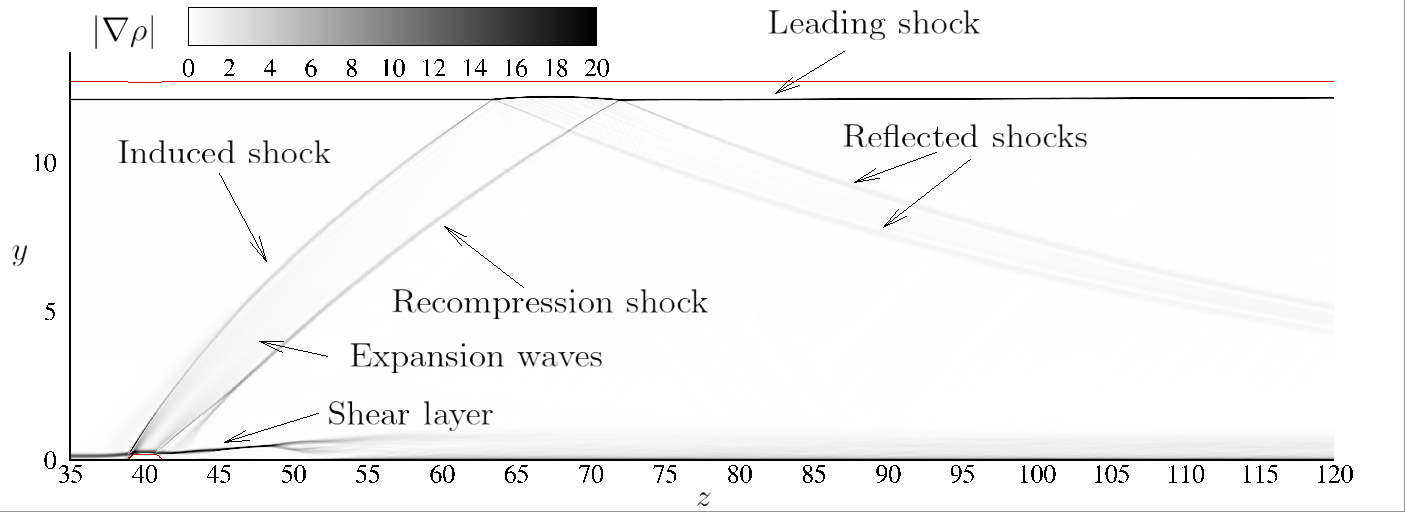}
  \put(1,45){$$}
  \end{overpic} 
  \caption{Density gradient magnitude contours of the case H0200, at attachment-line plane $x=0$. The red line stands for the computational domain.}
  \label{H02_Average_SliceX0}
\end{figure}

The magnitude contour of average density gradients for case H0200, at the attachment-line plane, is depicted in figure \ref{H02_Average_SliceX0}. 
This illustration provides a comprehensive view of the general flow field characteristics for both cases. 
The presence of surface roughness induces a shock slightly ahead of the roughness. 
As this shock evolves away from the surface and progresses downstream, it shapes into a curved shock surface under the influence of the incoming flow. 
The interaction of this induced shock with the leading shock of the blunt body, followed by its reflection back into the boundary layer downstream, results in a noticeable deformation of the leading shock. 
Subsequent to the roughness-induced shock, the compressed fluids undergo expansion and acceleration, leading to the formation of a recompression shock at the roughness's tail along the $z-$direction. 
Meanwhile, a shear layer develops behind the roughness, and the recompression shock once again impinges on the leading shock, reflecting back into the boundary layers. 
When the flow evolves further downstream, the high-shear region at the outside of the boundary layer becomes much weak, as reflected as the decrease of magnitudes for the density gradient $\left| \nabla \rho\right|$. 

\begin{figure}
  \centering
  \begin{overpic}[width=\textwidth]{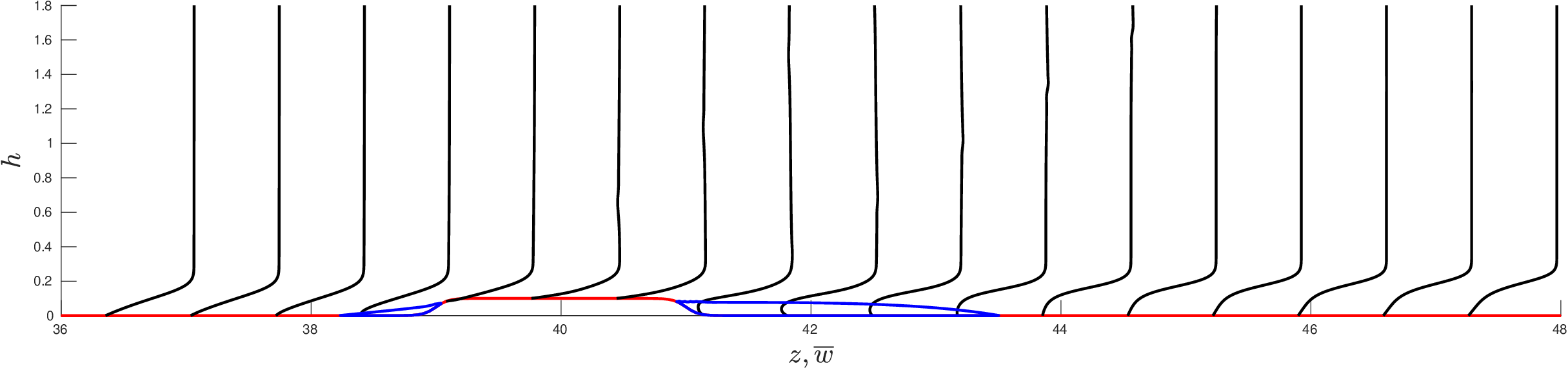}
  \put(-2,22){$(a)$}
  \end{overpic} 
  \begin{overpic}[width=\textwidth]{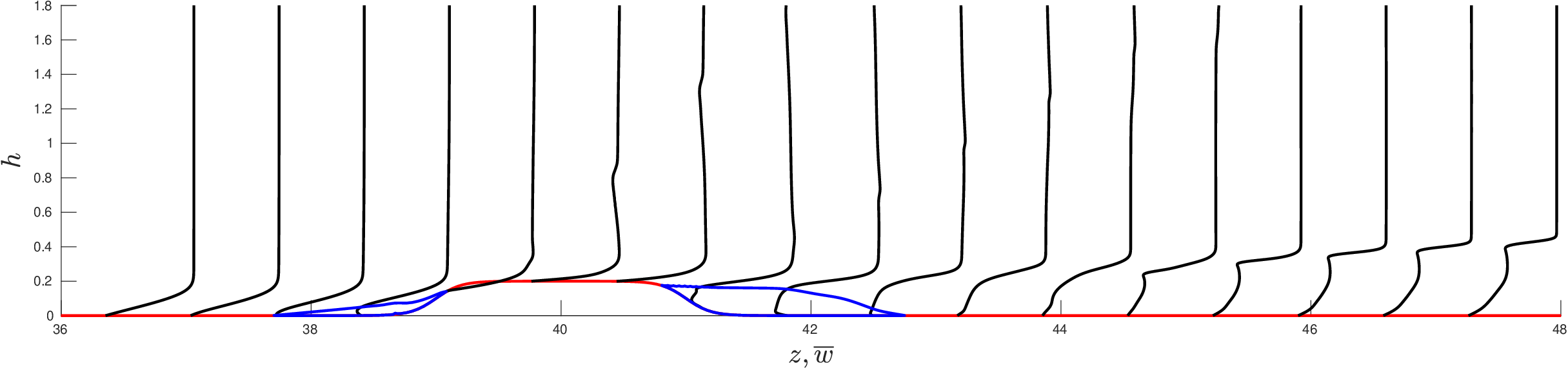}
  \put(-2,22){$(b)$}
  \end{overpic} 
  \caption{Line plots of average spanwise velocity $\overline{w}$ around the roughness for $(a)$ H0100 and $(b)$ H0200 cases. The red and blue lines stand for the wall surfaces and seperation bubbles, respectively.}
  \label{Average_W_CenterLine}
\end{figure}

\begin{figure}
  \centering
  \begin{overpic}[width=\textwidth]{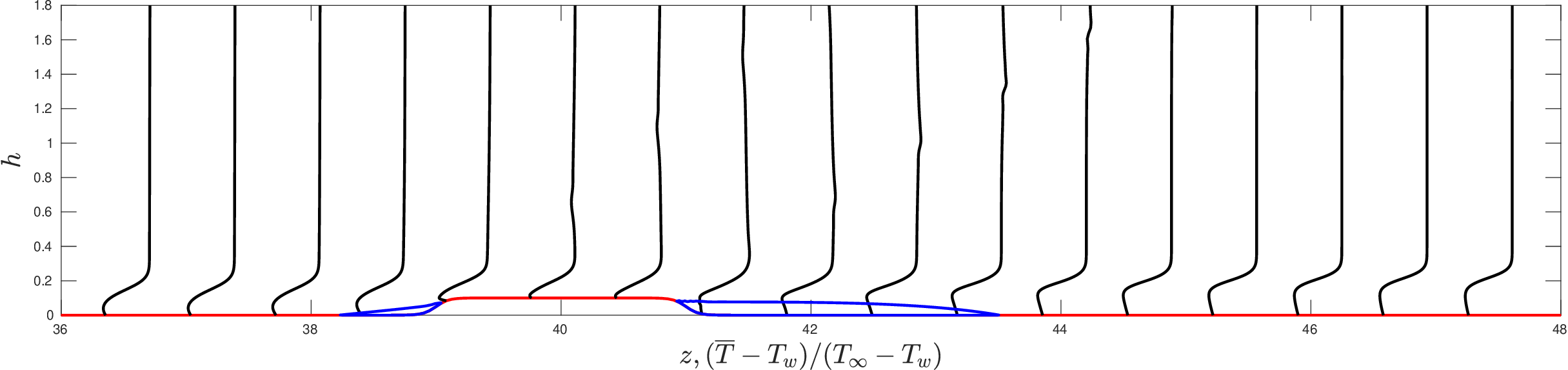}
  \put(-2,22){$(a)$}
  \end{overpic} 
  \begin{overpic}[width=\textwidth]{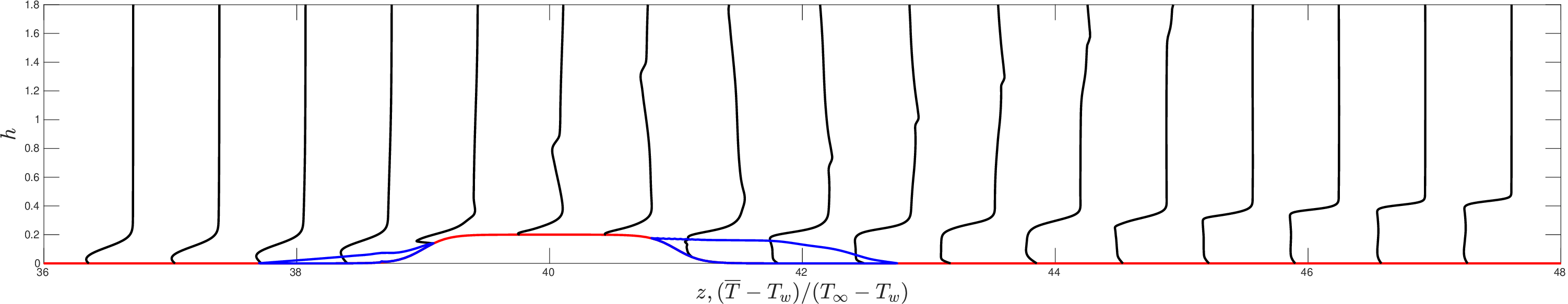}
  \put(-2,18){$(b)$}
  \end{overpic} 
  \caption{Line plots of average Temperature $(\overline{T} - T_w)/(T_{\infty} - T_w)$ at the attachment line around the roughness for $(a)$ H0100 and $(b)$ H0200 cases. The red and blue lines stand for the wall surfaces and seperation bubbles, respectively.}
  \label{Average_T_CenterLine}
\end{figure}

The figure \ref{Average_W_CenterLine} and \ref{Average_T_CenterLine} show the distributions of mean velocity and temperature along the attachment line in the wall-normal direction. 
Additionally, the size and specific location of the corresponding separation bubbles are indicated by blue lines in the figures. 
It is evident from the figure that the separation bubbles induced by small roughness elements are lower in height compared to those induced by large roughness elements, but extend farther downstream.
Conversely, the separation bubbles induced by larger roughness elements extend farther upstream. 
The corresponding velocity and temperature profiles highlight the approximate location of the shear layer and illustrate the process through which low-speed fluid, due to the lift-up effect, is elevated away from the wall.

\begin{figure}
  \centering
  \begin{tabular}{cc}
  \begin{overpic}[width=0.48\textwidth]{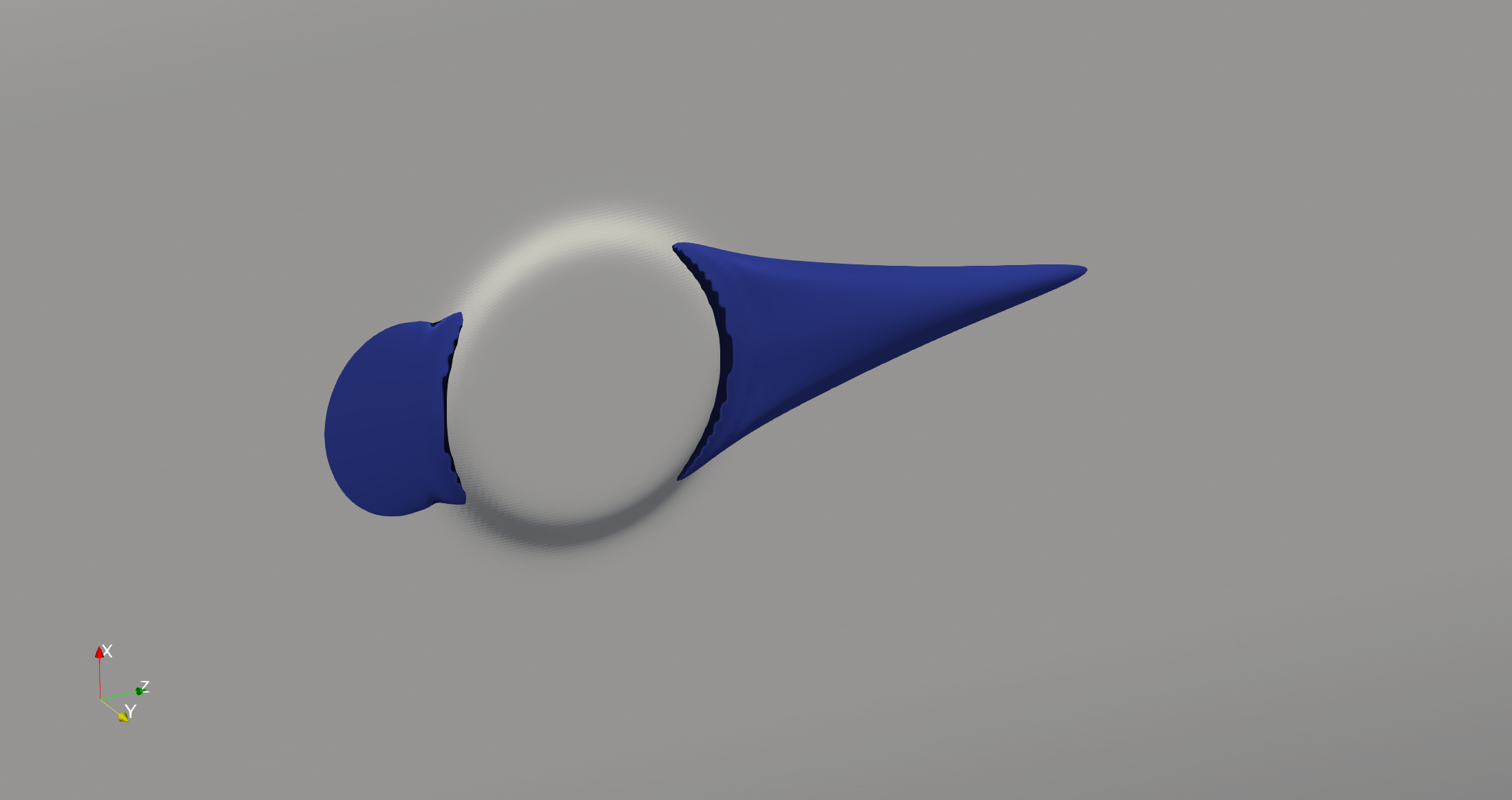}
  \put(1,42){$(a)$}
  \end{overpic} & 
  \begin{overpic}[width=0.48\textwidth]{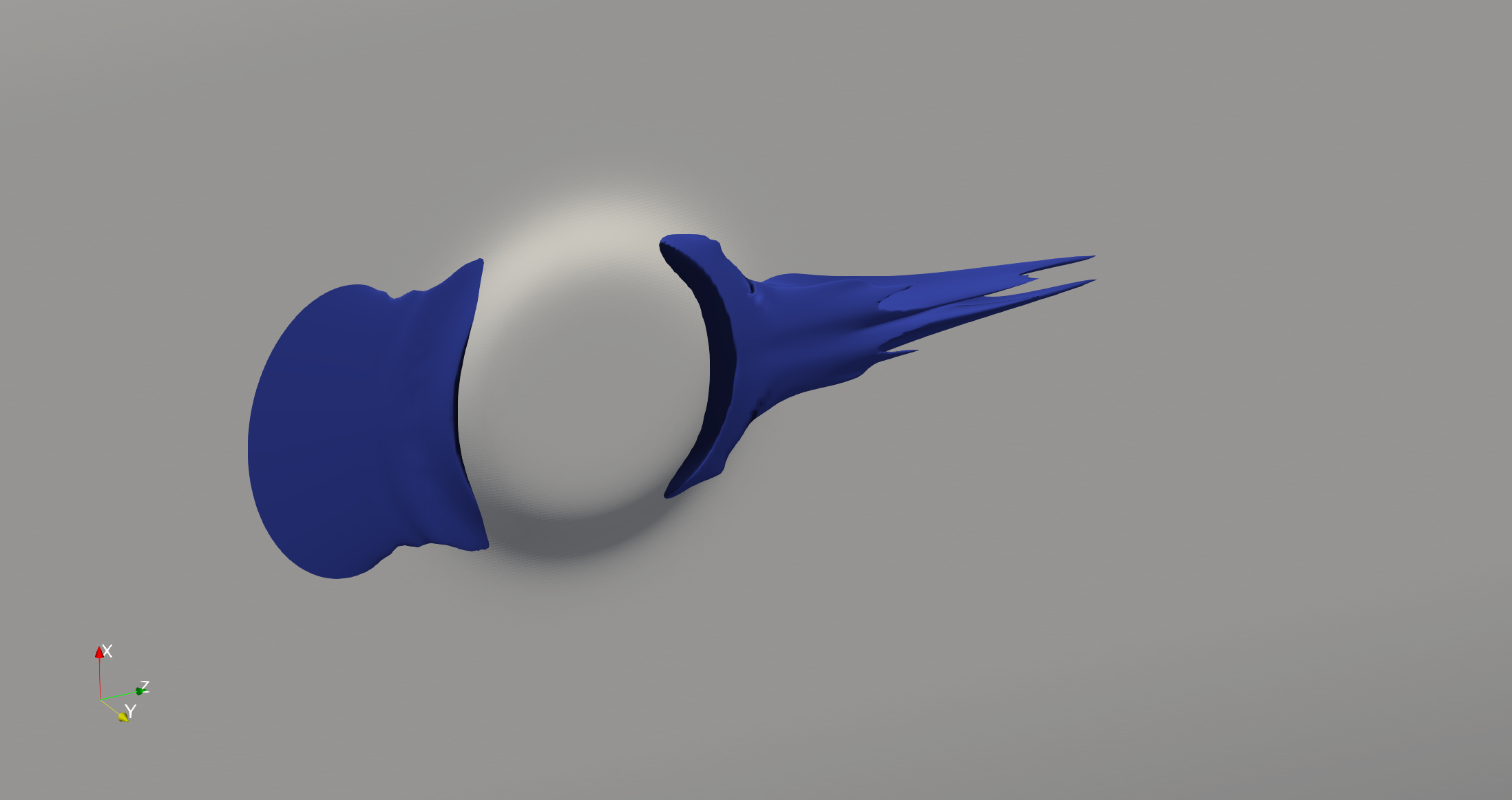}
  \put(1,42){$(b)$}
  \end{overpic} \\
  \begin{overpic}[width=0.48\textwidth]{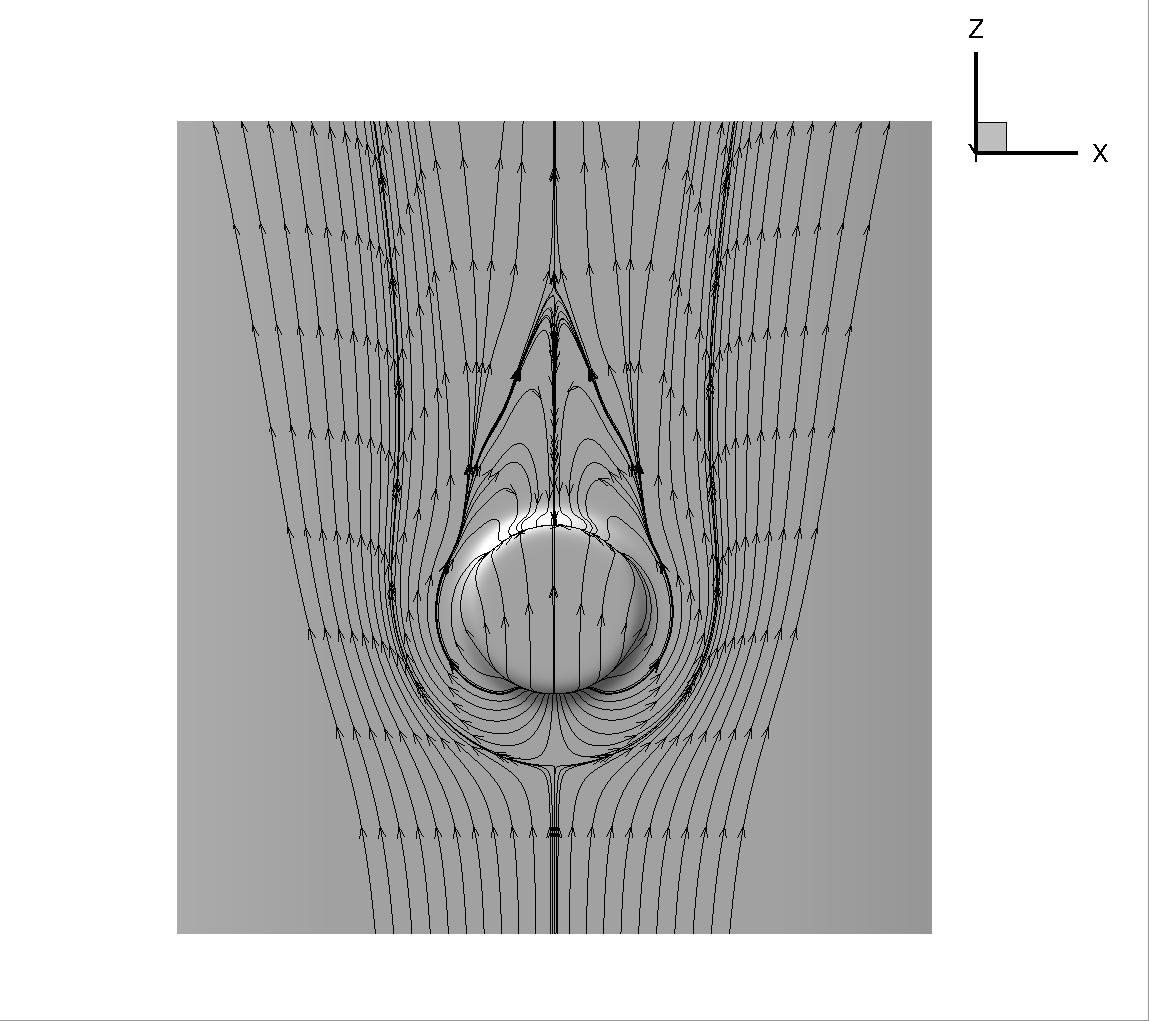}
  \put(1,76){$(c)$}
  \end{overpic} & 
  \begin{overpic}[width=0.48\textwidth]{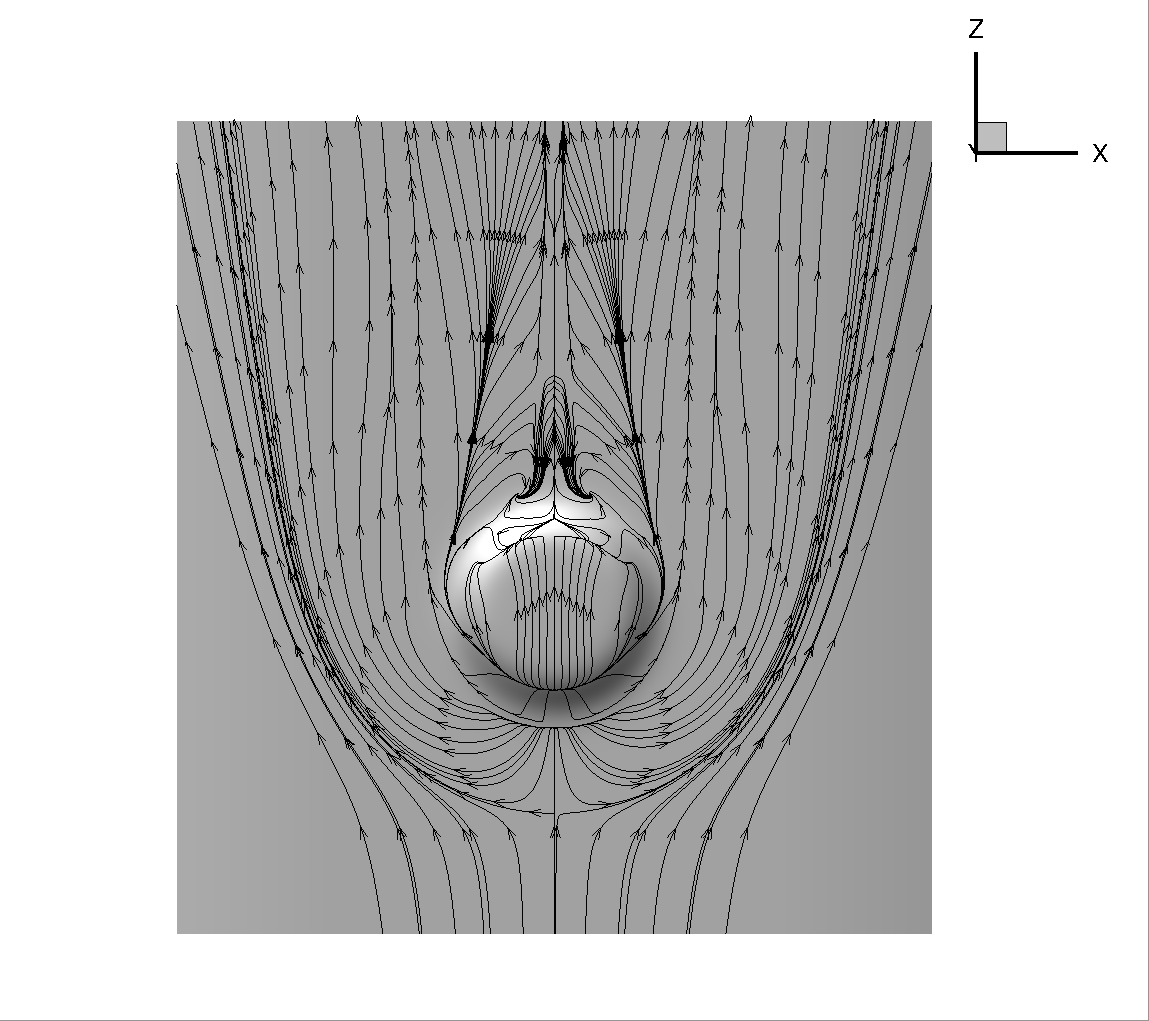}
  \put(1,76){$(d)$}
  \end{overpic} \\
  \begin{overpic}[width=0.48\textwidth]{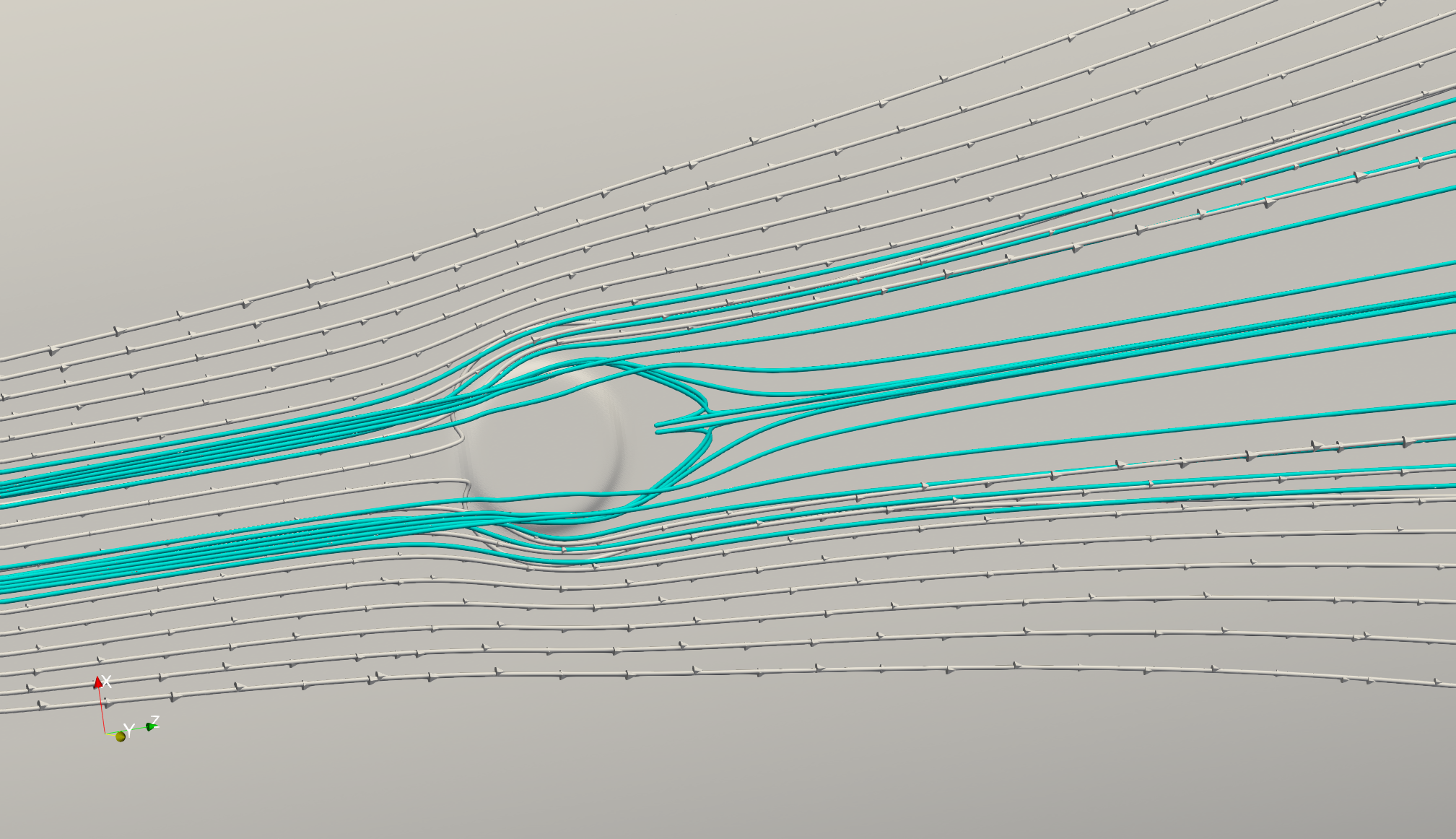}
  \put(1,52){$(e)$}
  \end{overpic} & 
  \begin{overpic}[width=0.48\textwidth]{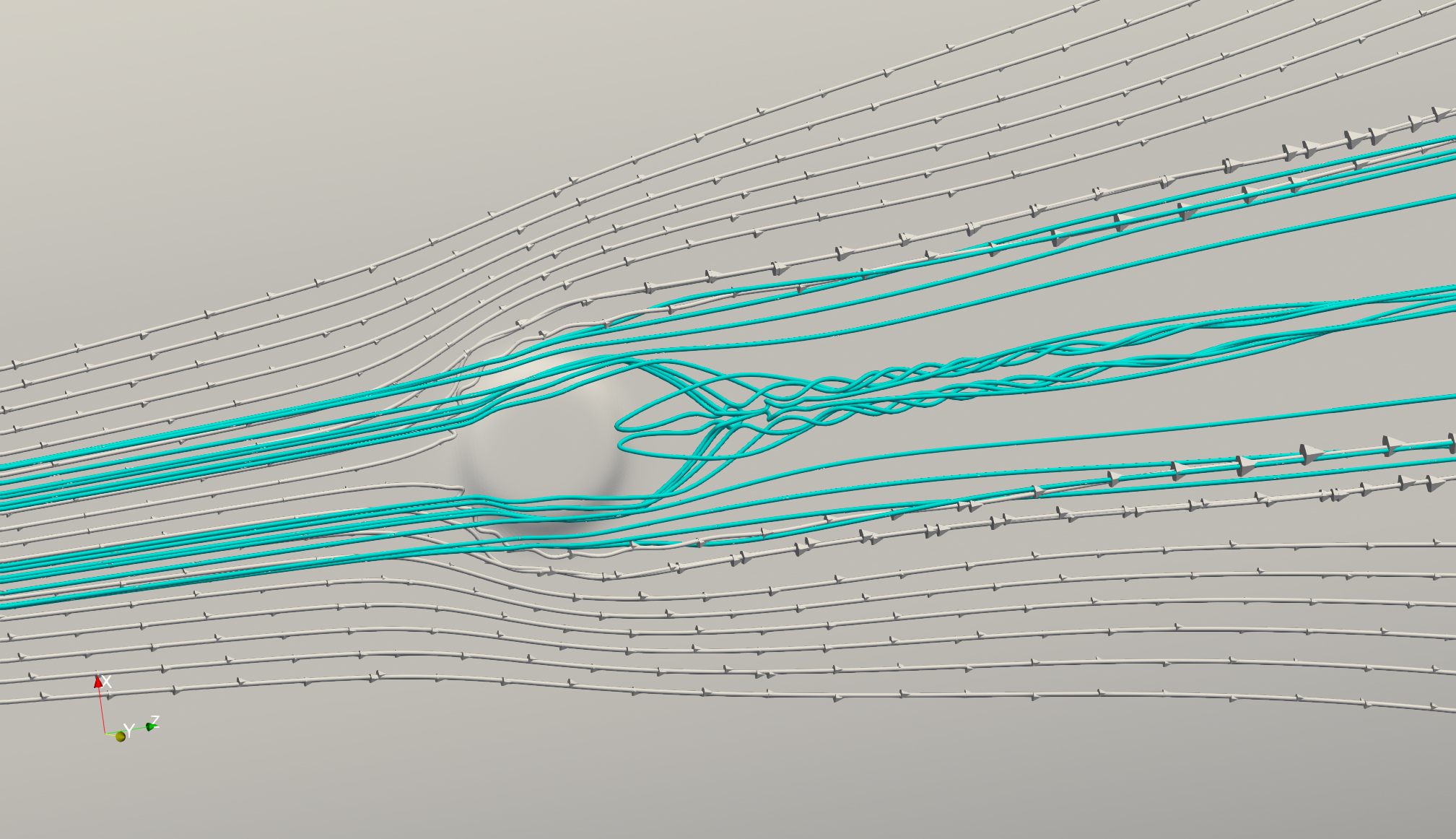}
  \put(1,52){$(f)$}
  \end{overpic} 
  \end{tabular}
  \caption{$(a)$ and $(b)$ stand for the iso-surface of spanwise velocity $\overline{w} = 0$ for H0100 and H0200 cases, respectively. $(c)$ and $(d)$ represent the limiting streamlines along the surfaces for H0100 and H0200 cases, respectively. $(e)$ and $(f)$ stand for the spatial streamlines around the roughness for H0100 and H0200 cases, respectively. The colors of the lines are used to distinguish the different height of the streamlines, upstream. The seeds of the white lines are locate at $h=0.02$, while those of the light blue lines are locate at $h=0.05$.}
  \label{Average_3D_Features}
\end{figure}

Since the flow field induced by roughness elements is inherently a complex three-dimensional flow, characteristics along the attachment line alone are insufficient to adequately reflect the corresponding flow features. 
To capture the fundamental three-dimensional nature of the flow, we present the distribution of limiting streamlines near the roughness elements and the iso-surfaces (Figure \ref{Average_3D_Features}) where the average spanwise velocity $\overline{w}$ is zero, which are used to characterize the three-dimensional flow properties. 
Upstream of the roughness elements, a distinct recirculation region is observed. 
For lower roughness elements, this recirculation region forms a relatively complete separation bubble, as evidenced by the upstream recirculation streamlines. 
For higher roughness elements, in addition to a more pronounced primary separation region, a secondary separation line is clearly visible in the limiting streamlines near the roughness elements. 
As the flow progresses downstream, different degrees of flow convergence and vortex characteristics appear behind the roughness elements. 
The downstream streamlines indicate that the wake of the lower roughness element quickly returns to a more orderly state, which is also reflected by the absence of distinct vortex structures in the wake. The corresponding spatial streamlines in figure \ref{Average_3D_Features}$(e)$ become very orderly, with no noticeable twists or entanglements. 
In contrast, the higher roughness element exhibits completely different characteristics, with the formation of a pair of counter-rotating vortices in the wake, clearly reflected in the corresponding separation lines on the wall in the limiting streamlines in \ref{Average_3D_Features}$(d)$. Also, noticeable twists or entanglements of the spatial streamlines can be found in \ref{Average_3D_Features}$(f)$. 

\subsection{Mechanisms of the roughness induced transition}

\subsubsection{One-point statistics of the transitional flow field}
\begin{figure}
  \centering
 \begin{overpic}[width=\textwidth]{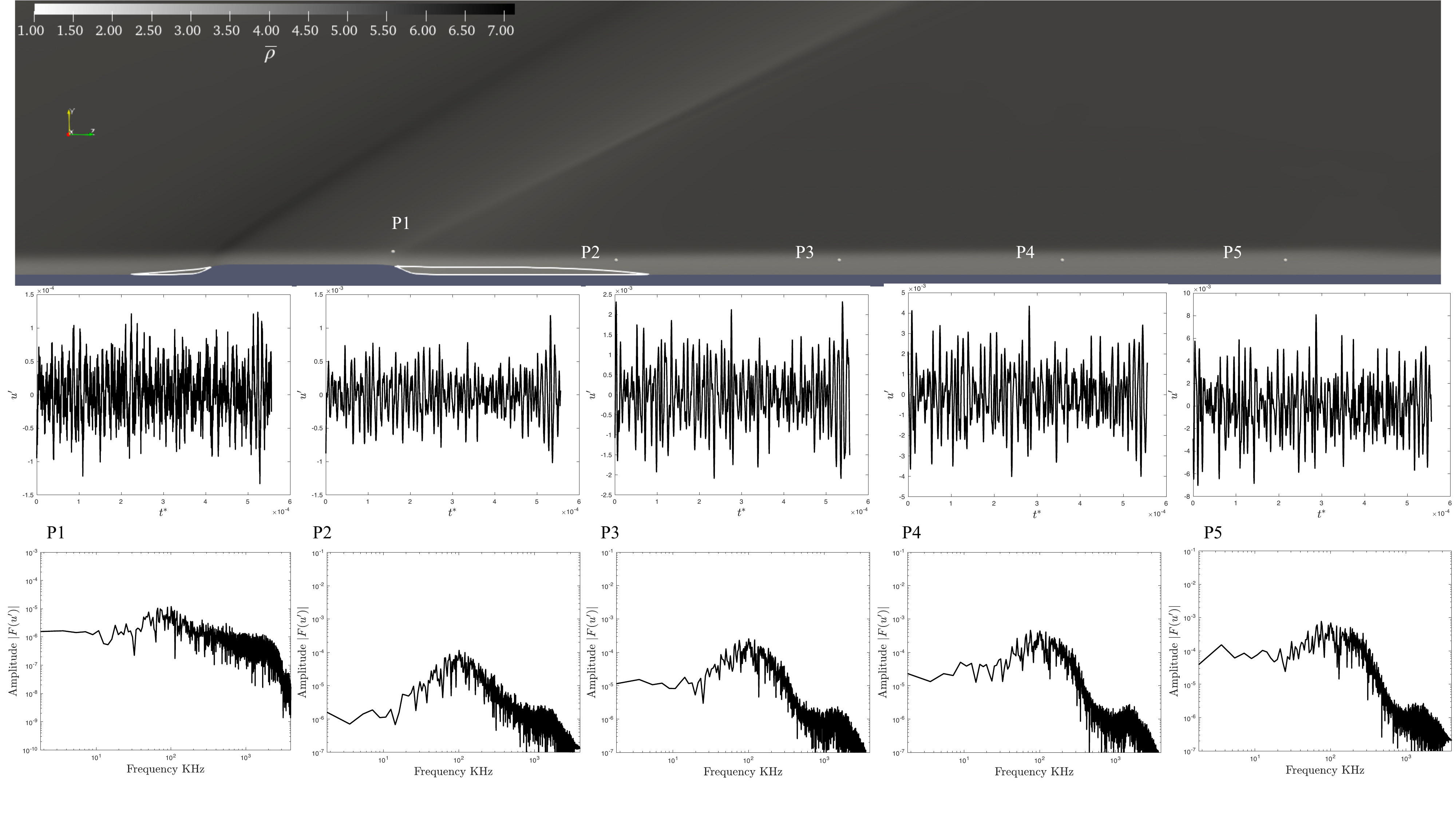}
 \put(-3,54){$(a)$}
 \put(-3,33){$(b)$}
 \put(-3,16){$(c)$}
  \end{overpic}
  \caption{$(a)$ The density contour at the attachment-line plane $x=0$ for case H0100. $(b)$ The temporal evolutions of instantaneous chordwise velocity perturbations $u^{\prime}$ of the selected points. $(c)$ The amplitudes for different Fourier modes $|F(u^{\prime})|$ for different frequency at the selected points.}
  \label{H01_Slices_TM}
\end{figure}

\begin{figure}
  \centering
 \begin{overpic}[width=\textwidth]{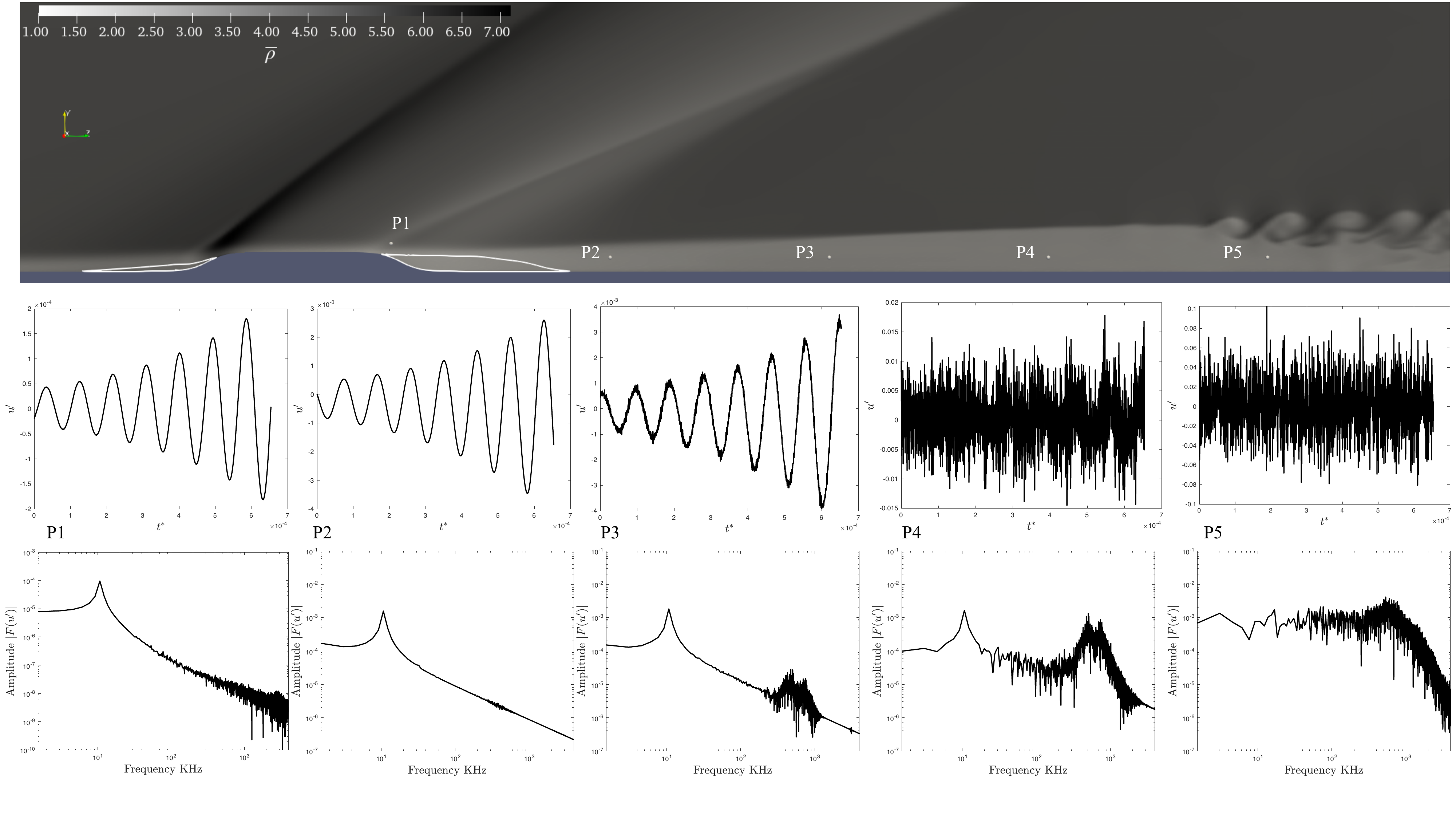}
  \put(-3,54){$(a)$}
 \put(-3,33){$(b)$}
 \put(-3,16){$(c)$}
  \end{overpic}
  \caption{$(a)$ The density contour at the attachment-line plane $x=0$ for case H0200. $(b)$ The temporal evolutions of instantaneous chordwise velocity perturbations $u^{\prime}$ of the selected points. $(c)$ The amplitudes for different Fourier modes $|F(u^{\prime})|$ for different frequency at the selected points.}
  \label{H02_Slices_TM}
\end{figure}

In this section, we try to understand the transition mechanisms of these flows by using long time one-point spectra statistics. At the beginning, all sampled points are located at the attachment-line plane($x = 0$). Figure \ref{H01_Slices_TM} and \ref{H02_Slices_TM} show the  flow fields at the exact attachment-line plane for the two cases, together with the statistical results of the selected points. 
The regions of the revease flow near the roughness are pointed by the white lines. 
As mentioned previous, the two distinct cases deliver two different dynamics with respect to the transition processes. 
In the first scenario, case H0100, the transition would not occur within the wake flow induced by the roughness elements, without the unsteady perturbations, even though the roughness Reynolds number $Re_{kk}$ or the height of the roughness $k_h$ are beyound the critical values in normal plate boundary layers\citep{Bernardini2012,Bernardini2014,Estruch2017}. 
Contrariwise, as the height of roughness element increases to 0.2, the transition will occur without employing forcing of any kind, which suggest a self-sustaining mechanism that causes the flow to transition.

This is also reflected in the corresponding measurement point signals of $u^{\prime}$. 
In H0100 case, we introduced random perturbations, resulting in the signal detected at point $P_1$ exhibiting typical broadband characteristics, with energy distributed relatively uniformly across a range of frequencies. 
Additionally, there is a slight increase in perturbation amplitude at the frequency around 100 KHz.
As the flow continues to develop downstream from $P_2$ to $P_5$, the perturbations gradually increase, and disturbances of the frequency around 100 KHz become the dominant perturbations. 
This leads to the frequency amplitude distribution characteristics evolving towards typical turbulent features.

When the height of the roughness element increases to 0.2 mm, prominent instability waves are observed at points $P_1, P_2$ and $P_3$. 
The figures indicate that the disturbances, which are amplified as they travel downstream from $P_1$ to $P_3$, not only grow in a convective manner but also exhibit characteristics of absolute instability, as they are amplified over time at fixed locations. 
Based on the results of the discrete Fourier transformation, the frequency of the most representative perturbations is around 10 KHz, referred to as low-frequency perturbations in this study. 
As these perturbations evolve downstream, some high-frequency components are gradually amplified. This amplification is evidenced by the increasing amplitude of disturbances in the region beyond 400 KHz. 
Simultaneously, the low-frequency perturbations appear to reach a saturation state as they evolve downstream, with their amplitude showing minimal growth. This is evidenced by the nearly constant amplitude of low-frequency disturbances from points $P_2$ to $P_4$ in the figures. 
As the flow continues to develop further downstream to point $P_5$, the overall disturbance spectrum exhibits typical broadband characteristics which indicates that the flow are stepping into full turbulent.

To better understand the mechanisms and identify possible nonlinear coupling features during the transition processes, higher-order spectral (HOS) analysis is employed. 
Specifically, bispectral analysis is utilized here to examine nonlinear signals. 
This method enables the detection and quantification of possible nonlinear interactions between different frequency components of the signals.
The bispectrum $\mathscr{B}(\omega_1, \omega_2)$ for a signal $f(t)$ is defined as
\begin{equation} \label{eq_bispectrum}
\mathscr{B}(\omega_1, \omega_2) =  E\left[F(\omega_1)F(\omega_2)F^c(\omega_1 + \omega_2)\right],
\end{equation}
where $F(\omega)$ is the Fourier transform of the temporal signal $f(t)$ and $\omega$ is the frequency. $E[.]$ stands for an expected value. The superscript $c$ represents the complex conjugate.   

\begin{figure}
  \centering
  \begin{tabular}{ccc}
 \begin{overpic}[width=0.32\textwidth]{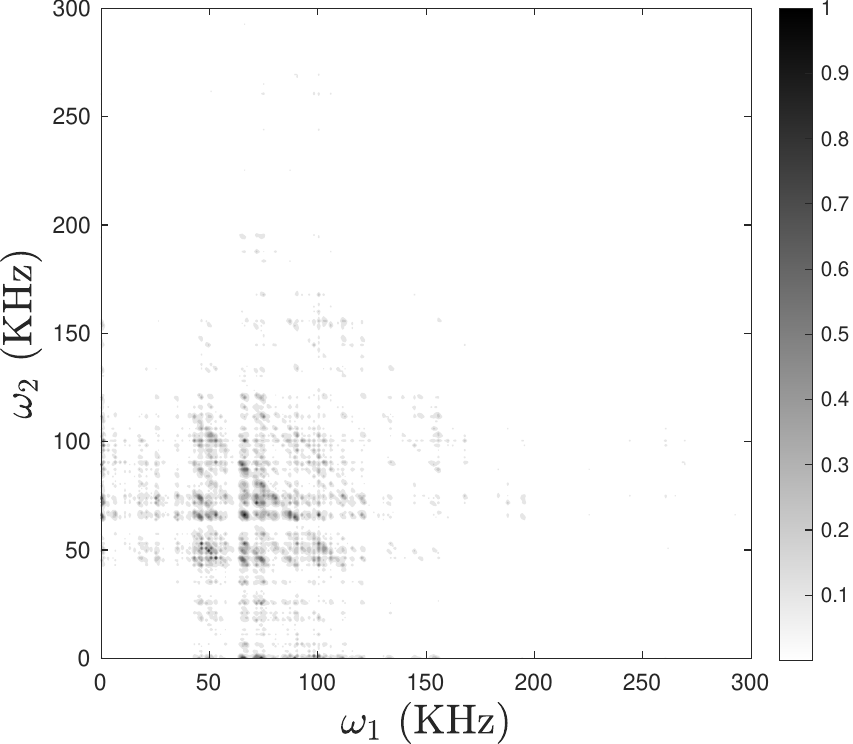}
 \put(18,75){$(a)$}
  \end{overpic} & 
 \begin{overpic}[width=0.32\textwidth]{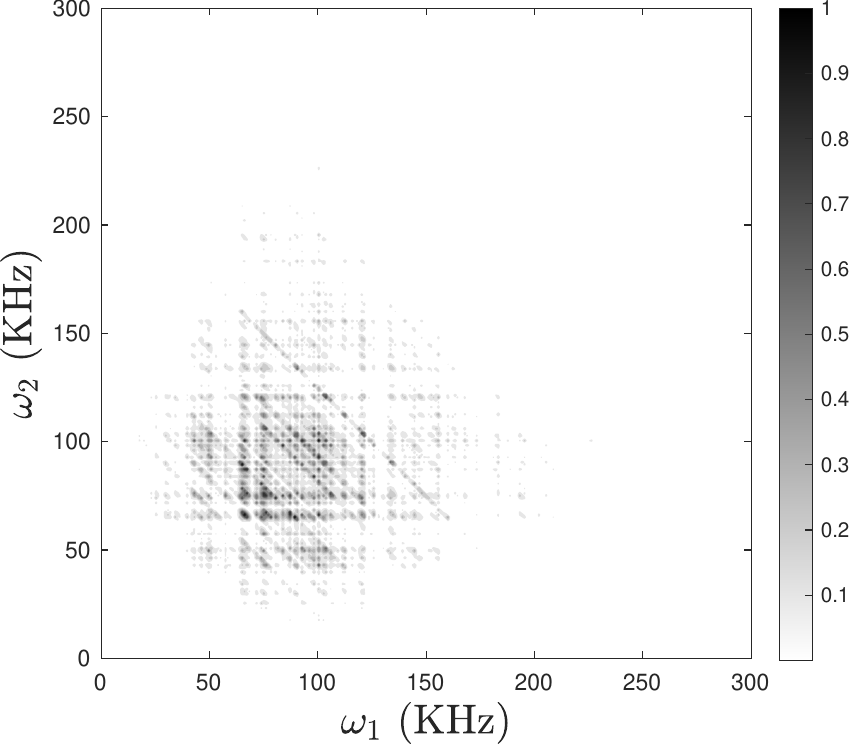}
 \put(18,75){$(b)$}
  \end{overpic} &
 \begin{overpic}[width=0.32\textwidth]{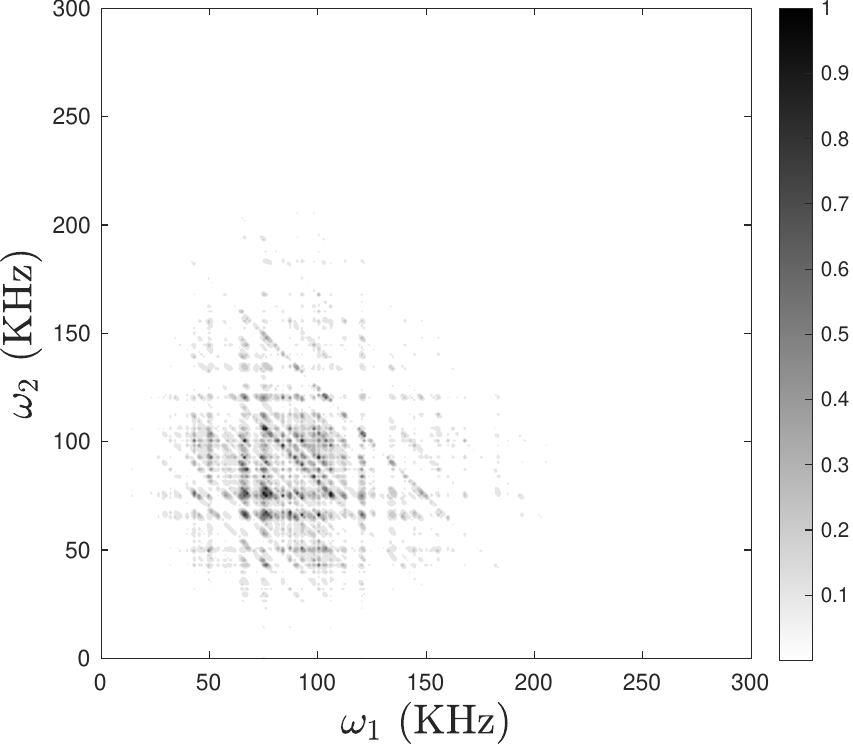}
 \put(18,75){$(c)$}
  \end{overpic} 
  \end{tabular}
  \begin{tabular}{cc}
 \begin{overpic}[width=0.32\textwidth]{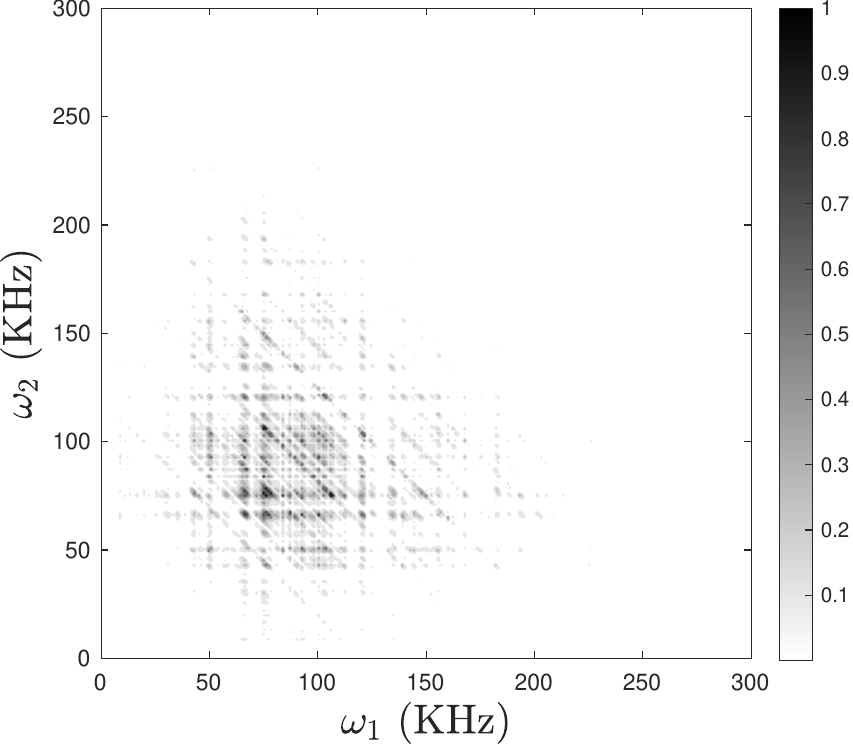}
 \put(18,75){$(c)$}
  \end{overpic} &
 \begin{overpic}[width=0.32 \textwidth]{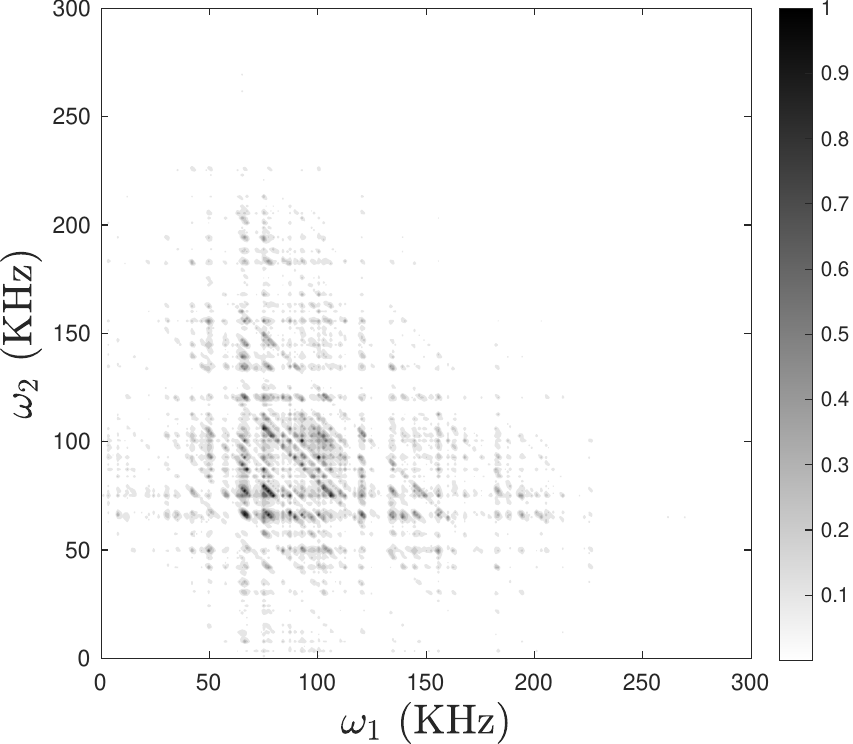}
 \put(18,75){$(d)$}
  \end{overpic}    
  \end{tabular}   
  \caption{The normalized bispectrum $|\mathscr{B}|$ of the perturbations $u^{\prime}$ at the points $P_1$ to $P_5$ for the H0100 case. Panels $(a)$ to $(d)$ correspond to points $P_1$ to $P_5$, respectively.}
  \label{H01_Bispectrum_TM}
\end{figure}

\begin{figure}
  \centering
  \begin{tabular}{ccc}
 \begin{overpic}[width=0.32\textwidth]{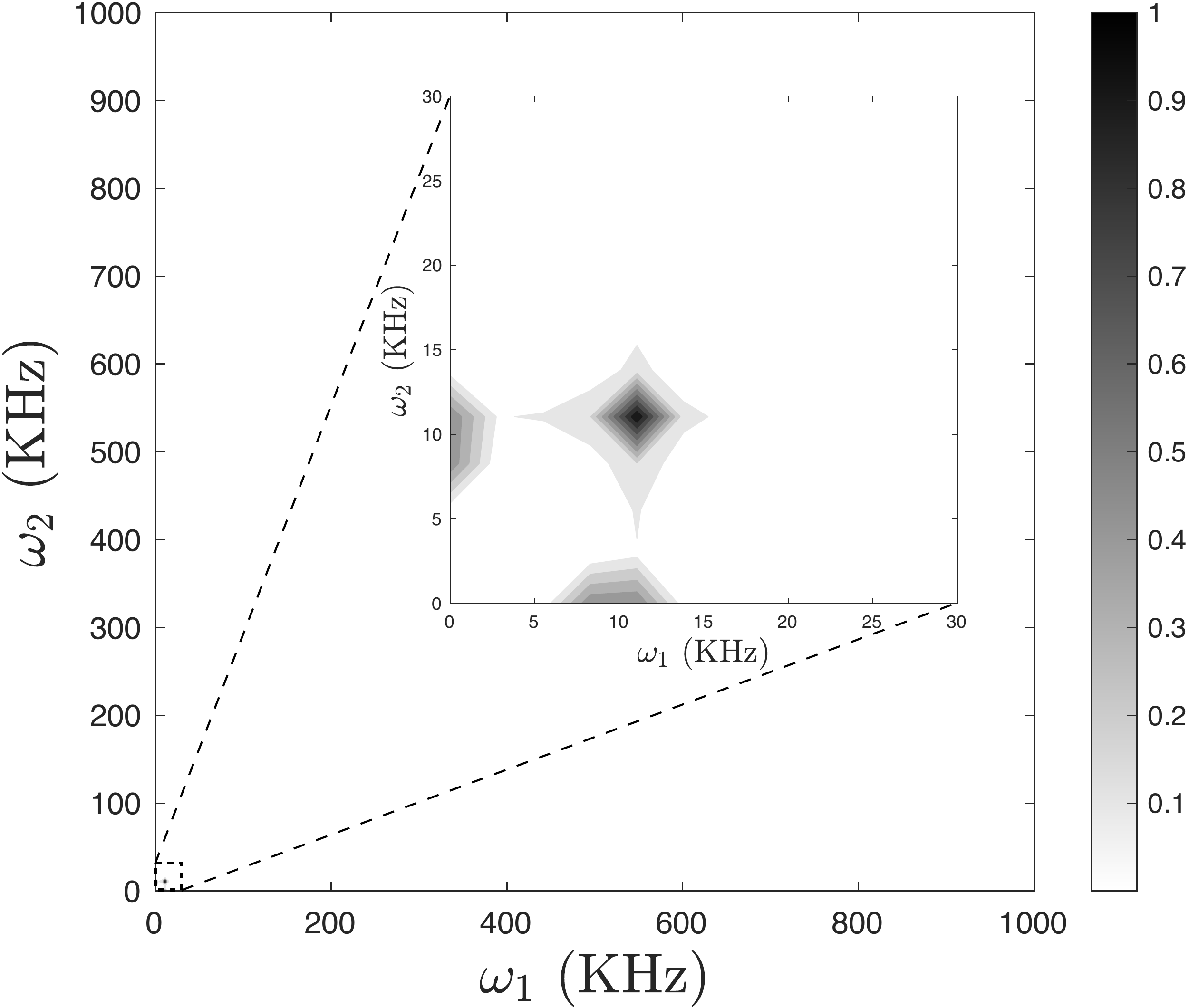}
 \put(18,75){$(a)$}
  \end{overpic} & 
 \begin{overpic}[width=0.32\textwidth]{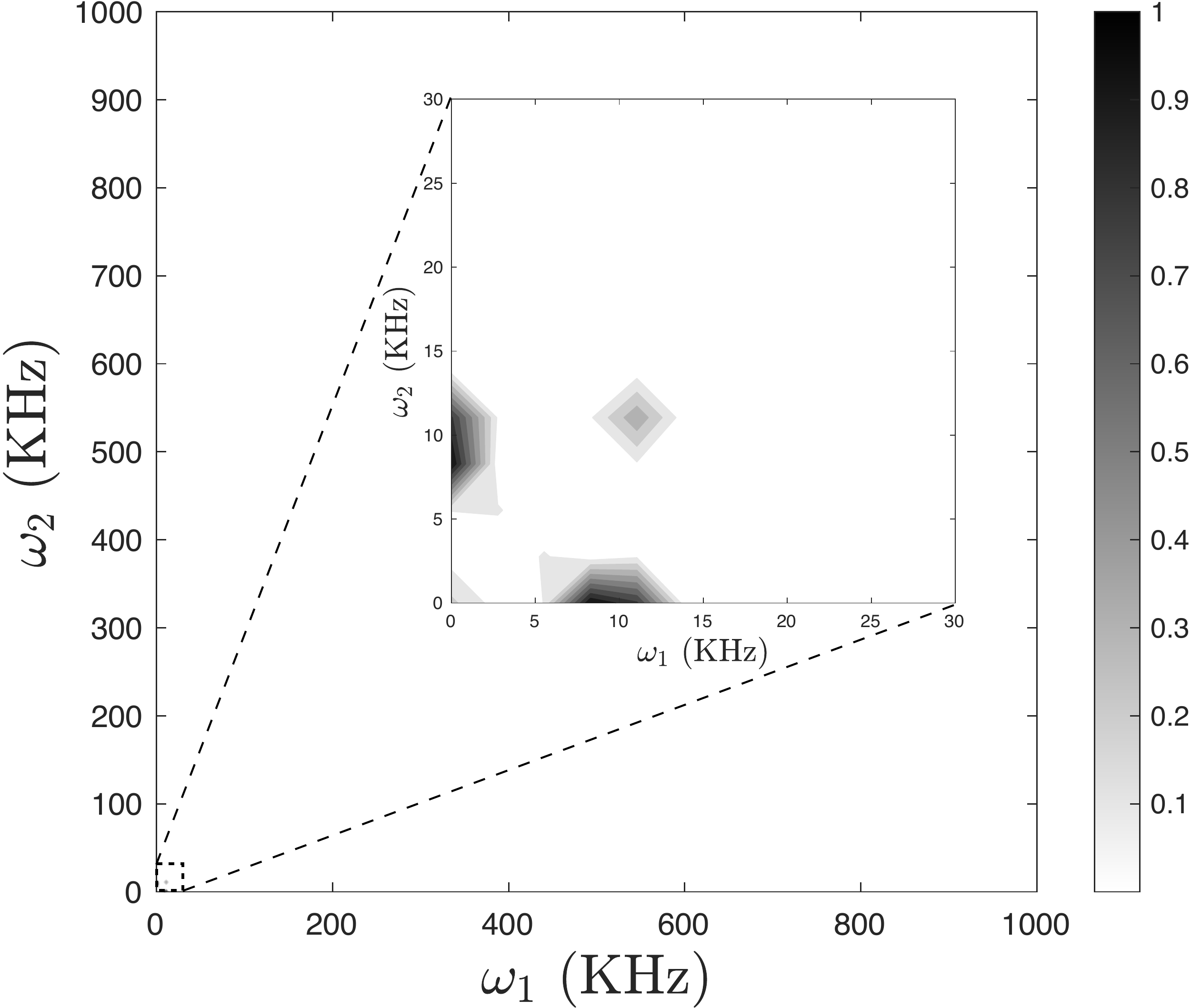}
 \put(18,75){$(b)$}
  \end{overpic} &
 \begin{overpic}[width=0.32\textwidth]{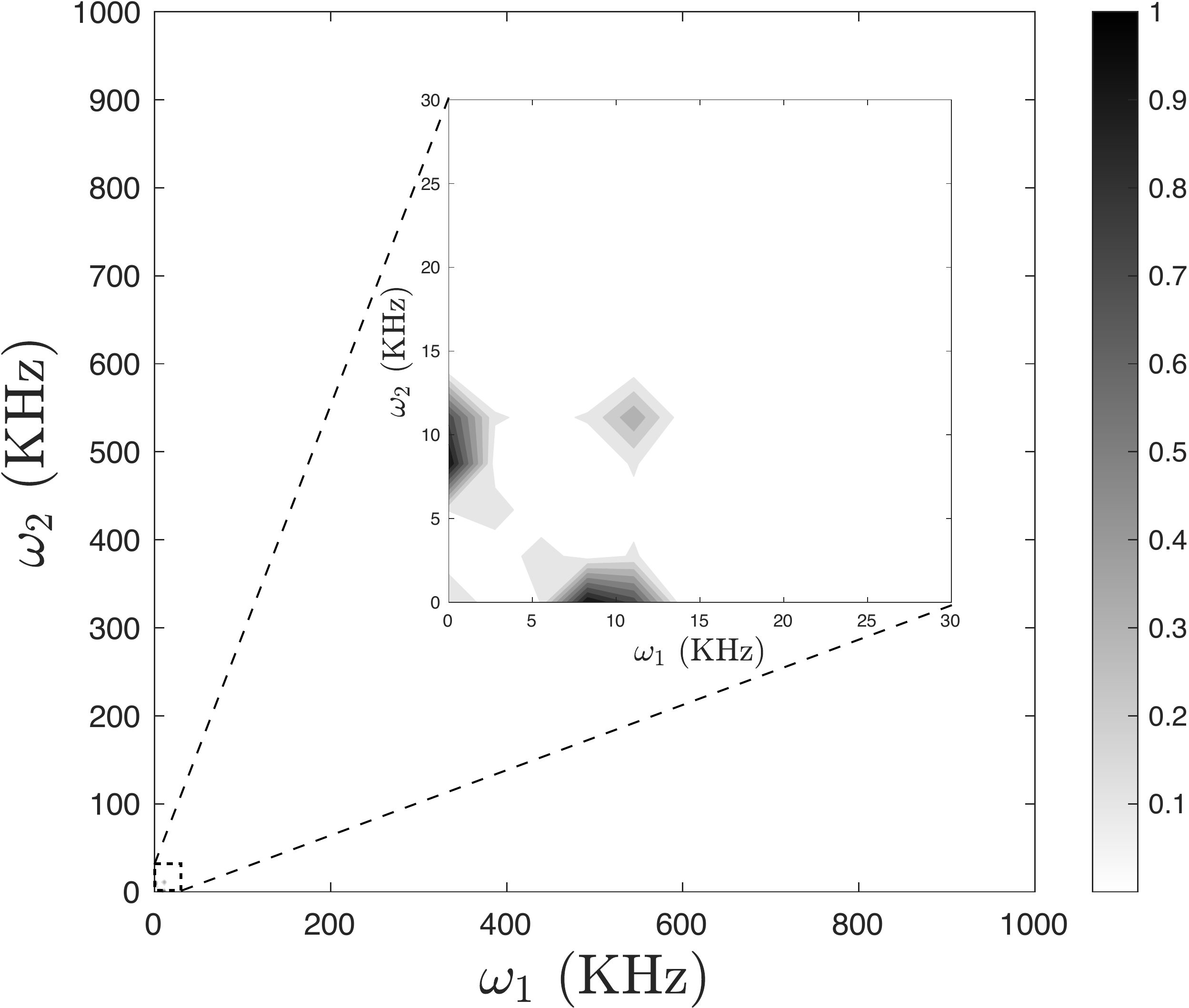}
 \put(18,75){$(c)$}
  \end{overpic} 
  \end{tabular}
  \begin{tabular}{cc}
 \begin{overpic}[width=0.32\textwidth]{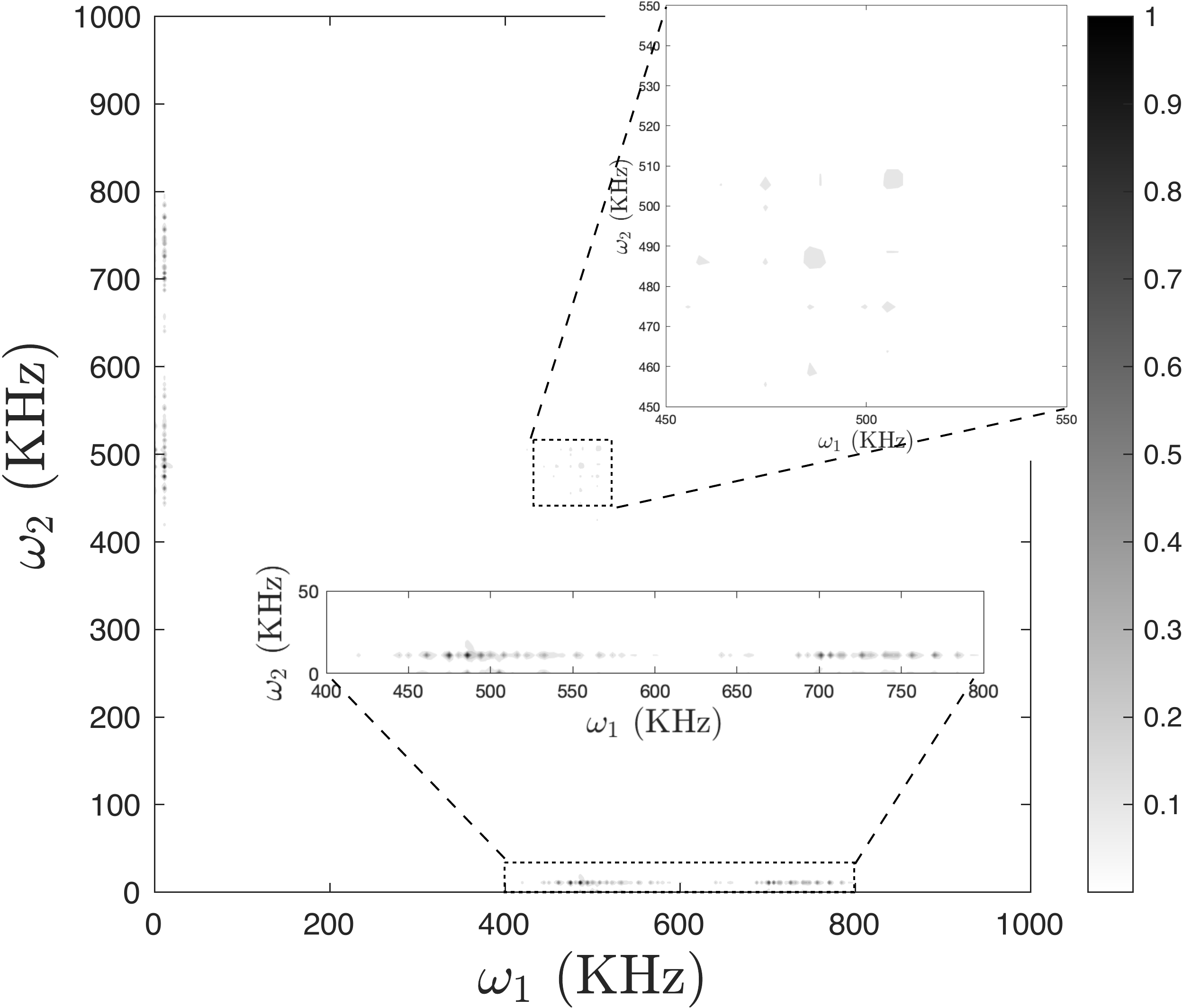}
 \put(18,75){$(c)$}
  \end{overpic} &
 \begin{overpic}[width=0.32\textwidth]{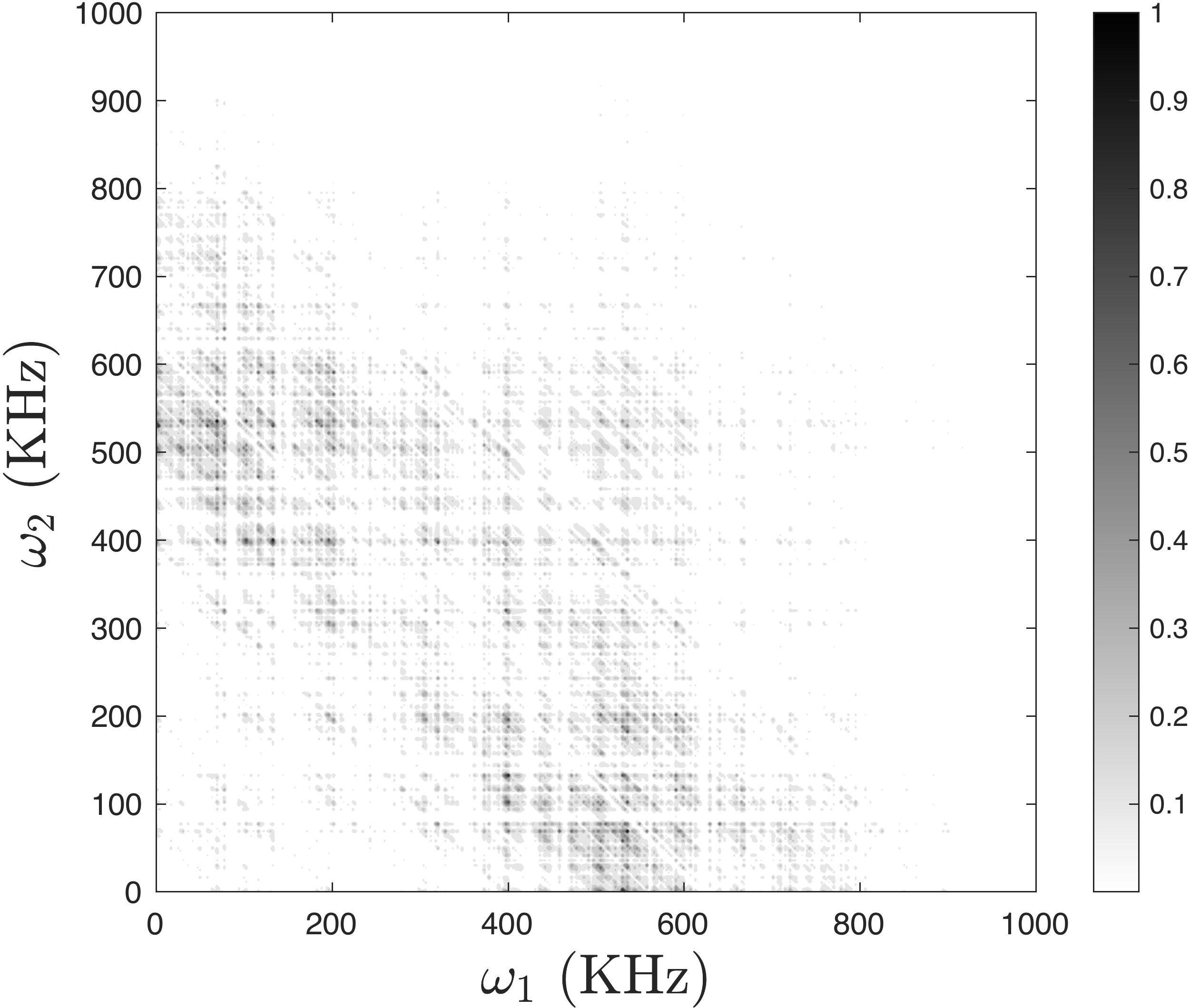}
 \put(18,75){$(d)$}
  \end{overpic}    
  \end{tabular}   
  \caption{The normalized bispectrum $|\mathscr{B}|$ of the perturbations $u^{\prime}$ at the points $P_1$ to $P_5$ for the H0200 case. Panels $(a)$ to $(d)$ correspond to points $P_1$ to $P_5$, respectively. Note that certain features have been magnified for enhanced readability and clarity.}
  \label{H02_Bispectrum_TM}
\end{figure}

The bispectrum of selected points in figure \ref{H01_Slices_TM} and \ref{H02_Slices_TM} are shown in figure \ref{H01_Bispectrum_TM} and \ref{H02_Bispectrum_TM}, respectively. 
For both cases, the most representative interactions between two waves with different frequencies are shown.
Based on the definition of bispectrum \eqref{eq_bispectrum}, it measures the nonlinear interactions between frequencies $\omega_1$ and $\omega_2$, as well as their sum $\omega_1 + \omega_2$. 
For a purely linear signal, the bispectrum theoretically should be zero or very close to zero. 
For general nonlinear signal, the diagonal elements (if $\omega_1 = \omega_2 = \omega_0$) of the bispectrum reflect the interactions between a frequency $\omega_0$ with itself and its double frequency $2\omega_0$. Significant values along the diagonal often indicate the presence of harmonic components in the signal. The off-diagonal elements (if $\omega_1 \neq \omega_2$) illustrate the nonlinear interactions between distinct frequencies $\omega_1$ and 
$\omega_2$, and their sum frequency $\omega_1 + \omega_2$. The obvious values off the diagonal suggest the nonlinear coupled phenomena between different frequency components. 

The basic behaviour of the perturbations, shown in figure \ref{H01_Bispectrum_TM} and \ref{H02_Bispectrum_TM}, are the same as described before. 
For the H0100 case, the results indicate that regions with larger magnitudes of low-frequency disturbances are primarily located on or near the diagonal. 
This suggests that in this condition, the excitation of higher-order harmonics plays a significant role in the nonlinear evolution of the corresponding disturbances. 
As the disturbances propagate downstream, the spectral distribution of disturbances at points $P_2$ through $P_4$ appears to remain relatively unchanged. 
This indicates that the composition of the disturbances remains nearly constant, suggesting that the disturbances have grown to a certain extent and have reached a nearly saturation stage.

For the H0200 case, the bispectrum exhibits characteristics that are markedly different from the previous condition. 
At the first three signal recording points ($P_1$, $P_2$, and $P_3$), the dominant disturbances in the overall disturbance spectrum appear only in the low-frequency region, consistent with the results from the previous power spectrum analysis. 
As the disturbances further develop to point $P_4$, some high-frequency disturbances begin to emerge, roughly within the $\left[400 \text{KHz}, 800 \text{KHz}\right]$ range. 
These high-frequency disturbances exhibit significant nonlinear interactions with the low-frequency disturbances as well as the zero-frequency disturbances, with the frequencies of the low-frequency disturbances remaining consistent with those recorded at the previous three points. 
This suggests that the low-frequency disturbances have evolved sufficiently, and the detected high-frequency disturbances correspond to a secondary instability arising in the already saturated low-frequency disturbances combined with the mean field.

\begin{figure}
  \centering
  \begin{tabular}{c}
  \begin{overpic}[width=0.98\textwidth]{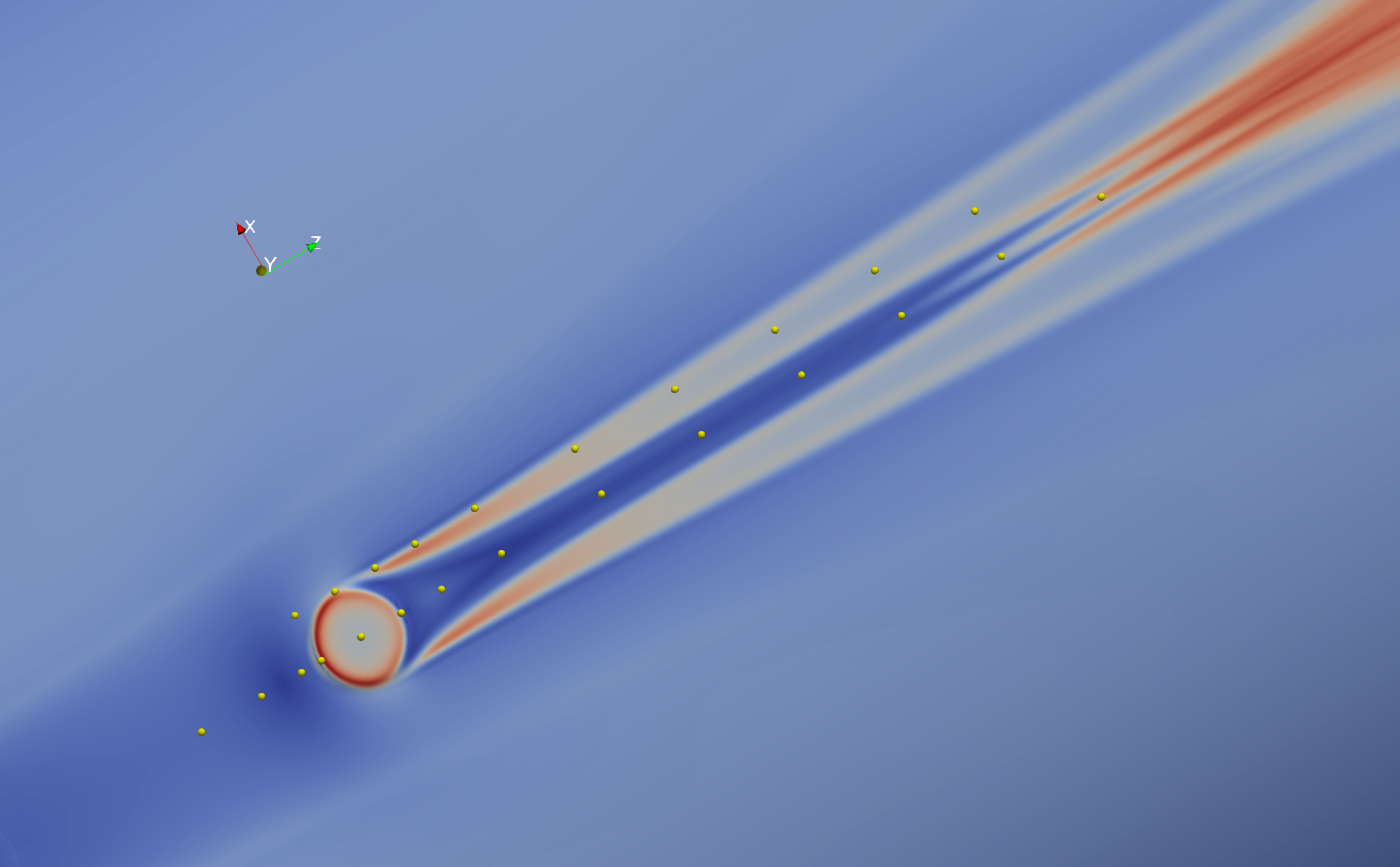}
  \put(1,55){$(a)$}
  \put(14,9){\vector(-1,-1){4}}
  \put(5,2){$P_{s_1,h_1}$}
  \put(79,47){\vector(2,-3){3}}
  \put(80,38){$P_{s_{14},h_1}$}
  \put(20.5,19){\vector(-1,1){4}}
  \put(12,25){$P_{s_{15},h_1}$}
  \put(69,48){\vector(-1,1){3}}
  \put(60,53){$P_{s_{24},h_1}$}
  \put(23.5,13.8){\vector(1,-2){3}}
  \put(28,5){$P_{s_{4},h_1}$}
  \put(32.5,19.5){\vector(1,-1){5}}
  \put(40,11){$P_{s_{7},h_1}$}
  \end{overpic} \\
  \begin{overpic}[width=0.98\textwidth]{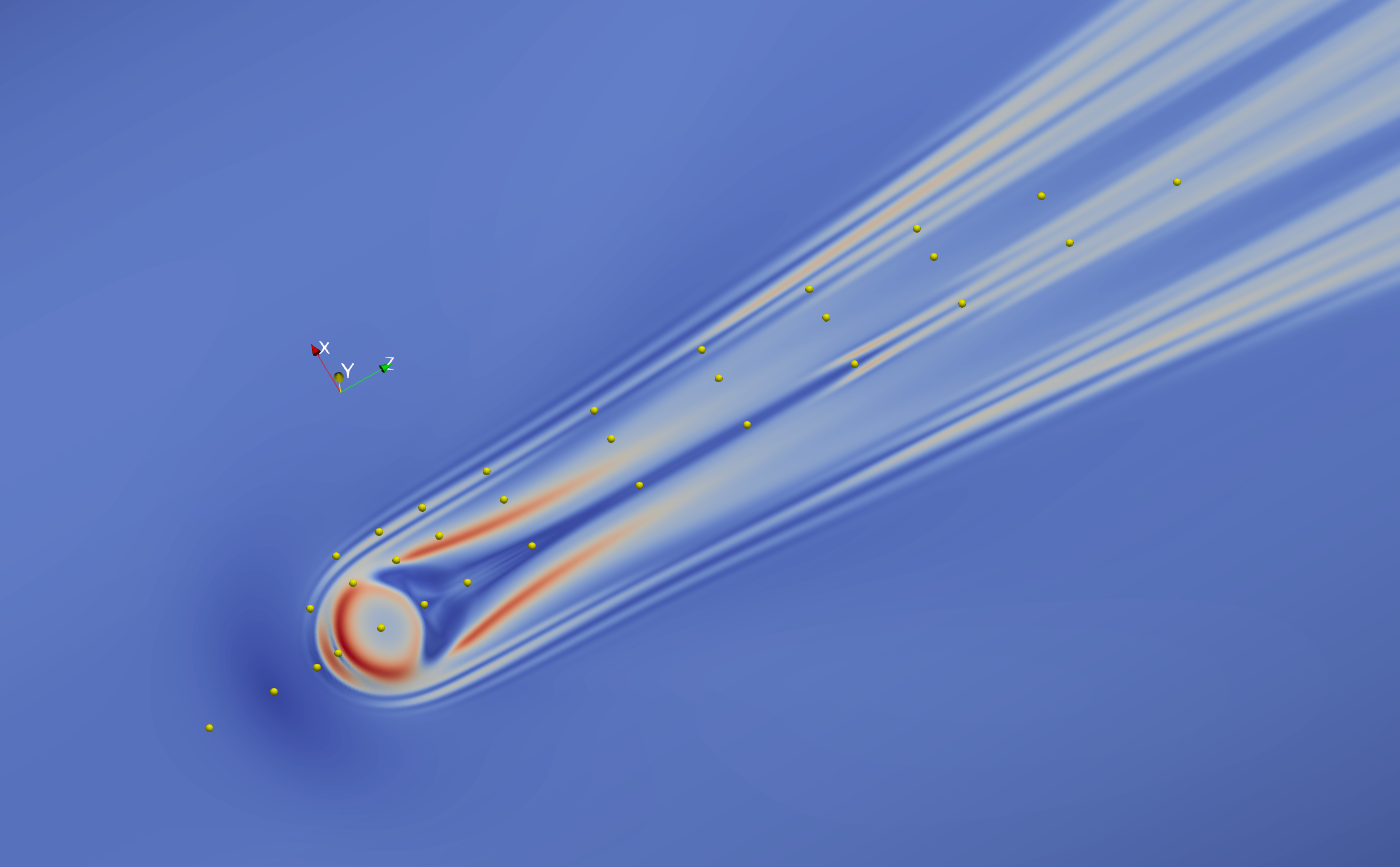}
  \put(1,55){$(b)$}
  \put(23.5,23){\vector(-1,1){3}}
  \put(15,28){$P_{s_{25},h_2}$}
   \put(65,46){\vector(-1,1){5}}
  \put(55,52){$P_{s_{32},h_2}$}
  \end{overpic}
  \end{tabular}
  \caption{The selected points for two cases. $(a)$ for H0100, $(b)$ for H0200. The corresponding points are sequentially recorded as $s_1,s_2,\cdots,s_{32}$ along the $z$-axis from upstream to downstream, starting from the attachment line to  chordwise downstream. The subscripts $h_1$ and $h_2$ are used to distinguish the points in different cases.}
  \label{Srf_SF_Select_Points}
\end{figure}

\begin{figure}
  \centering
  \begin{tabular}{c}
   \begin{overpic}[width=0.66\textwidth]{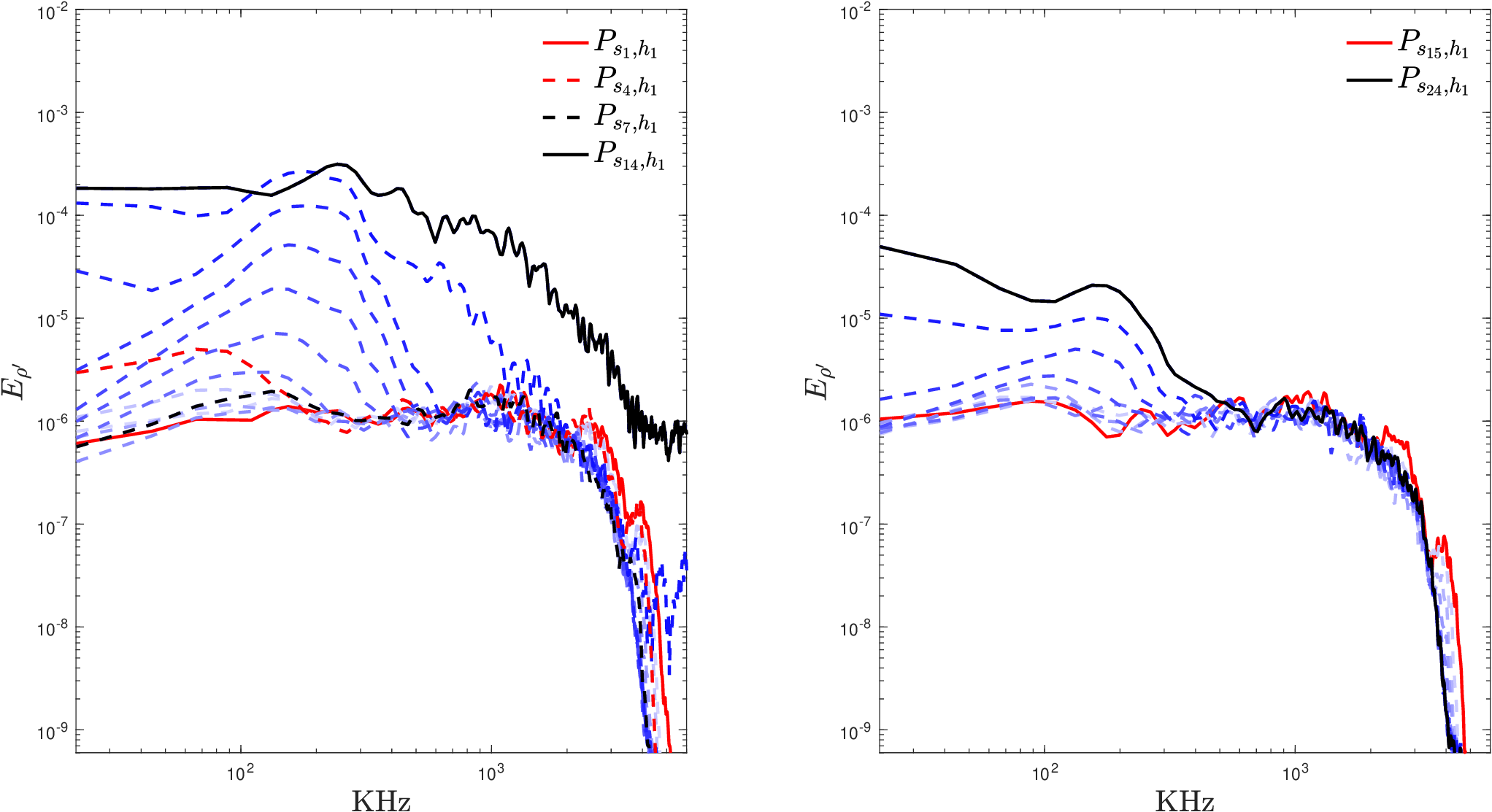}
   \put(8,47){$(a)$}
   \put(63,47){$(b)$}
   \end{overpic} \\
   \begin{overpic}[width=0.99\textwidth]{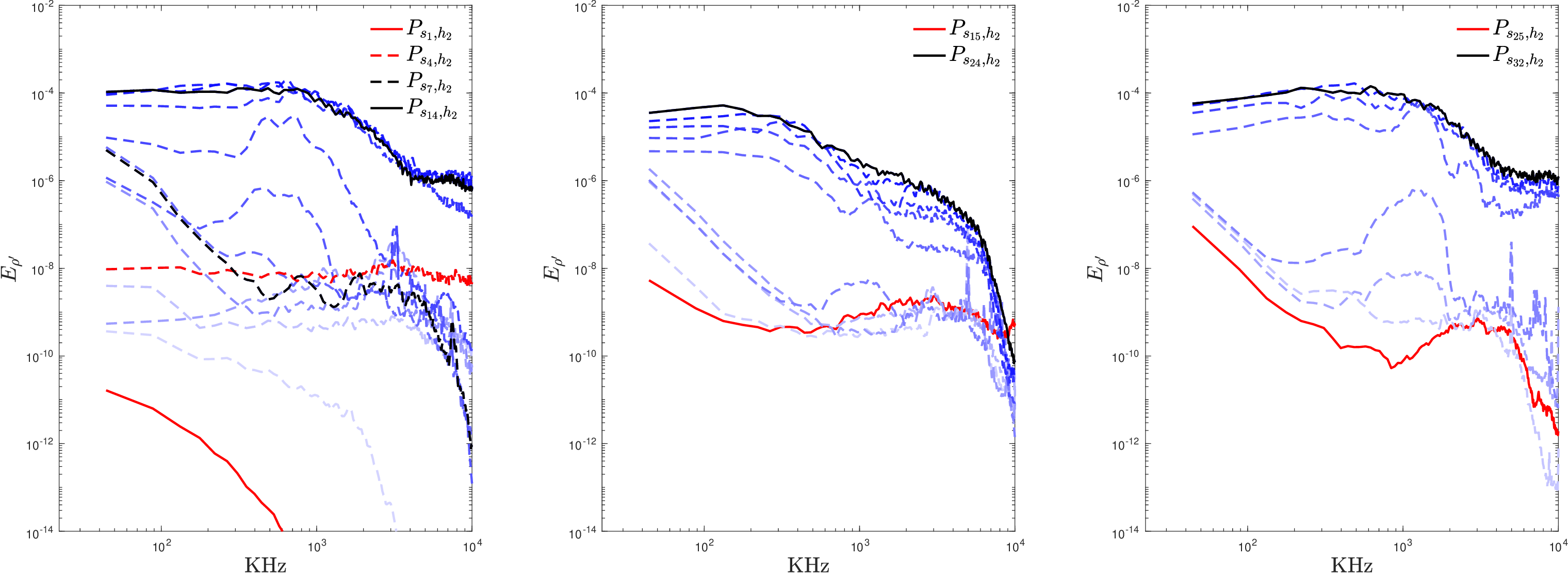}
   \put(8,32){$(c)$}
   \put(42,32){$(d)$}
   \put(76,32){$(e)$}
   \end{overpic}   
  \end{tabular}
  \caption{The spectra $E_{\rho^{\prime}}$ of perturbations density $\rho$ at the selected points in figure \ref{Srf_SF_Select_Points}. $(a)$ and $(b)$ stand for the two groups of selected points in H0100 case. $(c-e)$ represent the three groups of selected points in H0200 case. From light to dark blue dashed lines, the spectra represent points from spanwise upstream to downstream.}
  \label{Srf_SF_Select_Points_Spectra}
\end{figure}

To provide a more comprehensive understanding of the disturbance evolution process during transition, we shift our focus slightly away from the wake of roughness elements at the attachment-line region to analyze the characteristics of disturbance evolution around the entire roughness element, including the upstream and horseshoe vortex regions.
In order to trace the evolution characteristics of disturbances in the three-dimensional flow field, we select and record the fluctuating density on the three-dimensional surfaces around the roughness elements for analysis. The selected points for H0100 and H0200 cases are shown in figure \ref{Srf_SF_Select_Points} together with the surface skin friction. 
The selected points can roughly be divided into three groups. 
The first group is located along the exact attachment line, extending from upstream to downstream of the roughness element, labeled sequentially as $s_1, s_2, \cdots, s_{14}$. 
The second group is located on the side of the roughness element, also extending from upstream to downstream, labeled as $s_{15}, s_{16}, \cdots, s_{24}$. 
The third group, labeled as $s_{25}, s_{26}, \cdots, s_{32}$, is also on the side of the roughness element but is further from the second group of detection points. The first two groups are the same for both cases, while the third group appears only in the H0200 case to track the evolution of the corresponding horseshoe vortex.

The spectra of those points are shown in figure \ref{Srf_SF_Select_Points_Spectra}. 
In the scenarios of lower height roughness element, the incoming flow ahead of the roughness is subjected to the additional perturbations. Based on the spectra(the red lines in figure \ref{Srf_SF_Select_Points_Spectra}$(a)$ and $(b)$), no dominant frequency could be observed.
As the flow gradually approaches the roughness element, the overall amplitude of disturbances increases progressively (as indicated by the red dashed line in Figure \ref{Srf_SF_Select_Points_Spectra}$(a)$). After the fluid passes over the roughness element, the disturbances are somewhat suppressed due to the favourable pressure gradient resulting from the expansion effects of the high-pressure region induced by the shock wave ahead of the roughness. 
This suppression leads to a decrease in the amplitude of the disturbances (as indicated by the black dashed line in Figure \ref{Srf_SF_Select_Points_Spectra}$(a)$). 
However, once the flow moves past the roughness element, the disturbances exhibit a marked tendency to increase again. 
It is important to note that the frequency range of disturbances that first starts to grow significantly is approximately 100 KHz, which is consistent with previous analyses. 
As for the vortex structures formed on both sides of the roughness element, they also exhibit similar patterns of change (as indicated in Figure \ref{Srf_SF_Select_Points_Spectra}$(b)$).
However, the corresponding amplitudes have not grown significantly large (i.e., a distinct plateau region appears in the mid-to-low frequency range), which correlates with the previous observation that no significant transition phenomena were observed on both sides. 

In scenarios involving a higher roughness element, different phenomena are observed (as shown in Figures \ref{Srf_SF_Select_Points_Spectra}$(c-e)$). 
Due to the absence of additionally introduced artificial disturbances under this condition, the amplitude of disturbances upstream of the roughness element is relatively low, especially in the mid-to-high frequency range. 
As the disturbances approach the roughness element, the amplitude of low-frequency components is generally amplified. 
Additionally, within the separation bubbles both upstream and downstream of the roughness element, some peaks appear in the high-frequency range (indicated by the red dashed line and black dashed line, where the red dashed line represents the upstream separation bubble and the black dashed line represents the downstream separation bubble), reflecting the high-frequency characteristics of the separation bubbles. 
Similar to the lower roughness element, disturbances exhibit a certain degree of suppression after passing over the roughness element (the amplitude corresponding to the black dashed line is relatively low). 
As the flow further moves downstream, disturbances in the high-frequency range gradually increase, resulting in some peaks (these peaks correspond to the high-frequency range $[400KHz, 800KHz]$ identified in previous analyses), eventually leading to fully developed disturbances and the transition to turbulent flow. 
On the side stimulated by the horseshoe vortex generated by the higher roughness element, similar characteristics are observed, as shown in Figure \ref{Srf_SF_Select_Points_Spectra}$(e)$. 
As indicated by the previous analysis of wall heat flux and skin friction, there exists a relatively 'quiet' zone between the two transition peaks (as shown in Figure \ref{Srf_SF_Select_Points_Spectra}$(d)$). 
The corresponding disturbance amplitude in this region is smaller compared to the regions on either side. 
The disturbances in this zone only begin to grow once the disturbances on both sides have fully developed. 

\begin{figure}
\begin{center}
\begin{tabular}{cc}
\begin{overpic}[width=0.48\textwidth]{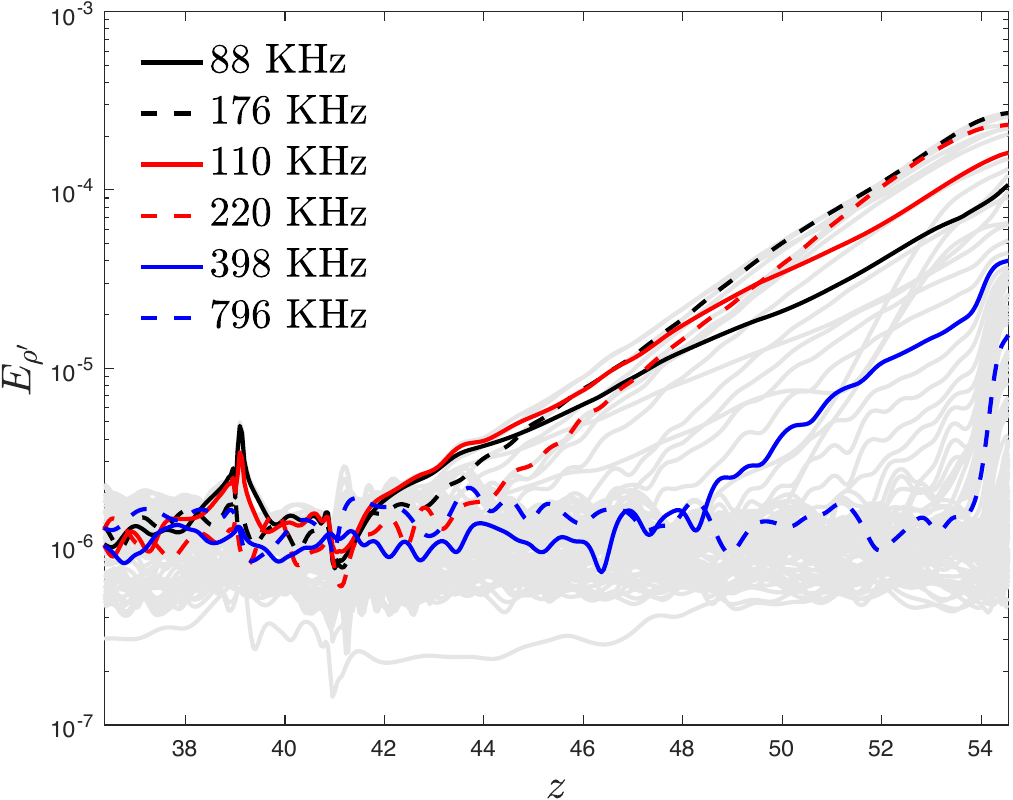}
\put(-3,74){$(a)$}
\end{overpic} &
\begin{overpic}[width=0.48\textwidth]{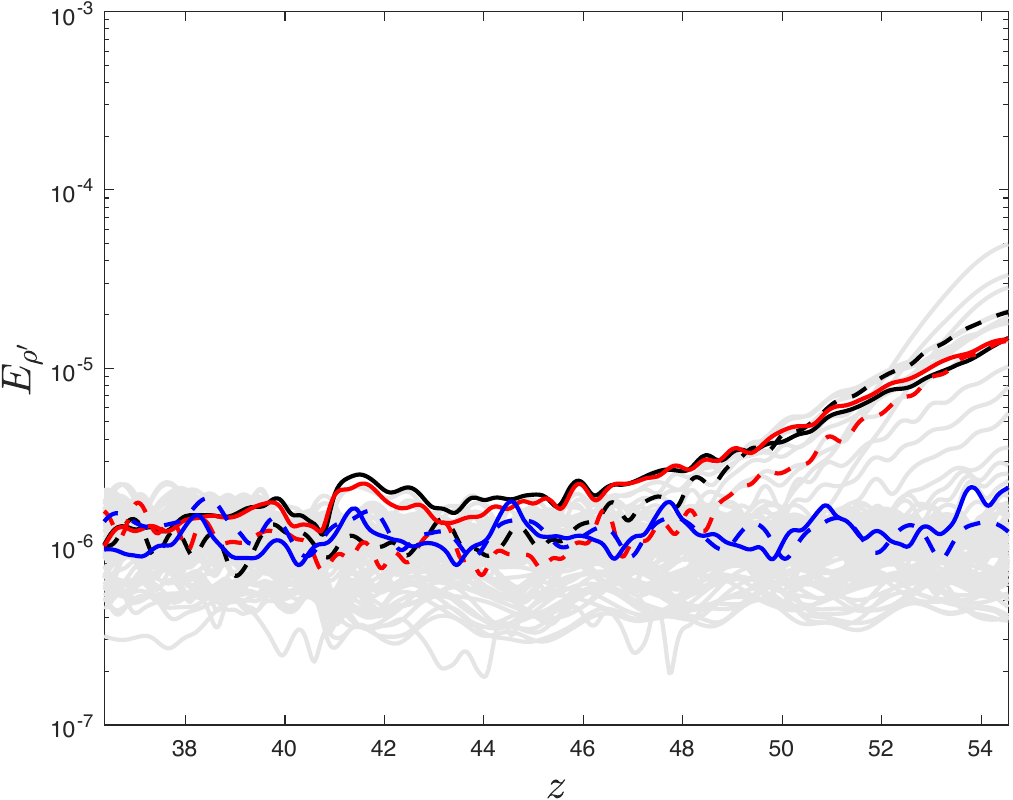}
\put(-3,74){$(b)$}
\end{overpic}
\end{tabular}
\end{center}
\caption{The spanwise evolutions of perturbations $\rho^{\prime}$ with different frequencies, around the roughness element for case H0100. $(a)$ and $(b)$ represent the perturbations along the surface lines of the first and second groups of selected points, respectively. The roughness is locate at the region $[38,42]$.}
\label{SpPRho_H01}
\end{figure}

\begin{figure}
\begin{center}
\begin{tabular}{cc}
\begin{overpic}[width=0.48\textwidth]{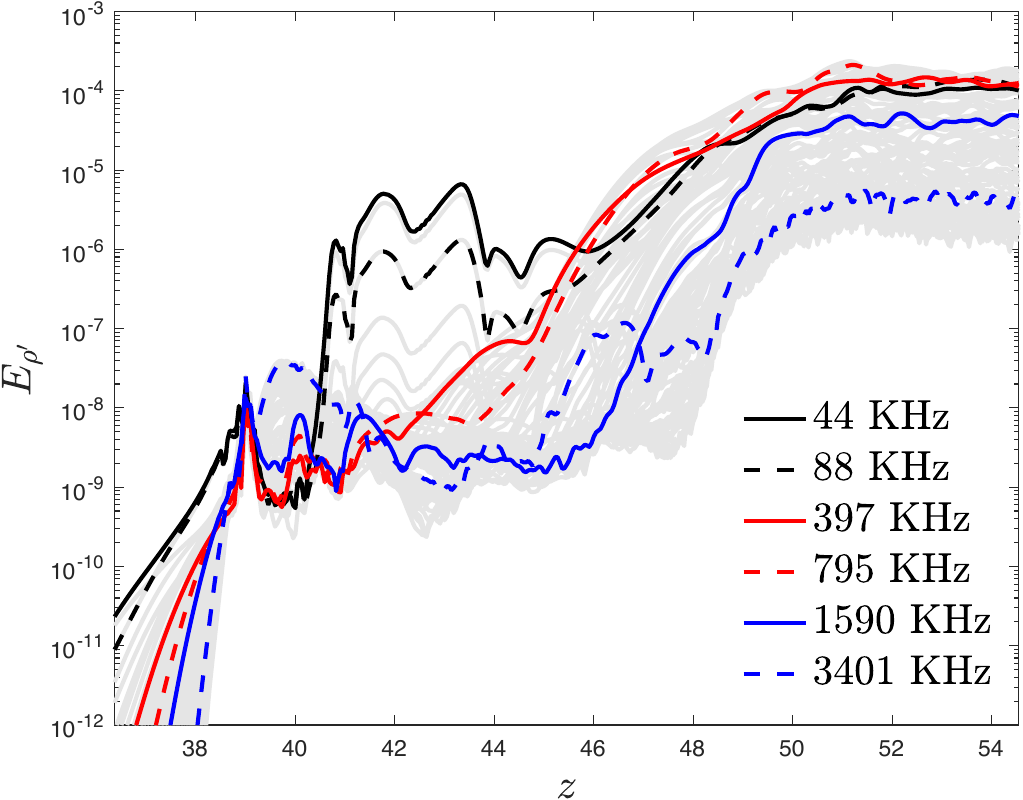}
\put(-2,74){$(a)$}
\end{overpic} &
\begin{overpic}[width=0.48\textwidth]{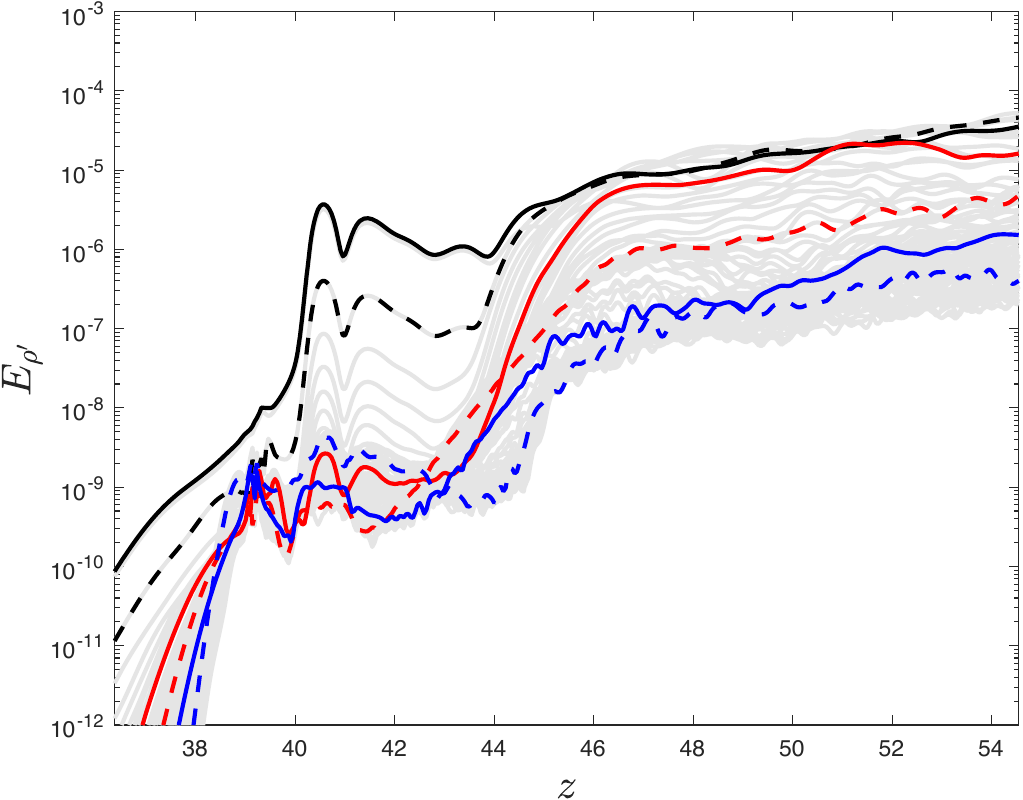}
\put(-2,74){$(b)$}
\end{overpic}
\end{tabular}
\begin{overpic}[width=0.48\textwidth]{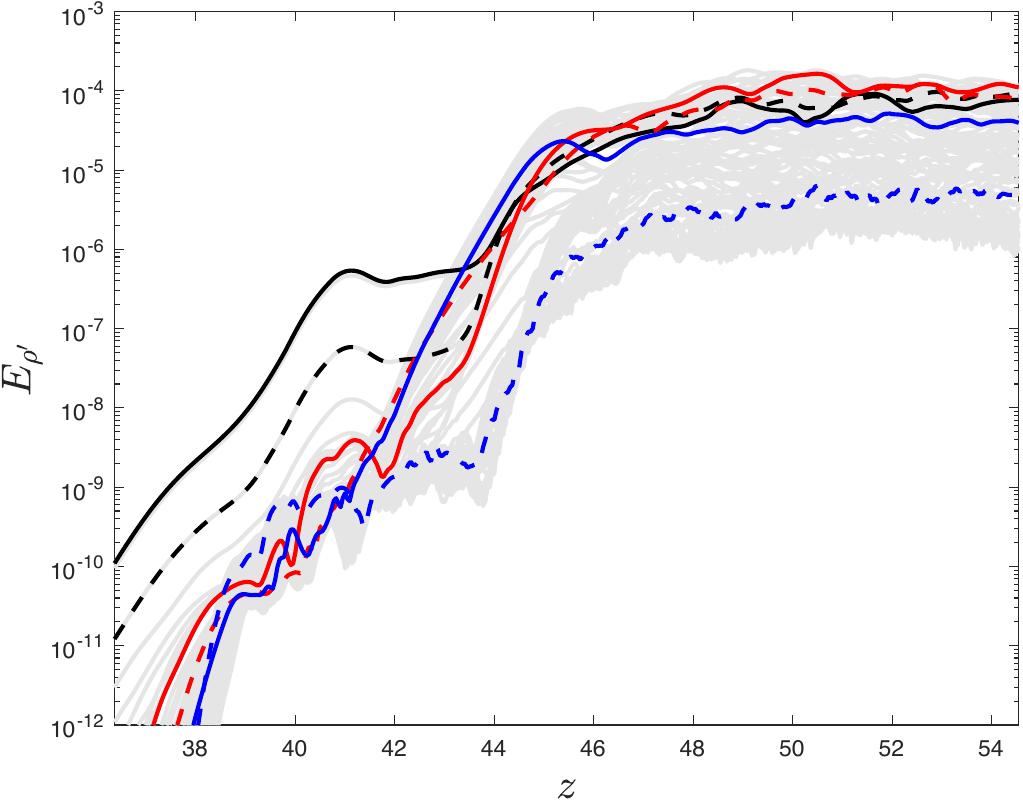}
\put(-2,74){$(c)$}
\end{overpic}
\end{center}
\caption{The spanwise evolutions of perturbations $\rho^{\prime}$ with different frequencies, around the roughness element for case H0200. $(a), (b)$ and $(c)$ represent the perturbations along the surface lines of the first, second and third groups of selected points, respectively. The roughness is locate at the region $[38,42]$.}
\label{SpPRho_H02}
\end{figure}

The evolution of amplitude along the spanwise direction for different frequency disturbances over the wall surface under two operating conditions is illustrated in the figure \ref{SpPRho_H01} and \ref{SpPRho_H02}. In the figures, all resolved frequencies are represented by semi-transparent grey lines, and the evolution processes of disturbances at several typical frequencies are prominently marked. These images provide a clearer depiction of the disturbance behavior described in our previous discussions.
    
Based on our existing data, it can be roughly determined that the triggering of transition downstream of smaller roughness elements is initially caused by disturbances around 100 KHz. These disturbances originate from the interaction between incoming flow disturbances and the separation bubbles at the leading and trailing edges induced by the roughness elements. Subsequently, disturbances corresponding to a frequency of around 200 KHz (second harmonic) grow rapidly, ultimately leading to turbulence formation. For higher roughness elements, the separation bubble at the trailing edge of the roughness element induces a distinct dominant frequency (Kelvin-Helmholtz instability of the shear layer). This dominant frequency, in the wake region, leads to the formation of corresponding low-frequency streaks. The secondary instability of these streaks further induces the generation of high-frequency secondary unstable waves, with components in the $[400 KHz, 800 KHz]$ range directly growing to saturation, causing transition. Additionally, the horse-shoe vortex structures induced by the leading-edge separation bubble and upstream flow for higher roughness elements develop downstream, forming corresponding sider streak structures. The secondary instability of these streaks is also one of the reasons for inducing transition. The high-frequency disturbance components corresponding to the leading-edge separation bubble (located between $[1 MHz, 2 MHz]$) also grow rapidly downstream. These higher frequency disturbances and the secondary instability of the streaks collectively induce the early occurrence of side-edge transition.     
 
\subsubsection{Mode decompositions of the transitional flow fields}
\begin{figure}
  \centering
  \begin{tabular}{cc}
  \begin{overpic}[width=0.49\textwidth]{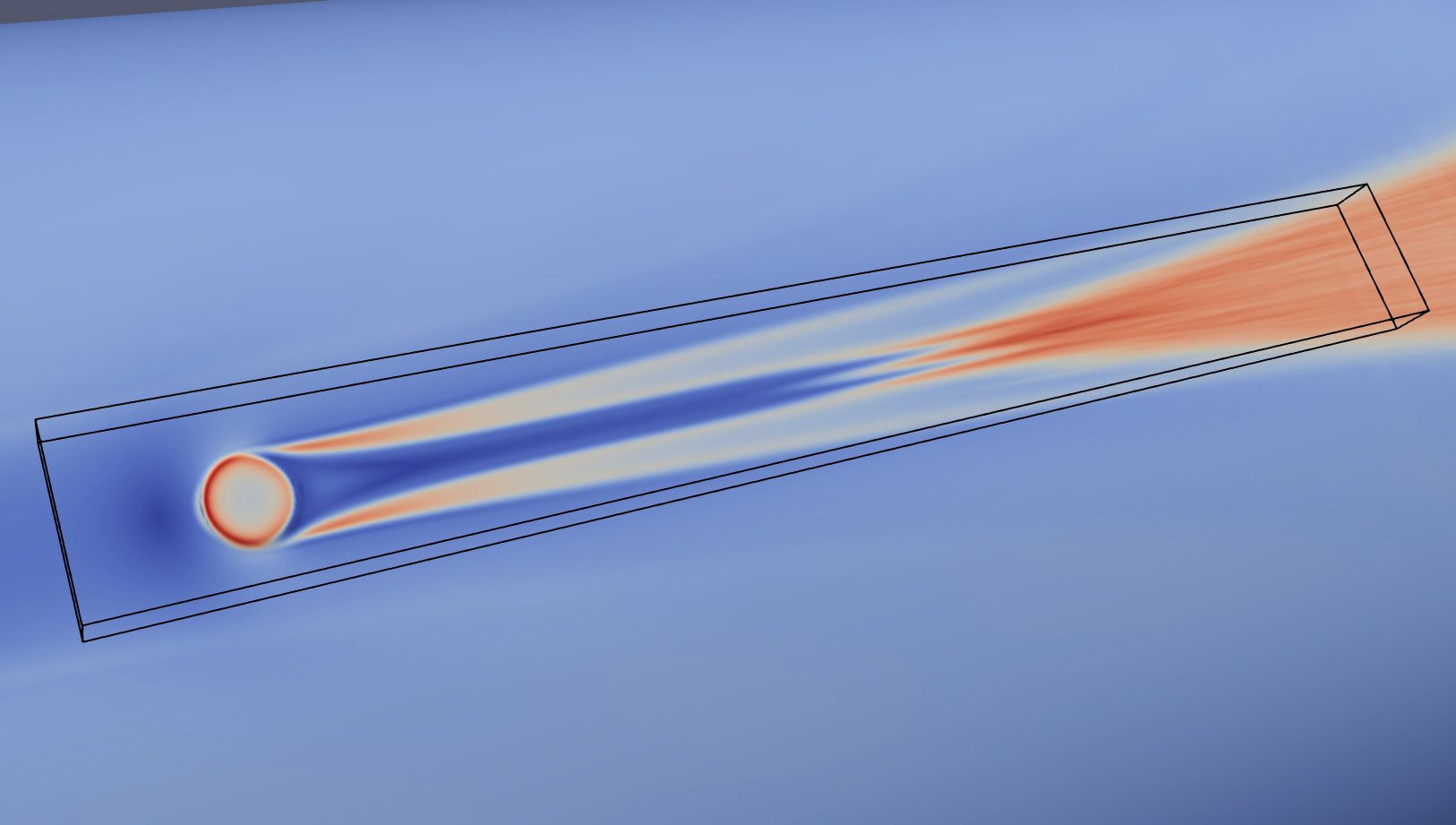}
  \put(1,50){$(a)$}
  \end{overpic} &
  \begin{overpic}[width=0.49\textwidth]{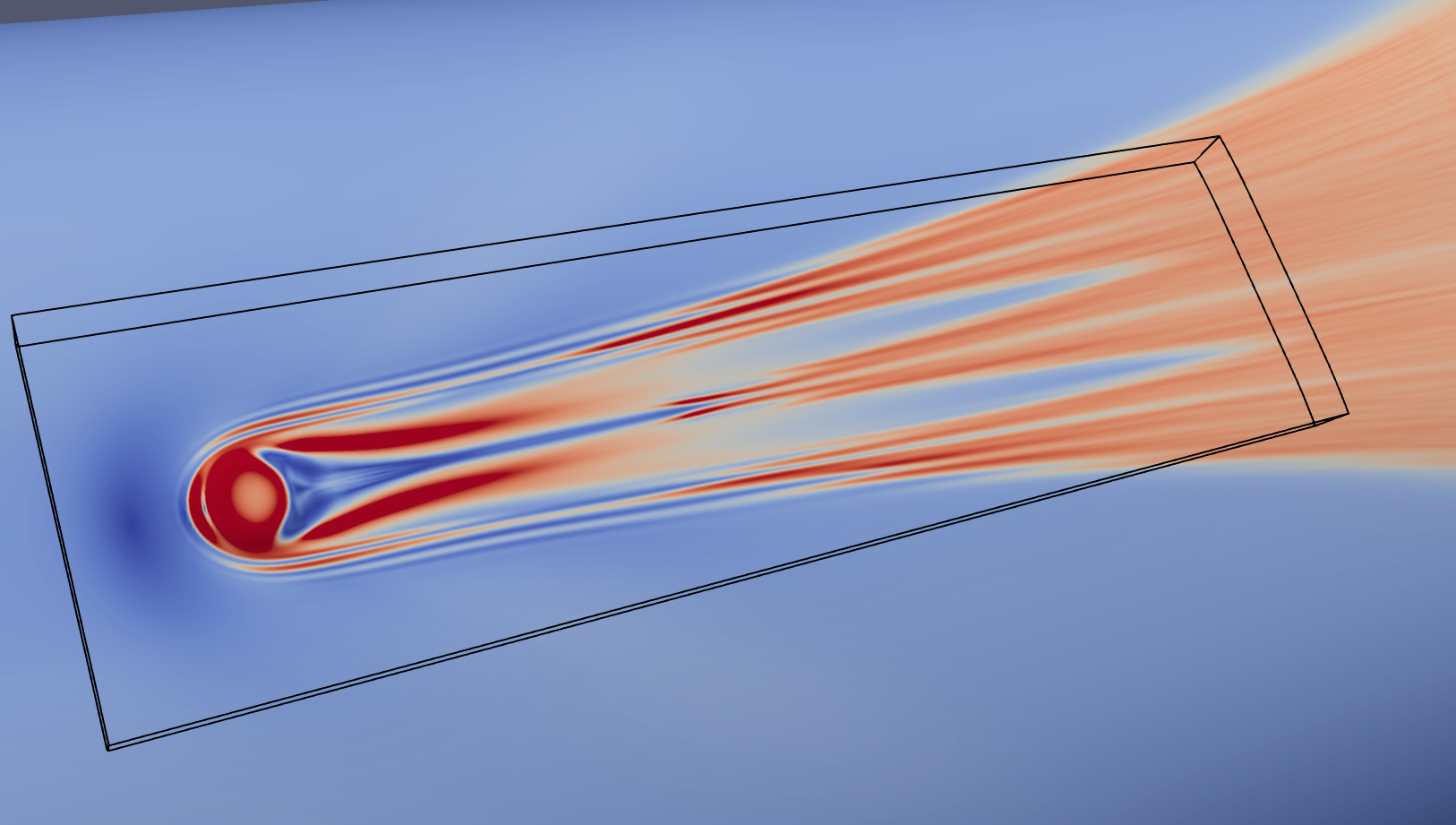}
  \put(1,50){$(b)$}
  \end{overpic}
  \end{tabular} 
  \caption{The sub-block regions used for spectra analysis. $(a)$ for H0100, $(b)$ for H0200. The surface skin-frictions are shown for clarity.}
  \label{SubBlocks}
\end{figure}

In this section, we aim to reveal the basic structures for later stage transition of these flows using two- and three-dimensional mode decompositions.
Given the extensive computational grid involved, documenting every instantaneous signal throughout the entire flow domain is virtually impossible. 
Therefore, to analyze specific flow characteristics, we strategically focused on capturing variable signals within targeted regions to reveal the featured flow structures.
Even though regarding the analysis of three-dimensional flow fields, it is important to acknowledge that the significant disk storage requirements for time-sequential data make it impractical to achieve the same level of precision across a broad frequency range(especially the low frequency region) as that obtained from one-point statistics analysis. 
Therefore, we intend to primarily utilize modal decomposition analysis to investigate the potential occurrence of high-frequency disturbances and the three-dimensional structural characteristics of later stage transition.
The sub-block regions shown in figure \ref{SubBlocks} are used to record the instantaneous signal of basic variables. 
The resolutions $(Ns_{\xi} \times Ns_{\eta} \times Ns_{\zeta})$ of the blocks for case H0100 and H0200 are $201 \times 301 \times 801$ and $401 \times 301 \times 701$, respectively. 
Additionally, $600$ time samples are recorded at a time step of $\Delta t = 0.1$, covering a total of 60 basic time units.  

Peoper Orthogonal Decomposition (POD) and Dynamic Mode Decomposition (DMD)\citep{Rowley2009,Schmid2010,Mezic2013,Schmid2022} are used to identify the representative 3D mode structures of the core regions for the three-dimensional transitional boundary layers, as depicted in figure \ref{SubBlocks}. For the completeness of the article, we briefly introduce the corresponding processes here. Both methods are started with the data sequence of the matrix form
\begin{equation}
\bm{V}_{1}^{N} = \left[ \bm{v}_1, \bm{v}_2, \cdots, \bm{v}_N
\right]^{T},
\end{equation}
with data samples \(( \bm{v}_i, i = 1,2,\cdots,N)\) at different temporal instants with a uniform time interval, $N$ is the number of temporal data samples and superscript $T$ represents the transpose.
 In this paper, $\bm{v}_i$ is a vector which can be any variable whthin the three-dimensional flow fields and matches exactly the resolution sizes of sub-block regions. Reduced Singular value decomposition (SVD)
\begin{equation}
\bm{V}_{1}^{N} = \boldsymbol{\varPhi} \boldsymbol{\Sigma} \boldsymbol{W}^{T},
\end{equation}
is the basic methods to determine the decomposition. 
The matrices $\boldsymbol{\varPhi} = \left[\Phi_1, \Phi_2, \cdots, \Phi_N \right]$ and $\boldsymbol{W}$ contain the left and right singular vectors, and matrix $\boldsymbol{\Sigma}$ holds the singular values \((\sigma_1, \sigma_2, \cdots, \sigma_N)\) along its diagonal. The columns of matrix $\boldsymbol{\varPhi}$ stands for the spatial distributions of the POD modes, and the average energy of the modes can be recoved by $\sigma_i^2 / M_{n}$ ($M_{n}$ is the grid numbers).

As the present POD modes generally contain multiple frequencies, to isolate the structure with a single characteristic frequency, DMD\citep{Rowley2009,Schmid2010,Mezic2013,Schmid2022} is used to find the structure, amplitude and dynamics. In DMD, the collected snapshots are divided into two groups $\bm{V}_{1}^{N-1}$ and $\bm{V}_{2}^{N}$. If a linear mapping matrix $\mathcal{M}_k$ connects two groups of snapshots is assumped,  we have
\begin{equation}
\bm{V}_{2}^{N} = \mathcal{M}_k \bm{V}_{1}^{N-1}.
\end{equation}
Replacing the snapshot matrix $\bm{V}_{1}^{N-1}$ by the singular values and vectors from SVD leads the relation
\begin{equation}
\bm{V}_{2}^{N} = \mathcal{M}_k \bm{U} \boldsymbol{\Sigma} \bm{V}^{T}.
\end{equation}
Then multiplying both sides by $\bm{U}^{T}$ produces
\begin{equation}
\bm{U}^{T}\mathcal{M}_k \bm{U} = \bm{U}^{T} \bm{V}_{2}^{N} \bm{V} \boldsymbol{\Sigma}^{-1} \equiv \tilde{\boldsymbol{S}},
\end{equation}
which is low-rank approximation of $\mathcal{M}_k$. The eigenvalues $\boldsymbol{\mu}$ and the eigenvectors $\boldsymbol{\Phi}$ of $\mathcal{M}_k $ can be approximately obtained by solving the eigenvalue problem: 
\begin{equation}
\tilde{\boldsymbol{S}} \boldsymbol{Y} = \boldsymbol{\mu} \boldsymbol{Y}, \boldsymbol{\mu} = \text{diag}\left(\mu_1, \mu_2, \mu_3, \cdots\right), \boldsymbol{Y} = \left( \bm{Y}_1, \bm{Y}_2, \bm{Y}_3, \cdots\right),
\end{equation}
and the eigenvectors can be expressed as
\begin{equation}
\boldsymbol{\Phi} = \bm{U}\bm{Y}
\end{equation}
The amplitudes $\alpha$ of the DMD modes are dertermined by solving a convec optimization problem
\begin{equation}
\underset{\alpha}{\operatorname{minimize}}\quad J(\alpha):=\left\|\boldsymbol{\Sigma} \boldsymbol{V}^T-\boldsymbol{Y} \boldsymbol{D}_\alpha \boldsymbol{V}_{and}\right\|_F^2,
\end{equation}
where $\left\| \cdot \right\|_F^2$ stands for the Frobenius norm of the matrix. $\boldsymbol{V}_{and}$ is the Vandermonde matrix 
\begin{equation}
\boldsymbol{V}_{and} = 
\left[\begin{array}{cccc}
1 & \mu_1 & \cdots & \mu_1^{N-1} \\
1 & \mu_2 & \cdots & \mu_2^{N-1} \\
\vdots & \vdots & \ddots & \vdots \\
1 & \mu_k & \cdots & \mu_k^{N-1}
\end{array}\right], 
\end{equation}
represents the temporal evolution of the dynamic modes. $\boldsymbol{D}_\alpha$ is the diagonal matrix
\begin{equation}
\boldsymbol{D}_\alpha = \left[\begin{array}{llll}
\alpha_1 & & & \\
& \alpha_2 & &\\
& & \ddots&\\
& & & \alpha_k
\end{array}\right]
\end{equation}
of the amplitude. The most important modes are selected using the sparsity promoting algorithm of \citet{Jov2014}. For specific modes $k$, its temporal growth rate and frequency can be represents by the real and imaginary parts of ${\log{(\mu_k)}}/{\Delta t}$, respectively.

\begin{table}
\centering
\begin{tabular}{cccccccc}
\toprule
 Case & Observations & $Ns_{\xi}$ & $ Ns_{\eta} $ & $Ns_{\zeta}$ & Dof & No. (snapshots) & Memory (TB)\\
\hline
H0100 & $ \lambda_2$ & 201 & 301 & 801 & $ 0.48\times10^8$ & 600 & 0.22\\
\midrule
H0200 & $ \lambda_2$ & 401 & 301 & 701 & $0.85\times10^8$ & 600 & 0.4\\
\bottomrule
\end{tabular}
\caption{Basic parameters of domains and the observations used for the decomposition. Degree of freedom(Dof) stands for the variables numbers per snapshots. Memory represents the memory requirements for storing all input data.}
\label{BasicParametersForDMD}
\end{table}%

\begin{figure}
  \centering
 \begin{overpic}[width=\textwidth]{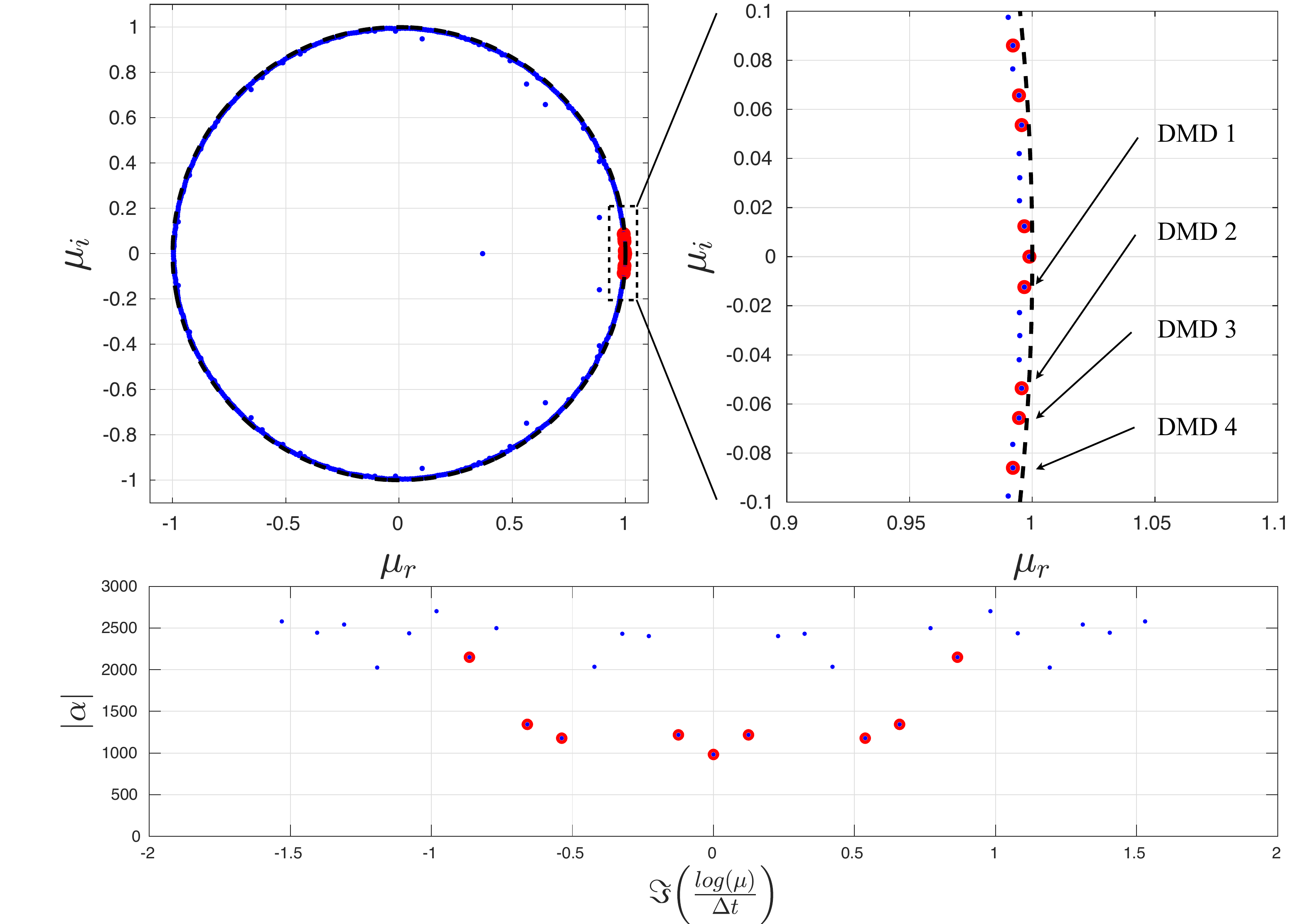}
 \put(0,68){$(a)$}
 \put(0,28){$(b)$}
  \end{overpic}
  \caption{The spectrum and selected modes of DMD for H0100 case in three-dimensional region. The $\lambda_2$ is used as observation variable. $(a)$ shows the spectrum. $(b)$ shows the frquency $\Im(\log(\mu)/\Delta t)$ and amplitude $|\alpha|$.}
  \label{DMD_Lambda2_H01_Spectrum}
\end{figure}

\begin{figure}
  \centering
 \begin{overpic}[width=\textwidth]{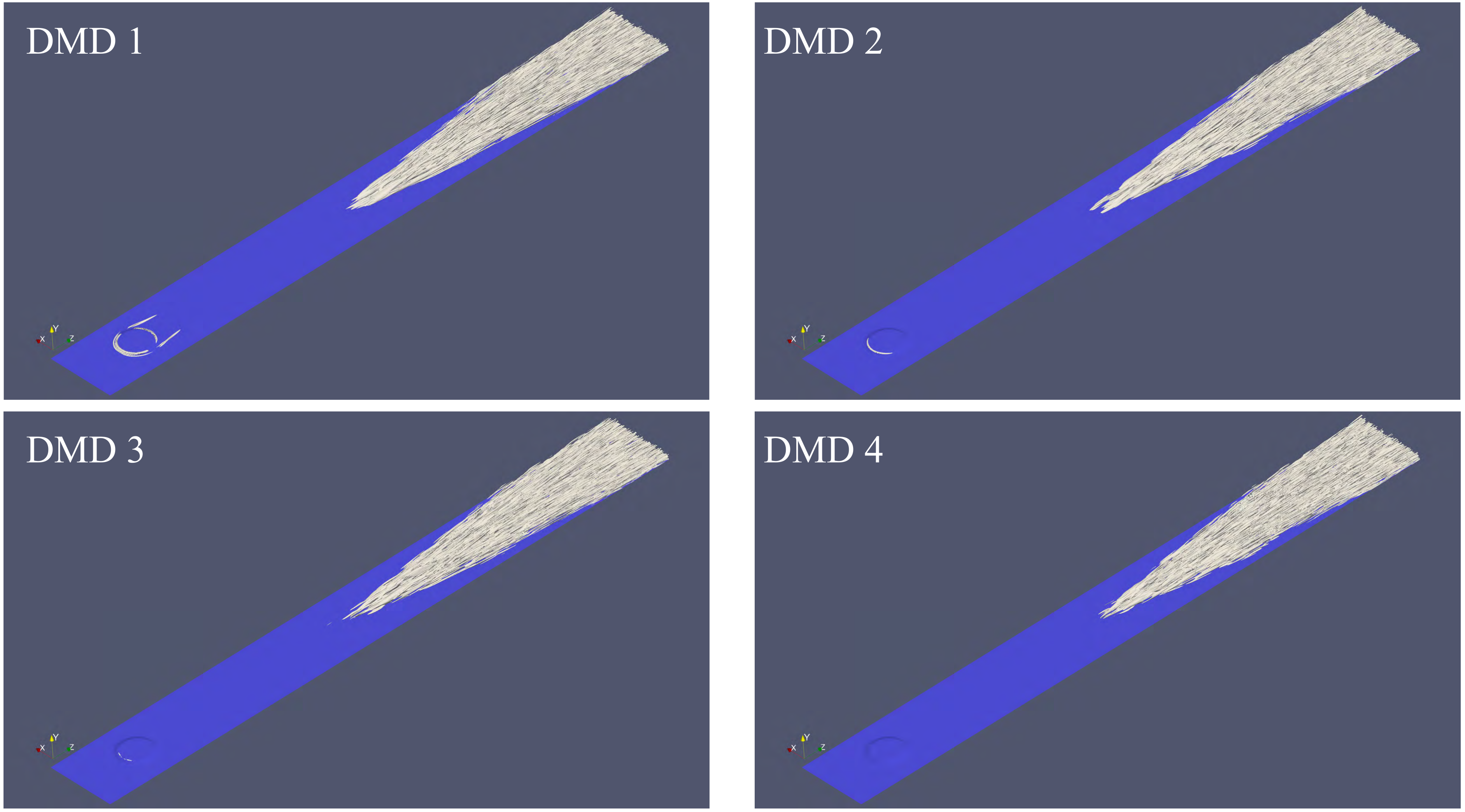}
  \end{overpic}
  \caption{Spatial structures of the selected DMD modes with iso-surfaces of $\lambda_2 = -5\times10^{-5}$, for case H0100.}
  \label{DMD_Lambda2_H01_Modes}
\end{figure}

In approximating the Koopman eigenfunctions of continuous systems associated with nonlinear NS equations through dynamic modes, selecting appropriate variables becomes crucial for generating significant spatio-temporal patterns. Koopman theory suggests that good observables might better capture the dynamics of nonlinear systems. Consequently, this study employs the $\lambda_2$, a variable derived from vortex identification, serving as a dynamic indicator for structures. The decomposition's fundamental parameters are detailed in Table \ref{BasicParametersForDMD}. Due to the prohibitive size of the input datasets, necessitating distributed memory high-performance computing, this work adopts and adapts a parallelized algorithm, as described by \citet{Sayadi2016}, to facilitate DMD.

The DMD results for case H0100 are shown in figure \ref{DMD_Lambda2_H01_Spectrum} and \ref{DMD_Lambda2_H01_Modes}. 
The most important modes are shown in the spectrum. 
Excluding the associated mean flow mode, the selected dominant modes all exhibit distinct streak characteristics. 
The corresponding disturbance structures are primarily distributed downstream of the roughness elements, highlighting the feature that in the later stages of transition and in turbulent states, higher-frequency disturbances gradually become more prominent. 
Additionally, the decomposed Modes 1, 2, and 3 all display significant disturbance characteristics near the roughness elements. 
This indicates that the transition is not simply and spontaneously occurring at a certain downstream distance from the roughness elements, but is instead connected with the upstream disturbances near the leading-edge of the roughness elements.

\begin{figure}
  \centering
 \begin{overpic}[width=\textwidth]{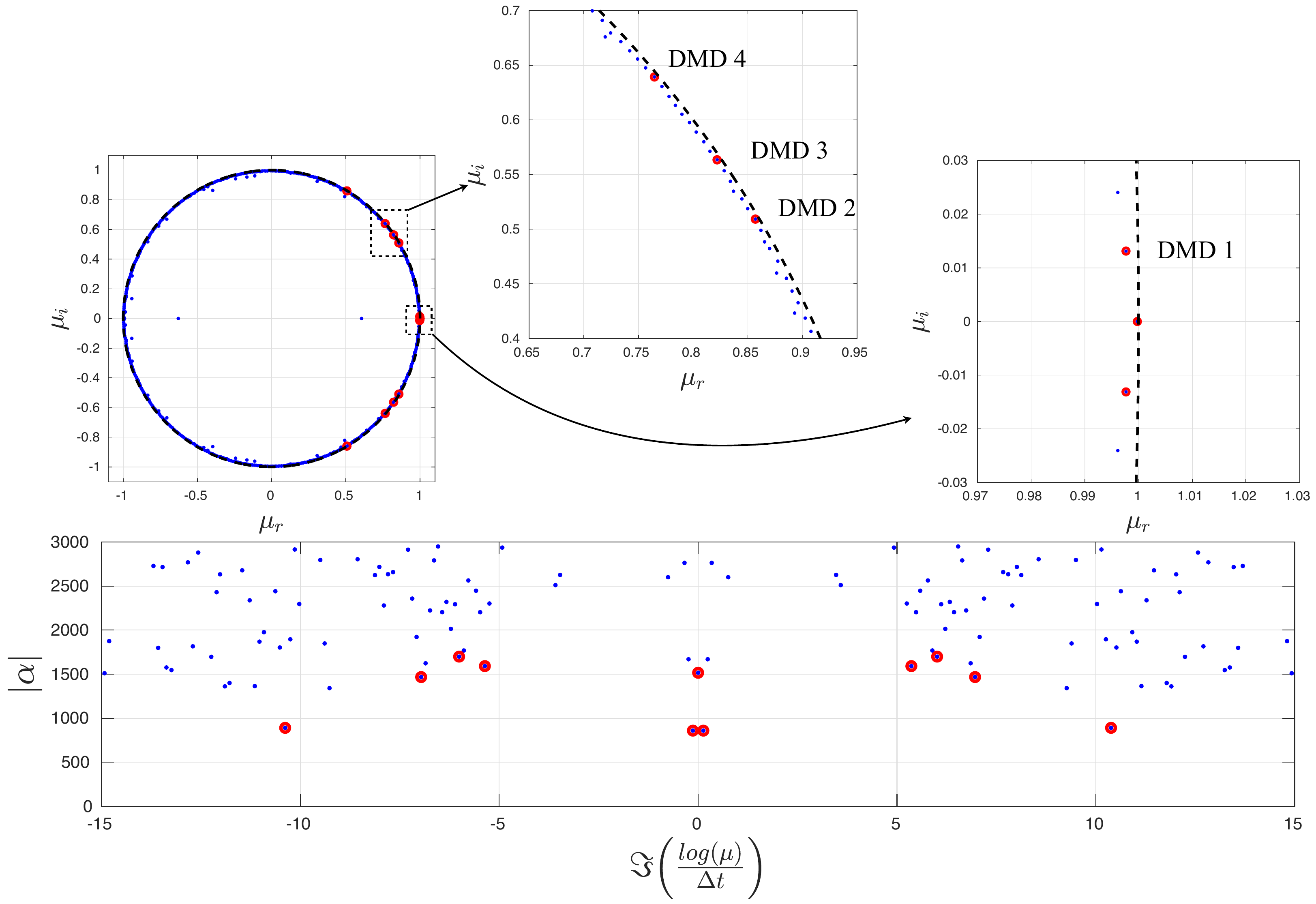}
 \put(0,65){$(a)$}
 \put(0,30){$(b)$}
  \end{overpic}
  \caption{The spectrum and selected modes of DMD for H0200 case in three-dimensional region. The $\lambda_2$ is used as observation variable. $(a)$ show the spectrum. $(b)$ shows the frquency $\Im(\log(\mu)/\Delta t)$ and amplitude $|\alpha|$.}
  \label{DMD_Lambda2_H02_Spectrum}
\end{figure}

\begin{figure}
  \centering
 \begin{overpic}[width=\textwidth]{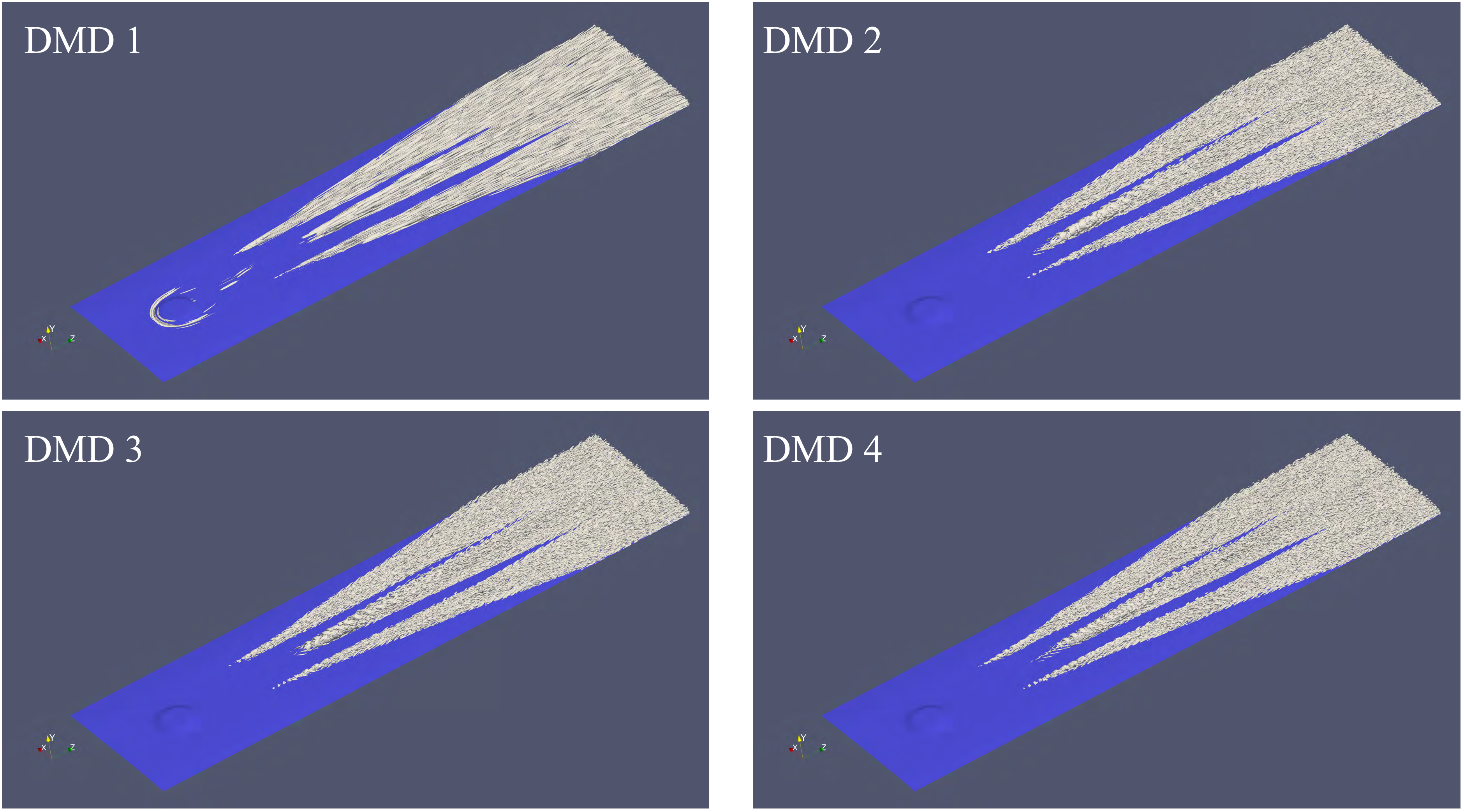}
  \end{overpic}
  \caption{Spatial structures of the selected DMD modes with iso-surfaces of $\lambda_2 = -5\times10^{-5}$, for case H0200.}
  \label{DMD_Lambda2_H02_Modes}
\end{figure} 

The results of case H0200 are presented in figure \ref{DMD_Lambda2_H02_Spectrum} and \ref{DMD_Lambda2_H02_Modes}, which exhibit distinct differences compared to case H0100. The Mode 1 exhibits typical streak structures, however, unlike previous conditions, in addition to the streaks at the central position (around $x=0$), there are also streak regions on either side. 
These lateral regions correspond to the vortex structures formed at the edges of the larger roughness elements, continuing their downstream development. 
Modes 2, 3, and 4 reflect the flow structures known as hairpin vortices in the typical transitional boundary layer induced by roughness elements. 
These structures represent the intense momentum mixing processes occurring during transition. 
These modes clearly illustrate that at certain distances from the roughness elements, the flow streaks evolving from the upstream horseshoe vortex enter a strong nonlinear phase earlier than those along the centerline, forming corresponding vortex structures.

\section{Conclusions and Discussions}\label{sec4}
In summary, this study presents simulations of the complete transition process from laminar flow to turbulence in a hypersonic three-dimensional swept leading-edge boundary layer over an experimental configuration. 
Based on the results from linear stability theory and the experimental observations, the subcritical state of the possible transition along the attachment line is confirmed. 
Two roughnesses based on the experimental tests are modeled and designed to trigger transitions. The main findings of this study can be summarized as the following points
\begin{itemize}
\item In the flow over a swept blunt body discussed in this paper, even without the assumption of infinite sweep, if the incoming boundary layer reaches an asymptotic state (for laminar flow), the subsequent flow state also satisfies the infinite sweep assumption—the boundary layer is homogeneous along the swept direction.

\item The two different heights of roughness elements in the configuration studied in this paper result in completely different transition characteristics. For lower-height roughness element, the element alone cannot directly induce the corresponding boundary layer transition. Certain random perturbations need to be introduced during the simulation. For higher-height roughness element, it can directly induce boundary layer transition by themselves without the need for additional perturbations. The flow near the roughness elements resembles that of a flat plate boundary layer, where vortex structures are triggered in their vicinity. In the transition phenomena induced by external disturbances and lower-height roughness elements, the transition mainly occurs directly downstream of the wake of the roughness elements, but the wake vortices induced by the roughness elements do not directly destabilize and lead to the final transition. In the transition simulation with higher-height roughness element, the horseshoe vortices generated by the roughness elements form corresponding streaks, which preferentially destabilize and lead to the transition to turbulence, while the wake vortices directly behind the roughness elements destabilize and transition further downstream.

\item One-point power spectral analysis and bispectral analysis, together with the DMD analysis, are used to identify the detailed transition mechanism. 
For the case of small roughness element, the wake flow induced by the roughness act as a disturbance selector and amplifier, selecting and amplifying the incoming disturbances from upstream. 
This causes disturbances with frequencies around 100 KHz to preferentially grow and lead to the final transition. 
Through DMD analysis, we can infer that the high frequency instability is strongly linked to the separation bubble upstream of the roughness elements. 
For larger roughness elements, there exists a low-frequency absolute instability in the wake induced by the roughness elements. 
This low-frequency disturbance, around 10 KHz,  generates corresponding low-frequency streaks, and the high frequency secondary instability of these low-frequency streaks is the primary reason for the transition in the wake. 
The high-frequency disturbance components corresponding to the leading-edge separation bubble (located between [1 MHz, 2 MHz]) also grow rapidly downstream. These higher frequency disturbances and the secondary instability of the streaks collectively induce the early occurrence of side-edge transition.
\end{itemize}

\backsection[Supplementary data]{}

\backsection[Acknowledgements]{We acknowledge Yancheng MetaStone Tech. Co. for prociding us with the computational resources required by this work. 
Useful discussions with Professor Qibing Li of Tsinghua University and Professor Jie Ren of Beijing Institute of Technology are gratefully acknowledged.}

\backsection[Funding]{This work received support from the NSFC Grants 12202242, 12172195 and 12388101. The authors are also grateful for the support from the National Key Research and Development Plan of China through project no. 2019YFA0405201, the National Key Project GJXM92579 and the Grants 20231001 in Supercomputing Center in Yancheng.}

\backsection[Declaration of interests]{The authors report no conflict of interest.}

\backsection[Data availability statement]{The full data set of the simulations is of the order of 80 thousands of gigabytes. By contacting the authors, a smaller subset can be made available.}

\backsection[Author ORCIDs]{

Youcheng Xi, https://orcid.org/0000-0002-6484-0231;

Bowen Yan, https://orcid.org/0009-0002-0655-9414;

Guangwen Yang, https://orcid.org/0000-0002-8673-8254

Song Fu, https://orcid.org/0000-0003-2052-7435}

\backsection[Author contributions]{
Youcheng Xi: Funding acquisition, Conceptualization, Data curation, Formal analysis, Coding, Investigation, Methodology, Validation, Writing original draft. 
Bowen Yan \& Guangwen YANG: Funding acquisition, Software and optimization on high performance computing, Computational Resources.
Xinguo Sha \& Dehua Zhu: Model designing, Wind Tunnel Testing, Experimental Investigation.  
Song Fu: Funding acquisition, Supervision, Resources, Writing-review \& editing.}

\appendix

 
\bibliographystyle{jfm}
\bibliography{jfm}

\begin{thebibliography}{50}
\expandafter\ifx\csname natexlab\endcsname\relax\def\natexlab#1{#1}\fi
\def\au#1{#1} \def\ed#1{#1} \def\yr#1{#1}\def\at#1{#1}\def\jt#1{\textit{#1}}
  \def\bt#1{#1}\def\bvol#1{\textbf{#1}} \def\vol#1{#1} \def\pg#1{#1}
  \def\publ#1{#1}\def\arxiv#1{#1}\def\org#1{#1}\def\st#1{\textit{#1}}

\bibitem[Arnal {\em et~al.\/}(1997)Arnal, Juillen, Reneaux \&
  Gasparian]{Arnal1997}
{\sc \au{Arnal, D.}, \au{Juillen, J.~C.}, \au{Reneaux, J.} \& \au{Gasparian,
  G.}} \yr{1997}  \at{Effect of wall suction on leading edge contamination}.
  \jt{Aerospace Science and Technology}  \bvol{1}~(8),  \pg{505--517}.

\bibitem[Beckwith \& Gallagher(1959)]{Beckwith1959}
{\sc \au{Beckwith, Ivan~E.} \& \au{Gallagher, James~J.}} \yr{1959}  \bt{Local
  heat transfer and recovery temperatures on a yawed cylinder at a mach number
  of 4.15 and high reynolds numbers}. Report R104.  \org{Langley Research
  Center}.

\bibitem[Bernardini {\em et~al.\/}(2012)Bernardini, Pirozzoli \&
  Orlandi]{Bernardini2012}
{\sc \au{Bernardini, Matteo}, \au{Pirozzoli, Sergio} \& \au{Orlandi, Paolo}}
  \yr{2012}  \at{Compressibility effects on roughness-induced boundary layer
  transition}.  \jt{International Journal of Heat and Fluid Flow}  \bvol{35},
  \pg{45--51}.

\bibitem[Bernardini {\em et~al.\/}(2014)Bernardini, Pirozzoli, Orlandi \&
  Lele]{Bernardini2014}
{\sc \au{Bernardini, Matteo}, \au{Pirozzoli, Sergio}, \au{Orlandi, Paolo} \&
  \au{Lele, Sanjiva~K.}} \yr{2014}  \at{Parameterization of boundary-layer
  transition induced by isolated roughness elements}.  \jt{AIAA Journal}
  \bvol{52}~(10),  \pg{2261--2269}.

\bibitem[Chen {\em et~al.\/}(1991)Chen, Creel \& Beckwith]{Chen1991}
{\sc \au{Chen, F.~J.}, \au{Creel, T.~R.} \& \au{Beckwith, I.~E.}} \yr{1991}
  \at{Transition on swept leading edges at mach 3.5}.  \jt{Journal of Aircraft}
   \bvol{24}~(10),  \pg{710--717}.

\bibitem[Chen {\em et~al.\/}(2022)Chen, Xi, Ren \& Fu]{Chen2022}
{\sc \au{Chen, Xianliang}, \au{Xi, Youcheng}, \au{Ren, Jie} \& \au{Fu, Song}}
  \yr{2022}  \at{Cross-flow vortices and their secondary instabilities in
  hypersonic and high-enthalpy boundary layers}.  \jt{Journal of Fluid
  Mechanics}  \bvol{947},  \pg{A25}.

\bibitem[Creel {\em et~al.\/}(1986)Creel, Beckwith \& Chen]{Creel1986}
{\sc \au{Creel, T.~R.}, \au{Beckwith, I.~E.} \& \au{Chen, F.~J.}} \yr{1986}
  {\em Effects of wind-tunnel noise on swept-cylinder transition at Mach
  3.5\/}.  \publ{Atlanta: American Institute of Aeronautics and Astronautics}.

\bibitem[Dang {\em et~al.\/}(2022)Dang, Liu, Guo, Duan \& Li]{Dang2022}
{\sc \au{Dang, Guanlin}, \au{Liu, Shiwei}, \au{Guo, Tongbiao}, \au{Duan, Junyi}
  \& \au{Li, Xinliang}} \yr{2022}  \at{Direct numerical simulation of
  compressible turbulence accelerated by graphics processing unit: An
  open-source high accuracy accelerated computational fluid dynamic software}.
  \jt{Physics of Fluids}  \bvol{34}~(12),  \pg{126106}.

\bibitem[De~Tullio {\em et~al.\/}(2013)De~Tullio, Paredes, Sandham \&
  Theofilis]{Tullio2013}
{\sc \au{De~Tullio, N.}, \au{Paredes, P.}, \au{Sandham, N.~D.} \&
  \au{Theofilis, V.}} \yr{2013}  \at{Laminar-turbulence transition induced by a
  discrete roughness element in a supersonic boundary layer}.  \jt{Journal of
  Fluid Mechanics}  \bvol{735},  \pg{613--646}.

\bibitem[De~Tullio \& Sandham(2015)]{Tullio2015}
{\sc \au{De~Tullio, Nicola} \& \au{Sandham, Neil~D.}} \yr{2015}  \at{Influence
  of boundary-layer disturbances on the instability of a roughness wake in a
  high-speed boundary layer}.  \jt{Journal of Fluid Mechanics}  \bvol{763},
  \pg{136--165}.

\bibitem[Di~Giovanni \& Stemmer(2018)]{Giovanni2018}
{\sc \au{Di~Giovanni, Antonio} \& \au{Stemmer, Christian}} \yr{2018}
  \at{Cross-flow-type breakdown induced by distributed roughness in the
  boundary layer of a hypersonic capsule configuration}.  \jt{Journal of Fluid
  Mechanics}  \bvol{856},  \pg{470--503}.

\bibitem[Estruch-Samper {\em et~al.\/}(2017)Estruch-Samper, Hillier, Vanstone
  \& Ganapathisubramani]{Estruch2017}
{\sc \au{Estruch-Samper, David}, \au{Hillier, Richard}, \au{Vanstone, Leon} \&
  \au{Ganapathisubramani, Bharathram}} \yr{2017}  \at{Effect of isolated
  roughness element height on high-speed laminar–turbulent transition}.
  \jt{Journal of Fluid Mechanics}  \bvol{818}.

\bibitem[Fedorov \& Egorov(2022)]{Fedorov2022}
{\sc \au{Fedorov, Alexander~V.} \& \au{Egorov, Ivan~V.}} \yr{2022}
  \at{Instability of the attachment line boundary layer in a supersonic swept
  flow}.  \jt{Journal of Fluid Mechanics}  \bvol{933},  \pg{A26}.

\bibitem[Gaster(1967)]{Gaster1967}
{\sc \au{Gaster, M.}} \yr{1967}  \at{On the flow along swept leading edges}.
  \jt{Aeronautical Quarterly}  \bvol{18}~(2),  \pg{165--184}.

\bibitem[Gennaro {\em et~al.\/}(2013)Gennaro, Rodríguez, Medeiros \&
  Theofilis]{Gennaro2013}
{\sc \au{Gennaro, E.~M.}, \au{Rodríguez, D.}, \au{Medeiros, M. A.~F.} \&
  \au{Theofilis, V.}} \yr{2013}  \at{Sparse techniques in global flow
  instability with application to compressible leading-edge flow}.  \jt{AIAA
  Journal}  \bvol{51}~(9),  \pg{2295--2303}.

\bibitem[Groskopf \& Kloker(2016)]{Groskopf2016}
{\sc \au{Groskopf, Gordon} \& \au{Kloker, Markus~J.}} \yr{2016}
  \at{Instability and transition mechanisms induced by skewed roughness
  elements in a high-speed laminar boundary layer}.  \jt{Journal of Fluid
  Mechanics}  \bvol{805},  \pg{262--302}.

\bibitem[Hader \& Fasel(2019)]{Hader2019}
{\sc \au{Hader, Christoph} \& \au{Fasel, Hermann~F.}} \yr{2019}  \at{Direct
  numerical simulations of hypersonic boundary-layer transition for a flared
  cone: fundamental breakdown}.  \jt{Journal of Fluid Mechanics}  \bvol{869},
  \pg{341--384}.

\bibitem[Hall {\em et~al.\/}(1984)Hall, Malik \& Poll]{Hall1984}
{\sc \au{Hall, Philip}, \au{Malik, M.~R.} \& \au{Poll, D. I.~A.}} \yr{1984}
  \at{On the stability of an infinite swept attachment line boundary layer}.
  \jt{Proceedings of the Royal Society of London. A. Mathematical and Physical
  Sciences}  \bvol{395}~(1809),  \pg{229--245}.

\bibitem[Huang {\em et~al.\/}(1995)Huang, Coleman \& Bradshaw]{Huang1995}
{\sc \au{Huang, P.~G.}, \au{Coleman, G.~N.} \& \au{Bradshaw, P.}} \yr{1995}
  \at{Compressible turbulent channel flows: Dns results and modelling}.
  \jt{Journal of Fluid Mechanics}  \bvol{305},  \pg{185--218}.

\bibitem[Jiang \& Shu(1996)]{Jiang1996}
{\sc \au{Jiang, G.~S.} \& \au{Shu, C.~W.}} \yr{1996}  \at{Efficient
  implementation of weighted eno schemes}.  \jt{Journal of Computational
  Physics}  \bvol{126}~(1),  \pg{202--228}.

\bibitem[John {\em et~al.\/}(2014)John, Obrist \& Kleiser]{John2014}
{\sc \au{John, Michael~O.}, \au{Obrist, Dominik} \& \au{Kleiser, Leonhard}}
  \yr{2014}  \at{Stabilisation of subcritical bypass transition in the
  leading-edge boundary layer by suction}.  \jt{Journal of Turbulence}
  \bvol{15}~(11),  \pg{795--805}.

\bibitem[John {\em et~al.\/}(2016)John, Obrist \& Kleiser]{John2016}
{\sc \au{John, Michael~O.}, \au{Obrist, Dominik} \& \au{Kleiser, Leonhard}}
  \yr{2016}  \at{Secondary instability and subcritical transition of the
  leading-edge boundary layer}.  \jt{Journal of Fluid Mechanics}  \bvol{792},
  \pg{682--711}.

\bibitem[Jovanović {\em et~al.\/}(2014)Jovanović, Schmid \& Nichols]{Jov2014}
{\sc \au{Jovanović, Mihailo~R.}, \au{Schmid, Peter~J.} \& \au{Nichols,
  Joseph~W.}} \yr{2014}  \at{Sparsity-promoting dynamic mode decomposition}.
  \jt{Physics of Fluids}  \bvol{26}~(2),  \pg{024103}.

\bibitem[Kurz \& Kloker(2014)]{Kurz2014}
{\sc \au{Kurz, Holger B.~E.} \& \au{Kloker, Markus~J.}} \yr{2014}
  \at{Receptivity of a swept-wing boundary layer to micron-sized discrete
  roughness elements}.  \jt{Journal of Fluid Mechanics}  \bvol{755},
  \pg{62--82}.

\bibitem[Kurz \& Kloker(2016)]{Kurz2016}
{\sc \au{Kurz, Holger B.~E.} \& \au{Kloker, Markus~J.}} \yr{2016}
  \at{Mechanisms of flow tripping by discrete roughness elements in a
  swept-wing boundary layer}.  \jt{Journal of Fluid Mechanics}  \bvol{796},
  \pg{158--194}.

\bibitem[Li \& Choudhari(2008)]{LiFei2008}
{\sc \au{Li, Fei} \& \au{Choudhari, Meelan}} \yr{2008} {\em Spatially
  Developing Secondary Instabilities and Attachment Line Instability in
  Supersonic Boundary Layers\/}.  \publ{American Institute of Aeronautics and
  Astronautics}.

\bibitem[Li {\em et~al.\/}(2008)Li, Fu \& Ma]{Li2008}
{\sc \au{Li, Xinliang}, \au{Fu, Dexun} \& \au{Ma, Yanwen}} \yr{2008}
  \at{Direct numerical simulation of hypersonic boundary layer transition over
  a blunt cone}.  \jt{AIAA Journal}  \bvol{46}~(11),  \pg{2899--2913}.

\bibitem[Li {\em et~al.\/}(2010)Li, Fu \& Ma]{Li2010}
{\sc \au{Li, Xinliang}, \au{Fu, Dexun} \& \au{Ma, Yanwen}} \yr{2010}
  \at{Direct numerical simulation of hypersonic boundary layer transition over
  a blunt cone with a small angle of attack}.  \jt{Physics of Fluids}
  \bvol{22}~(2),  \pg{025105}.

\bibitem[Liang {\em et~al.\/}(2010)Liang, Li, Fu \& Ma]{Liang2010}
{\sc \au{Liang, Xian}, \au{Li, Xinliang}, \au{Fu, Dexun} \& \au{Ma, Yanwen}}
  \yr{2010}  \at{Effects of wall temperature on boundary layer stability over a
  blunt cone at mach 7.99}.  \jt{Computers \& Fluids}  \bvol{39}~(2),
  \pg{359--371}.

\bibitem[Mack {\em et~al.\/}(2008)Mack, Schmid \& Sesterhenn]{Mack2008}
{\sc \au{Mack, Christoph~J.}, \au{Schmid, Peter~J.} \& \au{Sesterhenn,
  Jorn~L.}} \yr{2008}  \at{Global stability of swept flow around a parabolic
  body: connecting attachment-line and crossflow modes}.  \jt{Journal of Fluid
  Mechanics}  \bvol{611},  \pg{205--214}.

\bibitem[Mayer {\em et~al.\/}(2011)Mayer, von Terzi \& Fasel]{Mayer2011}
{\sc \au{Mayer, C. S.~J.}, \au{von Terzi, D.~A.} \& \au{Fasel, H.~F.}}
  \yr{2011}  \at{Direct numerical simulation of complete transition to
  turbulence via oblique breakdown at mach 3}.  \jt{Journal of Fluid Mechanics}
   \bvol{674},  \pg{5--42}.

\bibitem[Mezić(2013)]{Mezic2013}
{\sc \au{Mezić, Igor}} \yr{2013}  \at{Analysis of fluid flows via spectral
  properties of the koopman operator}.  \jt{Annual Review of Fluid Mechanics}
  \bvol{45}~(1),  \pg{357--378}.

\bibitem[Murakami {\em et~al.\/}(1996)Murakami, Stanewsky \&
  Krogmann]{Murakami1996}
{\sc \au{Murakami, Akira}, \au{Stanewsky, Egon} \& \au{Krogmann, Paul}}
  \yr{1996}  \at{Boundary-layer transition on swept cylinders at hypersonic
  speeds}.  \jt{AIAA Journal}  \bvol{34}~(4),  \pg{649--654}.

\bibitem[Obrist {\em et~al.\/}(2012)Obrist, Henniger \& Kleiser]{Obrist2012}
{\sc \au{Obrist, Dominik}, \au{Henniger, Rolf} \& \au{Kleiser, Leonhard}}
  \yr{2012}  \at{Subcritical spatial transition of swept hiemenz flow}.
  \jt{International Journal of Heat and Fluid Flow}  \bvol{35},  \pg{61--67}.

\bibitem[Poll(1979)]{Poll1979}
{\sc \au{Poll, D. I.~A.}} \yr{1979}  \at{Transition in the infinite swept
  attachment line boundary layer}.  \jt{Aeronautical Quarterly}  \bvol{30}~(4),
   \pg{607--629}.

\bibitem[Rosenhead(1963)]{Rosenhead1963}
{\sc \au{Rosenhead, L.}} \yr{1963} {\em Laminar Boundary Layers\/}, 1st edn.
  \publ{New York: Oxford University Press}.

\bibitem[Rowley {\em et~al.\/}(2009)Rowley, Mezić, Bagheri, Schlatter \&
  Henningson]{Rowley2009}
{\sc \au{Rowley, Clarence~W.}, \au{Mezić, Igor}, \au{Bagheri, Shervin},
  \au{Schlatter, Philipp} \& \au{Henningson, Dan~S.}} \yr{2009}  \at{Spectral
  analysis of nonlinear flows}.  \jt{Journal of Fluid Mechanics}  \bvol{641},
  \pg{115--127}.

\bibitem[Sayadi \& Schmid(2016)]{Sayadi2016}
{\sc \au{Sayadi, Taraneh} \& \au{Schmid, Peter~J.}} \yr{2016}  \at{Parallel
  data-driven decomposition algorithm for large-scale datasets: with
  application to transitional boundary layers}.  \jt{Theoretical and
  Computational Fluid Dynamics}  \bvol{30}~(5),  \pg{415--428}.

\bibitem[Schlichting \& Gersten(2017)]{Schlichting2017}
{\sc \au{Schlichting, Hermann} \& \au{Gersten, Klaus}} \yr{2017} {\em
  Boundary-Layer Theory\/}, 9th edn.  \publ{Berlin Heidelberg: Springer-Verlag
  Berlin Heidelberg}.

\bibitem[Schmid(2010)]{Schmid2010}
{\sc \au{Schmid, P.~J.}} \yr{2010}  \at{Dynamic mode decomposition of numerical
  and experimental data}.  \jt{Journal of Fluid Mechanics}  \bvol{656},
  \pg{5--28}.

\bibitem[Schmid(2022)]{Schmid2022}
{\sc \au{Schmid, Peter~J.}} \yr{2022}  \at{Dynamic mode decomposition and its
  variants}.  \jt{Annual Review of Fluid Mechanics}  \bvol{54}~(1),
  \pg{225--254}.

\bibitem[Shrestha \& Candler(2019)]{Shrestha2019}
{\sc \au{Shrestha, Prakash} \& \au{Candler, Graham~V.}} \yr{2019}  \at{Direct
  numerical simulation of high-speed transition due to roughness elements}.
  \jt{Journal of Fluid Mechanics}  \bvol{868},  \pg{762--788}.

\bibitem[Skuratov \& Fedorov(1991)]{Skuratov1991}
{\sc \au{Skuratov, A.~S.} \& \au{Fedorov, A.~V.}} \yr{1991}  \at{Supersonic
  boundary layer transition induced by roughness on the attachment line of a
  yawed cylinder}.  \jt{Fluid Dynamics}  \bvol{26}~(6),  \pg{816--822}.

\bibitem[Spalart(1988)]{Spalart1988}
{\sc \au{Spalart, P.~R.}} \yr{1988} Direct numerical study of leading-edgy
  contamination.  \bt{In {\em Proc. AGARD Symp. Fluid Dynamics of
  Three-Dimensional Turbulent Shear Flows and Transitions\/}}.  \publ{Çesme,
  Turkey: AGARD CP-438}.

\bibitem[Theofilis(1998)]{Theofilis1998}
{\sc \au{Theofilis, Vassilios}} \yr{1998}  \at{On linear and nonlinear
  instability of the incompressible swept attachment-line boundary layer}.
  \jt{Journal of Fluid Mechanics}  \bvol{355},  \pg{193--227}.

\bibitem[Theofilis {\em et~al.\/}(2003)Theofilis, Fedorov, Obrist \&
  Ch.~Dallmann]{Theofilis2003}
{\sc \au{Theofilis, Vassilios}, \au{Fedorov, Alexander}, \au{Obrist, Dominik}
  \& \au{Ch.~Dallmann, U. W.~E.}} \yr{2003}  \at{The extended gortler hammerlin
  model for linear instability of three-dimensional incompressible swept
  attachment-line boundary layer flow}.  \jt{Journal of Fluid Mechanics}
  \bvol{487},  \pg{271--313}.

\bibitem[Theofilis {\em et~al.\/}(2006)Theofilis, Fedorov \&
  Collis]{Theofilis2006}
{\sc \au{Theofilis, V.}, \au{Fedorov, A.~V.} \& \au{Collis, S.~S.}} \yr{2006}
  {\em Leading-Edge Boundary Layer Flow (Prandtl's Vision, Current Developments
  and Future Perspectives)\/}, book section Chapter 7,  \pg{pp. 73--82}.
  \publ{Springer, Dordrecht}.

\bibitem[Xi {\em et~al.\/}(2021{\natexlab{{\em a\/}}})Xi, Ren \& Fu]{Xi2021a}
{\sc \au{Xi, Youcheng}, \au{Ren, Jie} \& \au{Fu, Song}} \yr{2021{\natexlab{{\em
  a\/}}}}  \at{Hypersonic attachment-line instabilities with large sweep mach
  numbers}.  \jt{Journal of Fluid Mechanics}  \bvol{915},  \pg{A44}.

\bibitem[Xi {\em et~al.\/}(2021{\natexlab{{\em b\/}}})Xi, Ren, Wang \&
  Fu]{Xi2021b}
{\sc \au{Xi, Youcheng}, \au{Ren, Jie}, \au{Wang, Liang} \& \au{Fu, Song}}
  \yr{2021{\natexlab{{\em b\/}}}}  \at{Receptivity and stability of hypersonic
  leading-edge sweep flows around a blunt body}.  \jt{Journal of Fluid
  Mechanics}  \bvol{916},  \pg{R2}.

\bibitem[Zhong(1998)]{Zhong1998}
{\sc \au{Zhong, X.~L.}} \yr{1998}  \at{High-order finite-difference schemes for
  numerical simulation of hypersonic boundary-layer transition}.  \jt{Journal
  of Computational Physics}  \bvol{144}~(2),  \pg{662--709}.

\end{thebibliography}

\end{document}